\documentclass[11pt,a4paper]{article}

\usepackage{jcappub, natbib}

\allowdisplaybreaks

\usepackage{ifthen}
\usepackage{graphicx}% Include figure files
\usepackage{dcolumn}% Align table columns on decimal point
\usepackage{bm}% bold math
\usepackage{color}
\usepackage{amsmath}
\usepackage{amssymb}
\usepackage{bbold}

\newcommand{\mpc}{\, {\rm Mpc}}
\newcommand{\kpc}{\, {\rm kpc}}
\newcommand{\hmpc}{\, h^{-1} \mpc}

\newcommand{\kms}{\, {\rm km\, s}^{-1}}

\newcommand{\lya}{Ly$\alpha$\ }

\newcommand{\lyb}{Ly$\beta$\ }

\newcommand{\vk}{\mathbf{k}}

\newcommand{\cm}{\, {\rm cm}}

\newcommand{\msun}{\, M_\odot}

\begin{document}

\title{The large-scale cross-correlation of Damped Lyman Alpha Systems with
 the Lyman Alpha Forest: First Measurements from BOSS}

\author[a,b]{Andreu Font-Ribera}
\author[c,d]{, Jordi Miralda-Escud\'{e}}
\author[d]{, Eduard Arnau}
\author[b]{, Bill Carithers}
\author[e]{, Khee-Gan Lee}
\author[f]{, Pasquier Noterdaeme}
\author[f,g]{, Isabelle P\^aris}
\author[f]{, Patrick Petitjean}
\author[h]{, James Rich}
\author[f]{, Emmanuel Rollinde}
\author[b]{, Nicholas P. Ross}
\author[i,j]{, Donald P. Schneider}
\author[b,k]{, Martin White}
\author[l]{and Donald G. York}
%\author[]{Patrick McDonald,}
%\author[]{Uros Seljak,}
%\author[]{An\v{z}e Slosar}

\affiliation[a]{Institute of Theoretical Physics, University of Zurich, 
	8057 Zurich, Switzerland}
\affiliation[b]{Lawrence Berkeley National Laboratory, 
	University of California, Berkeley, California 94720, USA}
\affiliation[c]{Instituci\'o Catalana de Recerca i Estudis Avan\c cats,
	Catalonia}
\affiliation[d]{Institut de Ci\`{e}ncies del Cosmos (IEEC/UB),
	Barcelona, Catalonia}
\affiliation[e]{Max-Planck-Institut f\"ur Astronomie, K\"onigstuhl 17, D-69117
  	Heidelberg, Germany} 
\affiliation[f]{Universit\'e Paris 6 et CNRS, Institut d'Astrophysique
	de Paris, 98bis blvd. Arago, 75014 Paris, France}
\affiliation[g]{Departamento de Astronom\'ia, Universidad de Chile, 
	Casilla 36-D, Santiago, Chile}
\affiliation[h]{CEA, Centre de Saclay, IRFU, 91191 Gif-sur-Yvette,
  	France}
\affiliation[i]{Department of Astronomy and Astrophysics, 
 	The Pennsylvania State University, University Park, PA 16802}
\affiliation[j]{Institute for Gravitation and the Cosmos, 
	The Pennsylvania State University, University Park, PA 16802}
\affiliation[k]{Departments of Physics and Astronomy, 601 Campbell Hall,
	University of California Berkeley, CA 94720, USA}
\affiliation[l]{Deptartment of Astronomy and Astrophysics and The Fermi 
	Institute, University of Chicago, 5640 So. Ellis Ave., 
	Chicago, IL 60637, USA}

%\affiliation[g]{Physics Dept., Brookhaven National Laboratory,
%	Building 510A, Upton, NY 11973-5000, USA}
%\affiliation[h]{Institute for the Early Universe, Ewha Womans University, 
%	Seoul 120-750, S. Korea}

\emailAdd{font@physik.uzh.ch}

\abstract{ 
We present the first measurement of the large-scale cross-correlation 
of \lya forest absorption and Damped Lyman $\alpha$ systems (DLA),
using the 9th Data Release of the Baryon Oscillation 
Spectroscopic Survey (BOSS). The cross-correlation is clearly
detected on scales up to $40 \hmpc$ and is well fitted
by the linear theory prediction of the standard Cold Dark Matter model
of structure formation with the expected redshift distortions,
confirming its origin in the gravitational evolution of structure.
The amplitude of the DLA-\lya cross-correlation depends on only one free
parameter, the bias factor of the DLA systems, once the \lya forest bias
factors are known from independent \lya forest correlation measurements.
We measure the DLA bias factor to be $b_D = (2.17 \pm 0.20)
\beta_F^{0.22}$, where the \lya forest redshift distortion parameter
$\beta_F$ is expected to be above unity. This bias factor implies a
typical host halo mass for DLAs that is much larger than expected in
present DLA models, and is reproduced if the DLA cross section scales
with halo mass as $M_h^{\alpha}$, with $\alpha= 1.1 \pm 0.1$ for
$\beta_F=1$. Matching the observed DLA bias factor and rate of incidence
requires that atomic gas remains extended in massive halos over larger
areas than predicted in present simulations of galaxy formation, with
typical DLA proper sizes larger than $20 \kpc$ in host halos of masses
$\sim 10^{12} \msun$. We infer that typical galaxies at $z \simeq 2$ to
3 are surrounded by systems of atomic clouds that are much more extended
than the luminous parts of galaxies and contain $\sim 10\%$ of the
baryons in the host halo.
}


\maketitle

\section{Introduction}

%\dy{In the intro, you may want to add a few words (maybe I missed them) about 
%the notation. You seem to assume the reader knows that "halo" means DM halo. 
%But, with QSO absorption lines, and you comments about extended gas, the word 
%halo has another usage, that predates the DM halo notation.}
%\af{I've added ''dark matter'' in from of the first time we use halos.}

  Damped \lya systems, defined as absorption systems with neutral hydrogen
column density $N_{HI} \ge 2\times 10^{20}\cm^{-2}$ (hereafter referred to
as DLAs), are a powerful probe of the physical evolution of gas that
has condensed to high density to become self-shielded and atomic, and
is presumably in the process of forming galaxies. 
Spectroscopic surveys of quasars
to search for DLAs are usually performed %easiest to carry out 
at $z>2$, because the
absorption spectra can be observed from the ground at wavelengths longer
than the atmospheric cutoff. These surveys have revealed the rate of
incidence of DLAs to be $\sim 0.2$ per unit of redshift at $z=3$, slowly
increasing with redshift, and their total gas content to be 
close to $\Omega_{DLA} \simeq 10^{-3}$, or a few percent of all the
baryons in the universe
(see the review \cite{2005ARA&A..43..861W}).
This baryonic mass
of DLAs is about one third of all the mass that is contained today in
stars \cite{2001MNRAS.326..255C,2008MNRAS.388..945B}, and is comparable
to the mass of stars that had formed in the universe by $z=2$
(although the total stellar mass at this redshift is substantially
uncertain; see \cite{2011ARA&A..49..525S}, for a review).

  These observational facts suggest that DLAs are connected with the gas
clouds that are responsible for forming galaxies at high redshift.
Moreover, they imply that once the gas that has gravitationally collapsed 
into dark matter halos becomes self-shielded against ionizing radiation
and mostly atomic, it must either remain in atomic form in the DLAs for
a time comparable to the Hubble time before forming stars, or else be
ionized, expelled, and continuously replaced by other gas accreting onto
halos and recombining. The reason is that if atomic gas were to quickly
form molecular clouds and stars soon after recombining, then its
baryonic content would always be much smaller than that of stars during
the entire epoch of galaxy formation. At the same time, the total cross
section for intersecting DLAs is much larger than the fraction of the
sky covered by the starlight-emitting regions of observed galaxies.
%%\jr{Can you be more precise than "much larger"?}
Therefore, large reservoirs of atomic gas must remain orbiting in halos
for long periods of time, either covering wide areas around the
galaxies where star formation is most active, or remaining stable in
numerous low-mass halos with little star formation.

%\dy{Leaving aside assumptions, I do not think it is clear that the DLAs are in 
%the outer parts of the DM halo, or that they are very large. Normal ISM clouds 
%with these columns are only 1 pc in size and you are looking at a z range in 
%which the morphology of a galaxy could be quite different than what we see 
%today. Some clarification may be useful.}

  Despite these powerful observational constraints, the precise nature
of DLAs remains poorly known, primarily because the mass of
their host halos and the type of galaxies they are associated with has
not been observationally determined. The observed rate of incidence
tells us the product of the number density of DLA systems times their
cross section, but these two quantities are not separately known. In
reality, there can be a diverse population of objects with a wide
range of cross sections giving rise to the observed DLAs. In the context
of the Cold Dark Matter model of structure formation by hierarchical
merging of halos, we can define a mean proper cross section
$\Sigma(M,z)$ of a halo of mass $M$ at redshift $z$ for producing a DLA
in the spectrum of a background source.
The observed rate of
incidence of DLA absorbers per unit redshift, $R(z)$, is given by
\begin{equation}
 R(z)\, dz = {c (1+z)^2 \over H(z)}\,
 \int_0^\infty dM\, n(M,z) \Sigma(M,z) \, dz ~,
\label{eq:rate}
\end{equation}
where $n(M,z)\, dM$ is the comoving number density of halos of mass $M$
within the mass range $dM$, and $H(z)$ is the Hubble constant at
redshift $z$. The observational determination of any diagnostic of the
halo mass associated with specific DLAs has generally proved extremely
difficult. One of the possible avenues is to detect the galaxy associated
with a DLA absorber. Recent progress in the selection strategies  
and observational techniques has allowed for several detections 
(see
\cite{2012MNRAS.424L...1K} and references therein), 
%Krogager et al. 2012, MNRAS, 424, L1}, 
%\cite{2004A&A...422L..33M},% M{\o}ller et al. 2004, A\&A, 422, L33; 
%\cite{2010MNRAS.408.2128F},%Fynbo et al. 2010, MNRAS, 408, 2128;
%\cite{2011MNRAS.413.2481F},%Fynbo et al. 2011, MNRAS, 413, 2481;
%\cite{2012MNRAS.419....2B},%Bouch\'e et al. 2012, MNRAS 419, 2; 
%\cite{2012A&A...540A..63N},%Noterdaeme et al. 2012, A\&A, 540, A63; 
but the sample size remains very limited owing to the faintness of 
the associated galaxies and the additional difficulty involved in detecting 
them at a very small impact parameter from a bright quasar.

%\af{Do we really need 6 references for the associated galaxies detection?}
%\pn{These are basically {\bf all} detections of high-z DLA galaxies up to now, 
%so yes, I think we should cite them all.}

  Another method of characterizing the population of halos that are
hosting the DLA absorbers is through the large-scale clustering
amplitude. In the limit of large scales, any population of objects that
traces the primordial mass perturbations has a correlation function that
is proportional to the mass autocorrelation in the linear regime,
$\xi_m(r)$. Hence, the autocorrelation of DLA absorbers on large scales
in real space is $\xi_D(r) = b_D^2 \xi_m(r)$, where $b_D$ is the
{\it bias factor} of DLA absorbers. The
bias factor of halos of mass $M$, $b_h(M,z)$, can be computed by means
of approximate analytic models (e.g., \cite{1999MNRAS.308..119S}) and
accurately predicted with numerical simulations \cite{2010ApJ...724..878T}. 
The bias factor of the DLA absorbers is related to that of halos by
\begin{equation}
 b_D(z) = { \int_0^\infty dM\, n(M,z) \Sigma(M,z) b_h(M,z) \over
 \int_0^\infty dM\, n(M,z) \Sigma(M,z) } ~.
\label{eq:biash}
\end{equation}
In general, the bias factor of halos increases with their mass: halos
collapsing out of rare, high peaks, with a mass much higher than that of
a typical halo, are highly clustered \cite{1989MNRAS.237.1127C}.
Any measurement of the correlation amplitude of DLAs can be a powerful
probe of the characteristic halo mass hosting the DLAs: the higher the
value of the bias, the more massive their typical host halos need to be.

%\jr{I understand that this comes out of simulations but is there any direct 
%measurements that confirm this. 
%Perhaps the galaxy-galaxy lensing has something to say here.}
%\jm{We already give the reference to Cole and Kaiser which I think
%is right and we shouldn't overextend on this argument}

  Measuring the bias factor of DLA absorbers requires either measuring
their auto-correlation, or their cross-correlation with another tracer
population. The principal obstacle for measuring the clustering
amplitude of DLAs has been the sparseness of quasars at $z>2$ that are
bright enough to allow for spectroscopy to detect DLAs in absorption;
in addition, only $\sim 10\%$ of observed quasars yield a detected DLA.
The first measurement of
the clustering of DLAs was performed in \cite{2006ApJ...652..994C} using
the cross-correlation with luminous Lyman break galaxies, by measuring
redshifts of Lyman break galaxies identified in deep imaging of fields
around 9 quasars with 11 known DLAs in their absorption spectra. Their
result was that the bias factor of DLAs is in the range $1.3< b_D <4$.
The large uncertainty is due to the small size of their DLA sample,
which is difficult to increase because of the large observing time required
to map the area around each DLA.

  The Baryon Oscillations Spectroscopic Survey 
(BOSS,\cite{2012arXiv1208.0022D}) in the SDSS-III Collaboration 
\cite{2011AJ....142...72E} provides a new opportunity for an
accurate measurement of the clustering of DLAs. With a tenfold increase
of the number of known quasars at $z>2$ compared to previous surveys,
and a similar increase in the number of detected DLAs, we can attempt to
cross-correlate the DLAs with any other objects that are measured in the
available quasar spectra over the same redshift range: these are the
DLAs themselves, quasars, the \lya forest, and metal line absorbers.
Because of the low density of quasars, we can measure these correlations
more easily on large scales, corresponding to the typical separation
between neighboring BOSS quasars on the sky of 15', 
or about 15 $\hmpc$ (comoving), instead of
the smaller scales that are probed by deep imaging of Lyman break
galaxies in the area adjacent to targeted DLAs. Measurements on large
scales have the advantage that accurate predictions for the correlations
can be obtained from linear theory. These large-scale correlations can
also be measured for metal-line absorbers, and a first detection has
already been presented by \cite{Vikas2012}.

  In this paper, we measure the cross-correlation of DLAs with the \lya
forest transmitted flux fraction. This turns out to be the
cross-correlation that yields the most accurate measurement of the bias
factor of DLAs, because of the large number of independent \lya forest
fluctuations that are probed on every line of sight to a quasar. This
cross-correlation is proportional to the product of a bias factor for
DLAs and a bias factor for the \lya forest. The \lya forest
bias factor is independently derived from the observed autocorrelation
of the \lya forest transmitted flux \cite{2011JCAP...09..001S},
and therefore the DLA bias factor can be
robustly inferred from our measurement.
%\jr{Is it worth saying that this bias is ``measured'' by comparing the
%measured flux autocorrelation with the calculated matter autocorrelation,
%not with the measured matter autocorrelation?}
%\jm{I slightly changed the sentence to say that lya forest bias is
%derived, not observed directly, I think it's best not to go into more
%details here.}
We describe the sample of DLAs and \lya forest spectra
used for our analysis in section \ref{sec:data}. 
The detailed method for measuring the DLA-\lya forest cross-correlation is 
presented in section \ref{sec:method}, and the results for the
cross-correlation and the inferred bias factor of the DLAs are presented
in section \ref{sec:results}.
Finally the results are discussed in section \ref{sec:disc}.

%\af{In the introduction we focus on what a DLA is, but we do not talk much
%about Lyman alpha forest. It might be good to add a paragraph on what the 
%forest is and how we can use it to study large scale structure.}

Throughout this paper we use the flat $\, \rm\Lambda$CDM cosmology, with
$\Omega_m = 0.281$, $\Omega_b = 0.0462$, $h=0.71$, $n_s=0.963$ and
$\sigma_8=0.8$, similar with the best-fit parameters obtained from
the WMAP analysis in \cite{2011ApJS..192...18K}. We note that the values
of the bias factors we quote for the \lya forest and for DLAs vary in
inverse proportion to the assumed value of $\sigma_8$.

\section{Data}
\label{sec:data}

This study uses the quasar catalogue of the Data Release 9 
(DR9,\cite{2012arXiv1207.7137S}) of the Baryon Oscillation Spectroscopic 
Survey (BOSS, \cite{2012arXiv1208.0022D}), which is part of SDSS-III 
Collaboration (
\cite{2011AJ....142...72E}, % Eisenstein++ 2011, SDSS-III
\cite{Bolton2012}, % Bolton++ 2012, pipeline
\cite{1998AJ....116.3040G}, % Gunn++ 1998, SDSS camera
\cite{2006AJ....131.2332G}, % Gunn++ 2006, SDSS telescope
\cite{2012arXiv1208.2233S}, % Smee++ 2012, SDSS/BOSS spectrographs
\cite{2000AJ....120.1579Y} % York++ 2000, SDSS-I/II technical summary
).
The catalogue, described in detail in \cite{2012arXiv1210.5166P}, contains a 
total of 61931 quasars at $z>2.15$. The target selection procedure
used for identifying the quasar candidates for BOSS spectroscopy was
presented in \cite{2012ApJS..199....3R}, and uses the methods described
in \cite{2010A&A...523A..14Y}, \cite{2011ApJ...743..125K}, and
\cite{2011ApJ...729..141B}.

We now describe the sample of DLAs found in the spectra of these quasars
and the set of \lya forest spectra that we use for measuring the
DLA-Ly$\alpha$ cross-correlation.

\subsection{The DLA catalogue}
\label{sec:dlac}

Our analysis is based on a subset of the DLA catalogue of 
\cite{2012A&A...547L...1N}.
Here we briefly summarize the DLA detection method and refer the reader to
\cite{2009A&A...505.1087N} for further details.
DLAs are searched in each line of sight over a redshift range between
$z_{min}$, defined as the redshift where the spectral signal-to-noise ratio
per pixel, averaged over a $2000 \kms$ window, reaches 2, and the 
quasar redshift.
For the purpose of detecting DLAs, the quasar continuum is modeled by
fitting a modified power-law with smoothly
changing index plus Moffat profiles on top of the emission lines. 
The data is correlated with synthetic profiles of increasing column densities 
in order to detect DLA candidates and obtain a first guess of the $N_{HI}$. 
Whenever the system has associated metal lines, these are used to further 
improve the accuracy of the DLA redshift. 
Finally, the column density is obtained by fitting a Voigt-profile to 
the data.

The overall DLA catalogue contains a total of 12081 DLA candidates with
$\log N_{HI} \ge 20$. For the purpose of measuring the DLA incidence
rate and the total amount of neutral gas in the Universe, a statistical
sample optimized to achieve high completeness was defined by
\cite{2012A&A...547L...1N}, where systematic effects can be quantified
using mock data. Here, a high purity of the catalogue is our greatest
concern to measure the cross-correlation because any inclusion of
objects that are not real DLAs may systematically decrease the measured
amplitude of the cross-correlation. Completeness, on the other hand, is
less important because eliminating a fraction of the real DLAs will
increase the error of, but not systematically modify, the
cross-correlation.

  The first cut we apply is to eliminate DLAs outside the redshift range
$2.0 < z < 3.5$. The few DLAs that are outside this redshift range have very 
few nearby lines of sight with good signal-to-noise ratio in which the 
cross-correlation can be measured, so we eliminate them in order to have a 
well defined redshift interval of our systems. The standard definition of 
a DLA is an absorption system with a column density 
$N_{HI} \ge 10^{20.3}\cm^{-2}$ \cite{1986ApJS...61..249W}, so strictly
speaking, the systems we use below this column density are sub-DLAs.
We decide, however, to adopt a threshold of a factor of two below this
definition %(which is mostly historical)
because (i) systems down to $10^{20}\cm^{-2}$
are robustly identified and are not expected to sharply change their
nature with column density and (ii) our measurement of the cross-correlation 
increases in accuracy with the number of systems available.
In this paper we refer to all of the systems used
for our cross-correlation measurement, with $N_{HI} >
10^{20}\cm^{-2}$, as DLAs (however, when discussing rates of incidence and 
baryon contents of DLAs later in section \ref{sec:disc}, these quantities will
refer to systems with $N_{HI} > 10^{20.3}\cm^{-2}$). Systems of even
lower column density are less reliably detected. We shall see in section 
\ref{sec:results} that we do not detect any dependence of the clustering
properties with column density. The number of DLAs that satisfy these criteria
is 10512.

  The second cut requires that the continuum-to-noise ratio (CNR) of the
quasar spectrum is larger than three. The continuum-to-noise ratio is 
defined as the median value of the ratio of the fitted continuum in the DLA 
detection analysis in \cite{2012A&A...547L...1N} to the noise per pixel, 
over the observed \lya forest region for each quasar spectrum.
This provides a good estimate of the data quality over the region of interest, 
while being independent of the presence of an absorber.
We find this criterion to be a good threshold for ensuring the purity of our
sample without excessively reducing the number of systems.
The number of DLAs that survive this second cut is 9288.

%\jm{It would be good to be a bit more quantitative and specific here.
%What are the results of the analysis of mocks, in terms of the fraction of DLAs
%above this CNR that are still fake, and how this fake fraction increases
%if we try to go to lower CNR? I would say we want the fake fraction to
%be lower than 5\% or so, which is half of our statistical error for the bias.}.
%\pn{At CNR>3, fake should be already less than 3-4\% but keep in mind that 
%this could be tested only between Lyb and Lya emission, since mocks do not 
%contain Ly-beta forest. Note that, by "fake", I mean everything that is out 
%by 1dex or 0.05 in z. Purity does not increase much which stringer cuts. 
%Fake-detections increase at CNR<3 however.
%I will update text if you decide to cut on p-value. Can even provide a figure.}

\begin{figure}
 \begin{center}
  \includegraphics[scale=0.8, angle=-90]{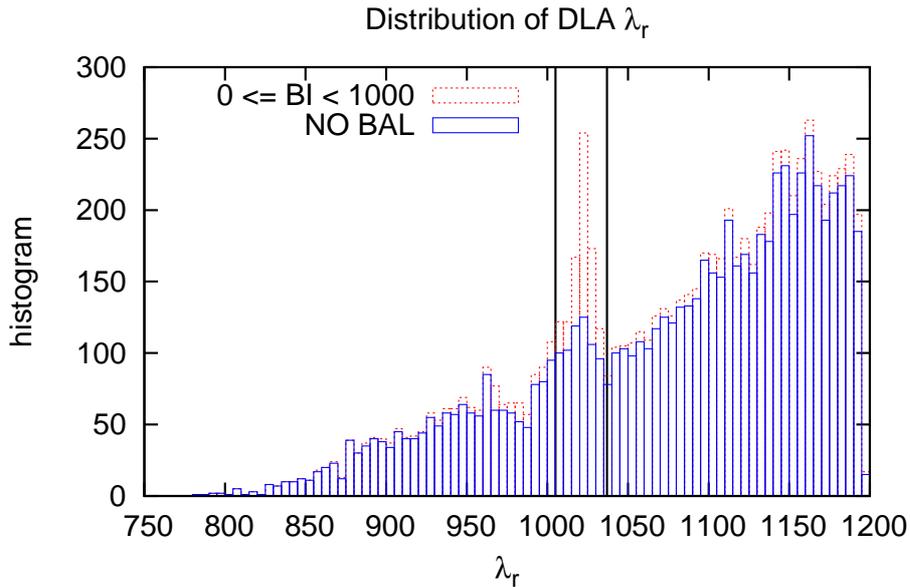}
 \end{center}
 \caption{
  Quasar rest-frame wavelength distribution of the DLAs that satisfy the 
  criteria of redshift and column density range, continuum to noise ratio
  and velocity separation from the background quasar.
  The red histogram (8189 systems) contains also quasars with BAL systems
  with a Balnicity index $BI < 1000 \kms$, 
  while the blue histogram (7458 systems) contains no systems flagged 
  as BAL.
  Vertical lines show the cut applied to our final sample, 
  $1005 {\, \rm\AA} \le \lambda_r \le 1037 {\, \rm\AA}$, reducing to 6780 
  the final number of DLAs used in this study.
 }
 \label{fig:lr}
\end{figure}

  The rest of the cuts we apply involve eliminating Broad Absorption Line 
systems (hereafter, BALs) that produce broad spectral troughs which, at the 
low signal-to-noise ratio of most of the BOSS spectra, are not easily 
distinguished from the Voigt profiles of DLAs with the superposed \lya forest.
Our third cut eliminates all DLAs found in quasars classified as BALs in the 
visual inspection of the DR9 quasars described in \cite{2012arXiv1210.5166P}, 
leaving 8469 DLAs.
In addition, in order to remove any BAL contaminants near the \lya emission 
line that may be too weak to have been identified, our fourth cut
eliminates all the DLAs that are within a velocity separation
$v < 5000 \kms$ from the quasar redshift, where
\begin{equation}
  \frac{v}{c} = \frac{z_q - z_D}{1+z_q} =
  \frac{\lambda_\alpha-\lambda_r}{\lambda_\alpha} ~,
\end{equation}
$z_q$ and $z_D$ are the redshifts of the quasar and the DLA,
$\lambda_\alpha$ is the \lya wavelength, and $\lambda_r$ is the
quasar rest-frame wavelength at which the DLA absorption line is
centered. Equivalently, this condition is $\lambda_r < 1196\, {\rm \AA}$.
Application of this constraint reduces the number of DLAs to 7458. 
We emphasize that most of the 1011 systems eliminated in this velocity range 
are probably real DLAs, but we prefer to eliminate them because they probably 
contain a substantial fraction of BAL absorbers among them.

  Figure \ref{fig:lr} shows the distribution of $\lambda_r$
for the 7458 DLAs passing all of the above cuts as the blue (solid)
histogram. The red (dotted) histogram shows the same distribution when
BALs with a Balnicity Index $BI < 1000 \kms$ (\cite{1991ApJ...373...23W}) are
included. The histograms clearly show an excess of DLAs in the interval
indicated by vertical black lines in figure \ref{fig:lr}. Moreover,
quasars with weak BALs are much more common in this interval than at
other values of $\lambda_r$. This strongly suggests that this excess is
due to BAL contamination from the \lyb and OVI absorption lines, and
that additional BALs that may be stronger in OVI than in
CIV and other detectable lines are probably lurking among the DLAs
identified in this interval of $\lambda_r$.
We therefore apply a fifth cut, eliminating all the DLAs in this
interval, with a maximum wavelength $\lambda_r < 1037 {\, \rm\AA}$
chosen equal to the longest wavelength member of the OVI doublet, and a minimum
wavelength $ \lambda_r > 1005 {\, \rm\AA}$, corresponding to a
velocity $v < 6000 \kms$ to the blue of the \lyb line. This restriction reduces
our DLA sample to 6780 systems. In section \ref{sec:results}, we shall show 
that the bias factor measured for the DLAs found in this interval is in 
fact much smaller than the ones outside, confirming a large degree of
contamination by BAL systems.

\begin{figure}
 \begin{center}
  \includegraphics[scale=0.5, angle=-90]{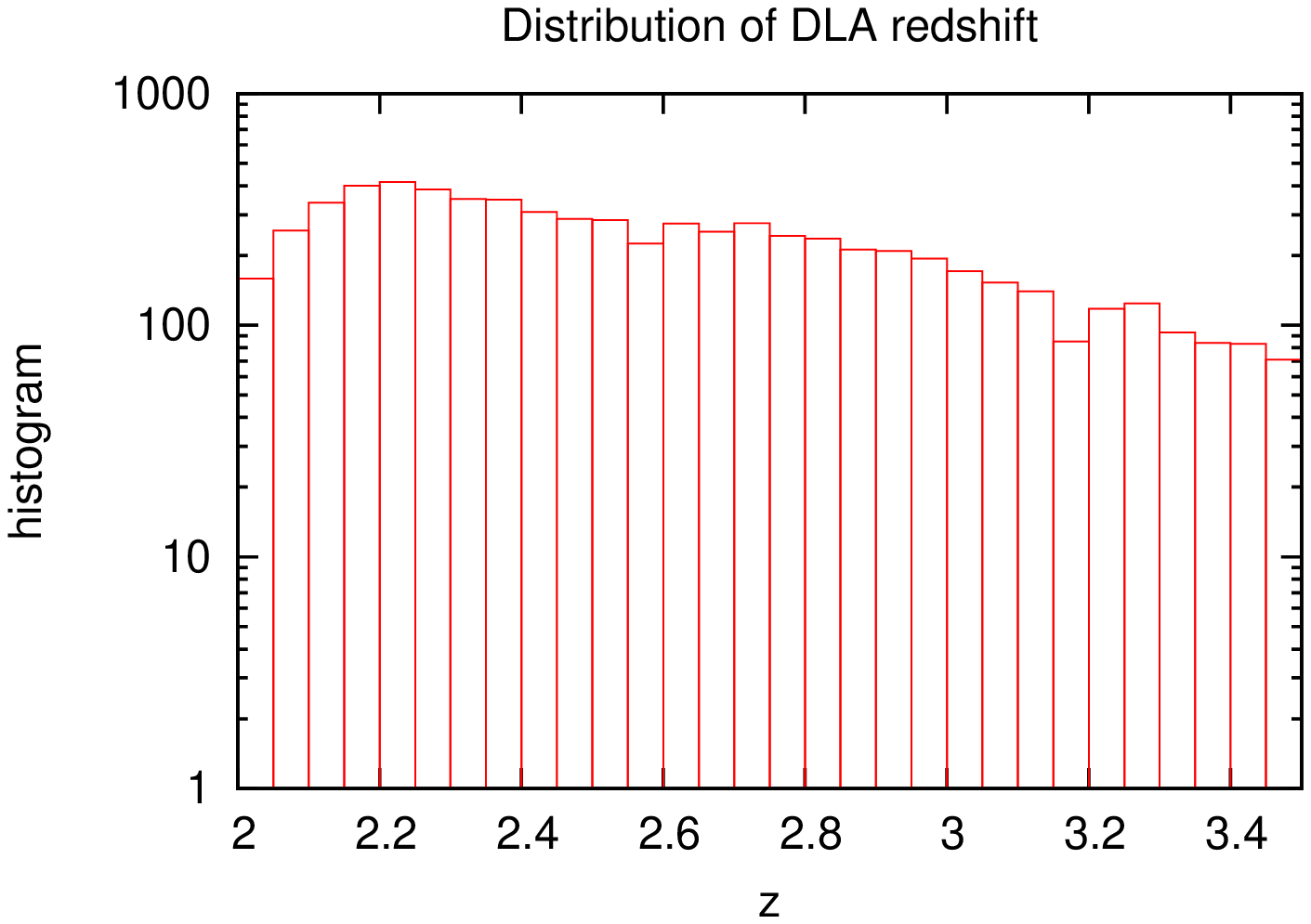}
  \includegraphics[scale=0.5, angle=-90]{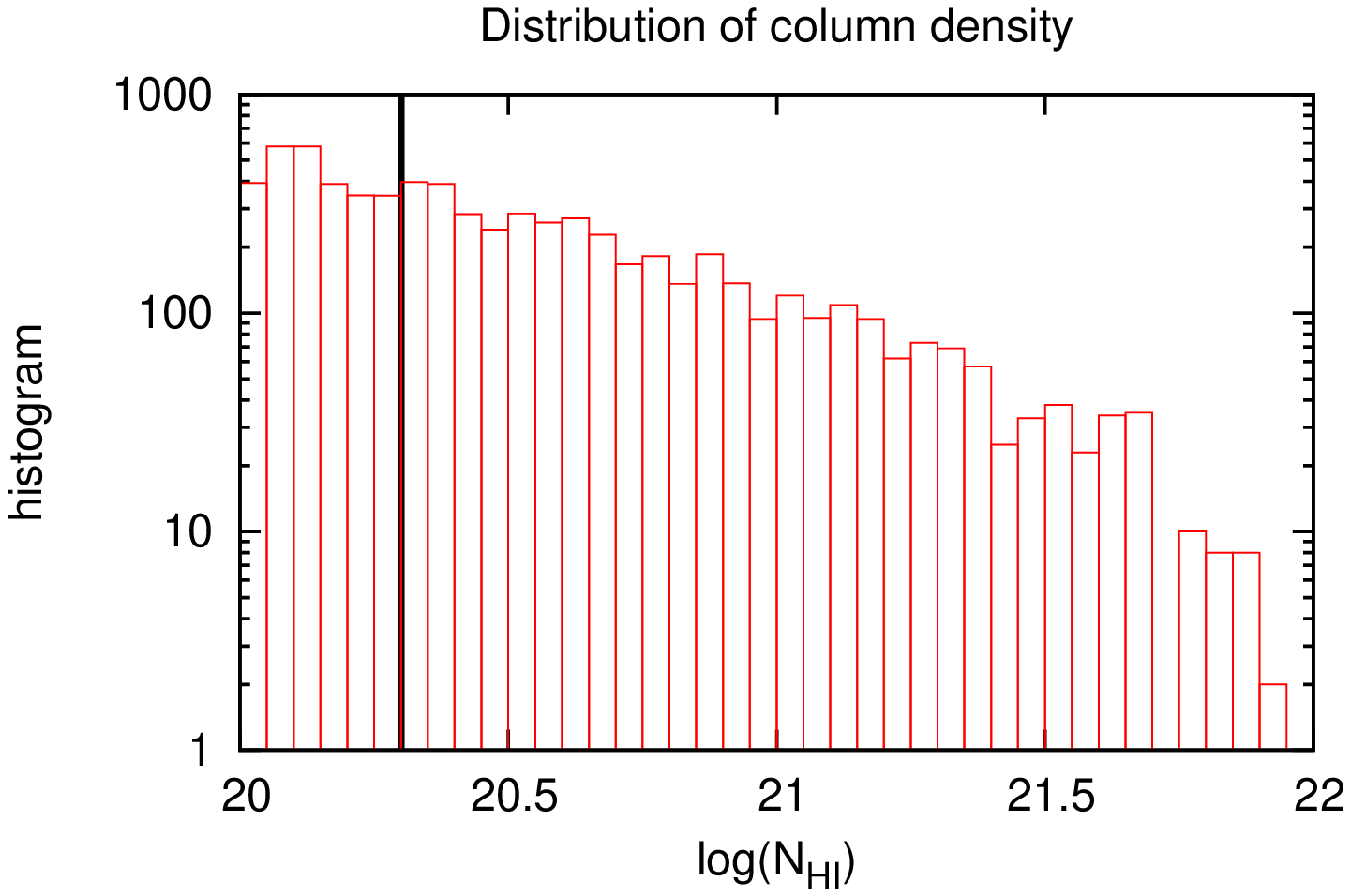}
 \end{center}
 \caption{Left panel: distribution of the 6780 DLA redshifts. Right panel:
 neutral hydrogen column density distribution.
 The vertical line on the right panel corresponds to the standard lower limit
 for DLAs, $\log N_{HI} > 20.3$.}
 \label{fig:dNdz}
\end{figure}

  The column density and redshift distributions of the final set of DLAs
selected for our study are plotted in figure \ref{fig:dNdz}. The redshift
distribution of the DLAs used for our measurement of the cross-correlation 
with the \lya forest peaks at $z\simeq 2.2$ and declines smoothly as a 
function of redshift.

%\kg{It might be worth mentioning the expected errors on the DLA absorption 
%redshift. Since Pasquier's method uses metal cross-correlations it should be 
%small, but it's worth noting that it's not a major source of error in the 
%cross-correlation measurement (e.g. unlike for QSO-LyaF)}

\subsection{\lya forest spectra}

  We select quasars in the redshift range $2.15 < z < 3.5$ for the
\lya forest spectra to be correlated with the DLA positions. The total
number of DR9 quasars in this redshift range is 58722. We eliminate
all the spectra that are identified as BAL, and we also make a
signal-to-noise ratio cut requiring a median $S/N > 0.5$ per pixel 
in the rest-frame wavelength interval $1220 {\, \rm\AA} \le \lambda_r \le 1600
{\, \rm\AA}$.
The number of spectra that are left after these cuts is $52449$.
%\af{Numbers updated, July 11TH.}

%  For most of our results, spectra that contain a DLA with
%$N_{HI}>10^{20} \cm^{-2}$ with a continuum-to-noise ratio
%${\rm CNR} > 4$ are also discarded from our \lya forest set. The number
%of quasar spectra with this cut is reduced to 48551. However, results
%that include all the spectra containing DLAs for the \lya forest
%measurements will also be shown.
%\af{We decided to keep the systems in the fiducial analysis. 
%should rewrite this. also, should explain that dla profiles are fitted by kg
%and corrected for.}

The co-added spectra released in DR9 are used, removing any pixels
that do not pass the bit mask and the sky mask %and the metal absorber mask 
as defined in \cite{Lee2012}. % metal mask was finally not implemented.
%\jm{Add one-line comments on masks here, and reference \cite{Lee2012}.}
For this study we define the \lya forest region as the pixels with a
rest-frame wavelength in the range 
$1041 {\, \rm\AA } \le \lambda_r \le 1185 {\, \rm\AA }$. 

 A total of 3047 of these lines of sight contain at least one DLA in the 
\lya forest region. For most of our results we include these lines of sight 
for our estimate of the cross-correlation of DLAs with the \lya absorption,
using the correction explained in \cite{Lee2012} for removing the DLA from
the \lya absorption, where the central region 
of the DLA is masked and the wings are corrected using a Voigt profile with 
the best estimate of the column density. 
As shown in section \ref{ss:noDLA}, rejecting these lines of sight does not 
significantly change the results.
Since DLAs are not easily identified in low S/N spectra, we expect some 
residual contamination of DLAs in our \lya sample.

%\dps{Show some DLA spectra!}

\section{Method}
\label{sec:method}

This section describes first the determination of a continuum model for
each quasar to infer the \lya absorption field. The method we use to compute
the cross-correlation of DLAs with the \lya forest and its covariance
matrix, and to fit a parameterized linear theory model to
the result, is then presented. We use the co-added quasar spectra of
DR9 selected as explained in section \ref{sec:data}, which contain
combined exposures
from a single plate after applying interpolated sky subtraction and flux
calibration corrections \cite{2012arXiv1208.0022D,Bolton2012}. 
The observed flux $f_i$ at each pixel $i$ at wavelength 
$\lambda_i = \lambda_\alpha \left(1 + z_i\right)$ (where
$\lambda_\alpha$ is the \lya wavelength and $z_i$ is the \lya absorption
redshift) is the product of the quasar continuum $C_i$ and the
transmitted flux fraction $F_i$,
plus the noise $N_i$ that we approximate as Gaussian with variance
$\left< N_i^2\right>$, as provided by the pipeline \cite{Bolton2012},
\begin{equation}
 f_i = C_i \, F_i + N_i 
     \equiv C_i \, \bar F(z_i) \left[1 + \delta_{Fi} \right] + N_i ~,
\label{eq:pflux}
\end{equation}
where $\bar F(z_i)$ is the mean value of $F$ as a function of redshift,
and $\delta_{Fi}$ is the \lya transmission perturbation. Our analysis
of the correlation will use the variable $\delta_{Fi}$, where often the
pixel subindex $i$ will be dropped for brevity.

%\jr{Why do you use $Q_i$ instead of $C_i$? Other papers (XiPush) used $C_i$.}
%\af{I agree. Even though $C$ is also the covariance matrix, I like it better
%than $Q_i$. I've changed this.}

\subsection{Continuum fitting}
\label{ss:cont}

%To extract this information we first need to model the continuum of the
%quasar spectrum $C_i$ at each pixel, i.e. the unabsorbed spectrum.
%This is a non-trivial job, specially in low resolution and low 
%signal-to-noise spectra as in BOSS. 
%However, uncorrelated errors in the continuum estimate will just introduce 
%noise to our measurement without biasing the result.

  We use the Principal Component Analysis (PCA) technique described in 
\cite{2012AJ....143...51L} for fitting a continuum to each quasar.
A set of PCA quasar templates with eight components is used to do a
least-squares fit on each quasar spectrum in the 
%Here we briefly summarize
%the method and the reader is referred to this publication for a detailed 
%description.
$1216 {\, \rm\AA} \le \lambda_r \le 1600 {\, \rm\AA}$ region to obtain
an estimate for the continuum, using equation (3) of
\cite{2012AJ....143...51L}. The total number of parameters used is 11
(eight for PCA eigenvalues, one for the flux normalization, and two for a
redshift and mean spectral slope corrections; see Table 1 of
\cite{2012AJ....143...51L}).
%\begin{equation}
%C_i = \mu_i + \sum^m_{j=1} c_j \, t_{ij} ~, 
%\end{equation}
%where $\mu_i$ is the mean quasar spectrum at pixel $i$, $t_{ij}$ is the
%$j$th template principal component (or eigenspectrum), and $c_j$ is the 
%corresponding weight derived for each individual quasar. 
%A total of $m=8$ principal components were used in this fit.
The PCA templates that are used span the wavelength interval
$1020 {\, \rm\AA} \le \lambda_r \le 1600 {\, \rm\AA}$, so a
predicted continuum in the Lyman $\alpha$ forest region can then be
obtained by extrapolating the fit to $\lambda_r < 1216 {\, \rm\AA}$.

This procedure is repeated for two sets of quasar templates. The first
was generated by \cite{2005ApJ...618..592S}
using HST ultraviolet spectra of 50 low-redshift quasars, where the
\lya forest continuum is easily estimated because of the low mean
absorption. The second was generated by \cite{2011A&A...530A..50P}
from 78 SDSS DR7 (\cite{2009ApJS..182..543A}) 
quasars with high signal-to-noise ratio, where the
continua were fitted with a low-order spline function.
The use of these different templates ensures full coverage of the
luminosity range spanned by the BOSS quasars.
%\dy{Do you mean "absolute luminosity range"? If so, specify the range of 
%your sample and the range of the two templates.}
The best of the two fits obtained from these two template sets is then
selected, which we designate as $C_i^{PCA}$. For the
DR9 quasars we use, the set of templates from \cite{2005ApJ...618..592S}
turns out to provide the best fit for $\sim$ 85\% of the quasars.

  An additional step was applied in \cite{2012AJ....143...51L},
referred to as {\it mean flux regulation}, where each quasar
continuum was multiplied by a second order polynomial with two
parameters that were fitted to match the mean flux evolution from
\cite{2008ApJ...681..831F}. This mean flux regulation substantially
reduces the variance of the \lya absorption field,
owing to the removal of spectrophotometric errors and of any quasar
intrinsic spectral diversity %variability 
that is not accounted for by the PCA
templates. However, large-scale power in the \lya forest is also
suppressed by this process in a way that is complex and difficult to
model. For this reason, we do not apply the mean flux regulation
procedure here. Instead, a more simple {\it mean transmission correction}
(MTC) is applied that is described below in section \ref{sec:mfc}, which
affects the measured cross-correlation with DLAs in a way that is easier
to correct for.

%\dps{Show some examples of the fits!}

\subsection{Mean transmitted fraction}
\label{ss:meanf}

  For the purpose of measuring the cross-correlation of the \lya forest
with DLAs, it is important to ensure that the average value of the 
transmission perturbation $\delta_F$ at each redshift is precisely zero.
If this condition is not imposed, the value of the cross-correlation will 
differ from zero in the limit of large scales simply because an incorrect 
value of $\bar F$ is being used. 

For this reason, we measure the mean transmitted fraction in 50
redshift bins of $\Delta z=0.03$ linearly spaced in redshift between
$z=1.9$ and $z=3.4$, by computing the weighted average of the flux in
all the pixels in each redshift bin $k$, centered at redshift $z'_k$,
\begin{equation}
 \bar F^{PCA}(z'_k) = \frac{ \sum_{i\in k} w^\prime_i \, f_i/C^{PCA}_i }
 {\sum_{i\in k} w^\prime_i} ~,
 \label{eq:meanflux}
\end{equation}
where the sums are done over all the pixels $i$ with a redshift $z_i$
within the redshift bin $k$.
The weights are equal to the total inverse variance of $f_i/C_i^{PCA}$,
\begin{equation}
 \label{eq:w_F}
  w^\prime_i  = \left[ \bar F_e (z_i)^2 \, \sigma_F^2(z_i) 
           + \frac{\left< N_i \right>^2}{ (C^{PCA}_i)^2 }  \right]^{-1} ~,
\end{equation}
where $\sigma_F^2=\left< \delta_F^2 \right>$ is the intrinsic variance
of the \lya forest, and $\bar F_e(z)$ is an externally determined value
of the mean transmitted fraction as a function of redshift, for which we
use the result of \cite{2008ApJ...681..831F}. Because these weights do
not need to be obtained to very high accuracy, we do not iterate
equations (\ref{eq:meanflux}) and (\ref{eq:w_F}) to calculate the
weights using the values of $\bar F^{PCA}$ instead of $\bar F_e$. For
the intrinsic variance as a function of redshift, we use the simple
expression
\begin{equation}
 \label{eq:intrinsic}
 \sigma^2_F(z) = 0.065 \left[(1+z)/3.25\right]^{3.8}~,
\end{equation}
based on the redshift evolution of the power spectrum measured in 
\cite{2006ApJS..163...80M}.

\begin{figure}[h!]
 \begin{center}
  \includegraphics[scale=0.8, angle=-90]{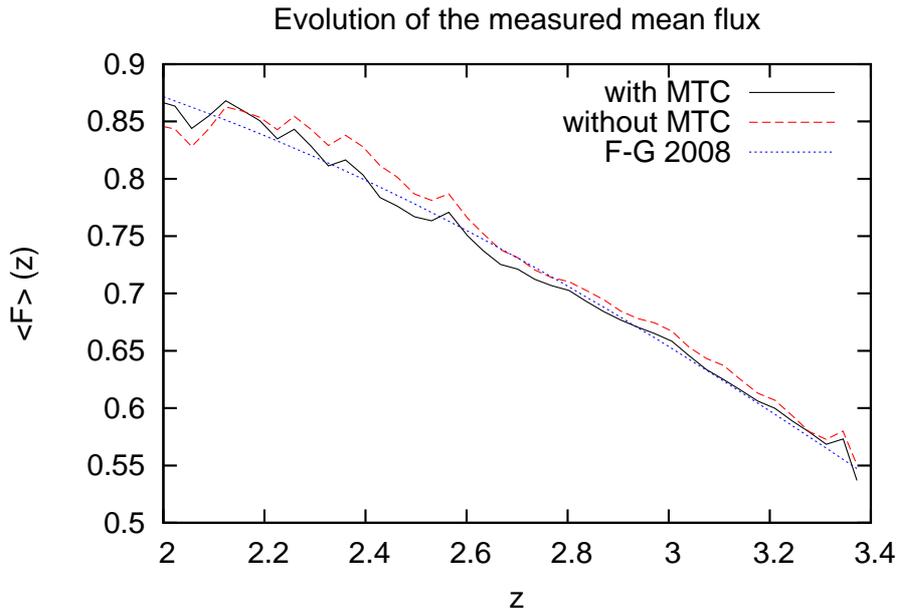}
 \end{center}
 \caption{Measured mean transmission as a function of redshift, 
  with (solid black line) and without (dashed red line) the Mean
  Transmission Correction, compared to the measurement of
  \cite{2008ApJ...681..831F} (blue dotted line).
 }
 \label{fig:mf}
\end{figure}

  Figure \ref{fig:mf} shows the mean transmitted
fraction $\bar F^{PCA}$ (red dashed line) compared to the result
of \cite{2008ApJ...681..831F} (blue dotted line). Our measurement is
very close to that of \cite{2008ApJ...681..831F}, and is typically
higher by only $\sim$ 1\%. In addition, there are sharp features in the
inferred $\bar F(z)$: the bump at $z=2.6$, and several other rapid
fluctuations at lower redshift. These features are a systematic error
that arises from the calibration of the spectra using F stars which
are particularly important in the Balmer stellar absorption lines.
Each observed plate in BOSS has a set of calibrations stars that
are used to translate photon counts in the CCD to flux units
\cite{2012arXiv1208.0022D}. The spectra of
these stars are masked near the Balmer lines, and this introduces an artifact
in the calibration vector. A similar artifact was present in earlier SDSS
data (\cite{2008ApJS..175..297A}), but the effect seems to be larger in BOSS,
as noted by \cite{Busca2012}. This systematic is being studied and will
be corrected in the future.

  The small difference between our measured $\bar F(z)$ and that in
\cite{2008ApJ...681..831F} may also be caused by errors in the
zero-flux level computed by interpolating from neighboring fibers 
used for sky subtraction (\cite{2012arXiv1210.5166P}), or to other
calibration systematic effects (\cite{Bolton2012}), but these errors
are not a problem for our study of the cross-correlation as long as
they are not correlated in any way with the presence of DLAs on nearby
lines of sight.
%The measured amplitude of the transmission fluctuation
%$\delta F$ depends only on the product of the continuum and the mean
%transmission fraction, which is not affected by any multiplicative error
%that arises from flux calibration. The zero flux level error is an
%additive one, and it affects the spectra in different ways depending on
%the quasar brightness, but it should not bias our correlation
%measurements provided its amplitude is not correlated in any way with
The black solid line in Figure \ref{fig:mf} is $\bar F(z)$ in our
redshift bins after the MTC is applied, which we describe next.

\subsection{Mean transmission correction}
\label{sec:mfc}

  The same 
%flux calibration, zero flux level, and continuum fitting
errors that can produce a difference in
the mean transmission $\bar F^{PCA}$ and the value $\bar F_e$ from 
\cite{2008ApJ...681..831F} imply the presence of stochastic errors 
in any single quasar spectrum that can systematically bias our estimated
value of $\delta_F$ on large scales along the line of sight. This may
substantially increase the noise in our cross-correlation measurement.
As a test of this effect, we measure the variance of the weighted
average of $\delta_F$ over the whole \lya spectrum of a single quasar,
which we find to be on average equal to $0.03$.

%\af{I test the effect of changing the minimum redshift of quasars and
%is negligible. Quasars at very low redshift have shorter forest, but they
%are very noisy so they do not contribute much to the average. 
%Even for $z_q > 2.5$ I get the same result.}
%
%\jm{I still don't understand. Quasars at low redshift with noisy spectra
%have a very high variance of their averaged $\delta_F$, so they will
%increase this variance, what do you mean that they do not contribute
%much to the average of this variance? Are you somehow weighting them down?} 
%
%\af{I compute what is the total weight of a given line of sight, but adding 
%the weight of each of these pixels. This total weight is what tells us how 
%important is this spectrum for the cross-correlation measurement. That's why 
%I weight the variance in each spectrum by its total weight, since that's also
%whay I do in the cross-correlation measurement. Low z quasars have less pixels
%and each of them has less noise, so the variance measured in these spectra do
%not contribute much to the weight-averaged variance.}

The expected variance from the \lya forest alone depends on the quasar redshift,
both because of the different comoving length of the forest and because of the
evolution of the line of sight power with redshift. However, for $z<2.8$ the 
expected variance is always below $ 0.001$.
Even though metal lines and Lyman limit systems should increase this variance, 
they cannot account for a value as large as $0.03$, which we infer mostly 
reflects the random variation from quasar to quasar of the 
systematic miscalibration. A variance of $0.03$ on the mean of $\delta_F$ 
over a spectrum can be induced by an rms variation of the continuum amplitude
of $\sim 17 \%$. 

%\kg{I'm pretty sure the additional variance is from continuum fitting errors 
%and not from 'systematic miscalibrations'. You seem to imply the same thing 
%in the last sentence anyway.}
%\af{Well, they are continuum fitting errors probably caused by systematic 
%miscalibrations.}

%In the right panel of Figure \ref{fig:mf} we also plot the variance of 
%$\delta_F$ at a pixel level. At this level the effect of coherent errors in 
%the continuum fit is smaller, but since they are correlated along the 
%whole spectrum they become dominant when the whole line of sight is averaged.

\begin{figure}[h!]
 \begin{center}
  \includegraphics[scale=0.8, angle=-90]{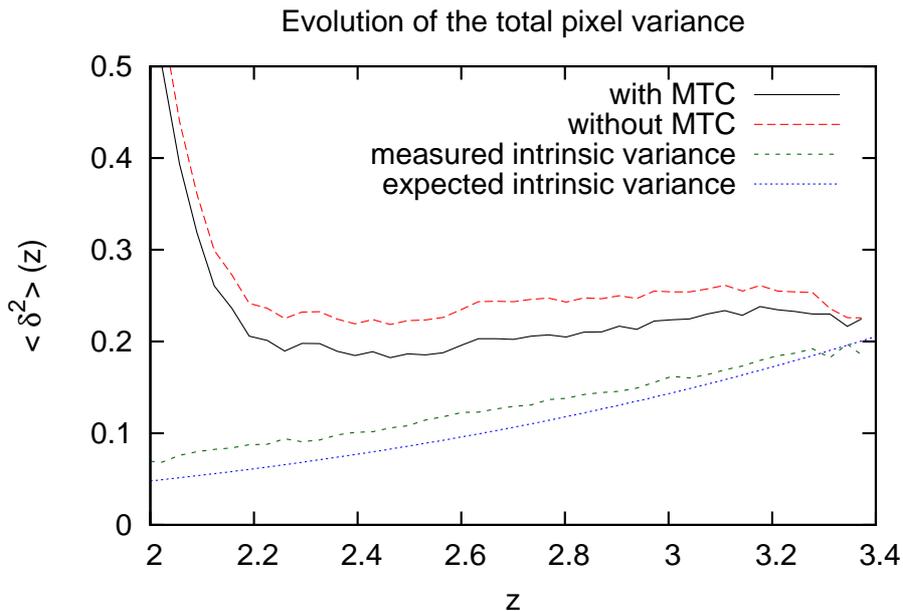}
 \end{center}
 \caption{
  Measured variance per pixel with MTC (solid black line) and without 
  (long-dashed red line). The inferred intrinsic variance using the MTC 
  assuming the corrected noise variance provided by the pipeline is shown 
  by the short-dashed green line, and the intrinsic variance assumed for 
  the weights [eq. (\ref{eq:intrinsic})] is the dotted blue line. }
 \label{fig:dfd}
\end{figure}

  With the aim of eliminating this additional source of noise, we apply
a {\it Mean Transmission Correction} (MTC) to the PCA continuum fit,
which we compute for each quasar spectrum as
\begin{equation}
 A_q = \frac{\sum_{i\in q} \, w^\prime_i \, f_i /
  \left[ C_i^{PCA} \bar F_e(z_i) \right] } {\sum_{i\in q} \, w^\prime_i} ~,
\label{eq:aqfac}
\end{equation}
where the sums are performed over all pixels $i$ in the \lya forest
region (as defined in section 2) of a quasar $q$, and the weights are
computed with equation (\ref{eq:w_F}). The corrected quasar continuum is
\begin{equation}
 C_i \equiv C_i^{PCA} A_q ~,
\end{equation}
and the corrected transmission fraction at each pixel is
\begin{equation}
 F_i \equiv {f_i \over C_i} = {f_i \over C_i^{PCA}  A_q} ~. 
\label{eq:fimtc}
\end{equation}
The corrected average mean transmission
$\bar F(z^\prime_k)$, computed
as in equation (\ref{eq:meanflux}) but with the corrected continua,
is now even closer to $\bar F_e(z)$, as shown in Figure \ref{fig:mf}. 
%The average of the correction factor $A_q$ over
%all the quasars results in the reduction of a few percent in the mean
%transmission.
The correction also reduces the variance of $\delta_F$,
even at the pixel level. This effect is shown in Figure
\ref{fig:dfd}, where the variance before the MTC correction is shown
as the dashed red line, and after the MTC correction as the black
solid line. When the contribution to the variance from the noise
provided by the BOSS pipeline, after we apply a wavelength-dependent correction
as explained in \cite{Lee2012}, is subtracted from this MTC-corrected
variance, the short-dash green line is obtained. This line should correspond
to the intrinsic variance of the \lya forest. The value of the variance
we obtain in this way is still higher than the expected intrinsic
variance from equation (\ref{eq:intrinsic}), shown as the dotted blue line,
but the difference is small and is probably
explained by residual errors in the continuum fit, metal lines and Lyman
limit systems.
%This result is probably due to a wavelength-dependent underestimate of the
%observational noise in the BOSS pipeline that is more severe at
%short wavelengths (and therefore has a stronger effect on the intrinsic
%variance estimate at low redshift), which was found also in
%\cite{2011JCAP...09..001S} and is discussed in more detail in (\cite{Lee2012}).

%\kg{I think the differences between the green-dashed and blue-dotted curves 
%can mostly be accounted for by the remaining continuum error, which the MTC 
%reduces but doesn't completely eliminate. My mean-flux regulation continua 
%has RMS errors of 3-4\%, and I'm guessing MTC will be a bit larger, 
%maybe 5-7\%. This means the contribution from the continuum error will be 
%~0.03-0.05 which will account for most of the residual variance. 
%The rise at z<2.2 is clearly due to instrumental systematics, however...}
%\af{With the new figure, once the noise correction has been fixed, it looks
%more as you say. I added this in the text.}

  Our estimator for the fluctuation in the transmitted flux fraction,
$\delta_F$, is obtained at every pixel as
\begin{equation}
 \delta_{Fi} = \frac{f_i}{C_i \bar F(z_i)} - 1 ~,
\end{equation}
where $\bar F(z_i)$ is the mean transmission that has been obtained in
the redshift bin which contains the pixel redshift $z_i$. By
construction, the average of $\delta_{Fi}$ at a given redshift bin is
zero. Because $\bar F(z_i)$ is close to $\bar F_e(z_i)$, the
average of $\delta_{Fi}$ over each individual quasar spectrum is also
nearly zero. The MTC is therefore approximately the same operation as
forcing the weighted average of $\delta_F$ to be zero on each individual
quasar spectrum by subtracting its mean value.

  Subtracting the mean value of $\delta_F$ in each line of sight removes
some of the large-scale power of the \lya transmission field. This
subtraction was also applied in \cite{2011JCAP...09..001S} for measuring
the \lya transmission autocorrelation, and an analytical expression to
correct for its effect on the correlation function was presented in
their Appendix A. We compute an equivalent expression of this correction
to the measured cross-correlation of DLAs and the \lya forest in our 
Appendix \ref{app:correc}.

  Unless otherwise specified, our results will incorporate the MTC, and 
the DLA bias parameter will be fitted including the MTC correction of
our theoretical models that is derived in Appendix \ref{app:correc}.
However, results without including the MTC will also be presented and
discussed.

\subsection{Estimator and covariance matrix of the cross-correlation}
\label{ss:cov}

  The simplest method is used in this paper to obtain an estimator for
the cross-correlation of DLAs and the \lya forest, $\hat\xi_A$: we
compute the weighted
average of $\delta_F$ over all the pixels located at a separation from
a DLA that is within a bin $A$ of $\mathbf{r}_A$,
%within a bin $A$ of the
%parallel and perpendicular components of the separation, and then
%we average the result over all the DLAs,
%(as a function of both the parallel and perpendicular components
%We can estimate the cross-correlation in a bin
%with a weighted average of $\delta_F$ in all pixels at a separation from 
%a DLA that lies inside the bin:
\begin{equation}
 \hat\xi_{A} = \frac{\sum_{i \in A}  w_i\, \delta_{Fi}}{\sum_{i \in A} w_i} ~.
 \label{eq:xiA}
\end{equation}
The summation symbols in this equation indicate a double sum: first over all
the DLAs, and then over all the pixels at a separation within the $A$
bin from the DLA. Consequently, the value of $w_i\, \delta_{Fi}$ of one
pixel appears repeated several times in the sum whenever the pixel is
within the separation bin $A$ from several DLAs. The DLAs are all
weighted equally, and the \lya pixels are weighted as the inverse of
their total variance, including the noise contribution and the intrinsic
variance: 

%\jm{please check this, the previous equality is no longer true}:
%\af{Well, it is true that each DLA gets a weight 1 for each pair DLA-pixel. 
%The number of pixels that are correlated with a given DLA is a different q
%uestion, but you could also complain then that the weight that every \lya 
%pixel gets is proportional to the number of DLAs that are at a given distance.}
%\jm{I just wanted to say here that you should check that this is the actual
%expression for the weights that you are really using in the calculation. If
%not it should be corrected. I think it has changed slightly from what was
%previously here because of the specification of what we mean by $Q_i$ and
%$F_i$ in equations 3.6 and 3.7}
%\af{I see. Actually this is the correct form, but using $\bar F_e(z)$ instead
%of $\bar F(z)$, but since they are almost identical I wouldn't bother changing
%it}
\begin{equation}
 \label{eq:w_delta}
  w_i
% = w^\prime_i \, \bar F (z_i)^2 
      = \left[ \sigma_F^2(z_i) + \dfrac{\left< N_i^2\right>}
              { C_i^2 \, \bar F_e^2 (z_i) }  \right]^{-1} ~.
\end{equation}
The method is similar to the one used by
\cite{2011JCAP...09..001S} for measuring the \lya forest transmission
autocorrelation. There are a number of assumptions and simplifications
involved in the use of the simple equation (\ref{eq:xiA}) for the
estimator of the cross-correlation:
\begin{enumerate}
 \item The correlations among the values of $\delta_F$ in neighboring
pixels (both along a single line of sight and along nearby, parallel
lines of sight) can be neglected for the purpose of optimizing the
weights. We take into account these correlations below for the purpose
of computing the covariance matrix of the
cross-correlation, but we neglect them for obtaining an optimal
estimator by choosing the simple weights in equation (\ref{eq:w_delta}).
 \item We assume that the DLAs can be detected with a
probability that is independent of their large-scale environment, and
therefore independent of the mean \lya forest absorption that is
superposed on their damped wings.
\end{enumerate}
Neither of these two assumptions is precisely correct. Failure of the first
one makes our measurement suboptimal, but does not bias it in any way.
The optimal manner to estimate the cross-correlation is discussed in
Appendix \ref{app:estim}, where the assumptions involved in
obtaining the simplified estimator in equation (\ref{eq:xiA}) are
analyzed and discussed in more detail. The second assumption is
required to avoid a systematic bias of this estimator: if DLAs are more
likely to be detected when the forest absorption is higher, then the
cross-correlation is artificially enhanced. We believe this second effect
is small enough to be neglected, because tests have shown that the
cross-correlation amplitude is not affected above the continuum-to-noise
ratio that we impose on our sample (section \ref{sec:dlac}).
We nevertheless plan to make a correction for this effect in the future
by testing how the probability of DLA detection with a specific
algorithm is modified by the superposition of the \lya forest on the
damped wings, using mock \lya spectra that include DLAs with a biased
distribution relative to the \lya forest \cite{2012JCAP...07..028F}. 

  The measurement of the cross-correlation with equation (\ref{eq:xiA})
is done using 16 bins in the radial separation
$\pi$, and 8 bins in the transverse separation $\sigma$. To
test for the symmetry of the cross-correlation under a sign change of
$\pi$, we use different bins for negative and positive $\pi$, where
$\pi$ is positive when the pixel of the \lya forest transmission
perturbation is at higher redshift than the DLA. The bins in $\pi$ are,
in units of comoving $\hmpc$, bounded by the values ($-60$, $-40$,
$-30$, $-20$, $-15$, $-10$, $-6$, $-3$, $0$) and the same positive
values, while the bins in $\sigma$ are bounded, in the same units, by
($1$, $4$, $7$, $10$, $15$, $20$, $30$, $40$, $60$). Pairs at
transverse separation $\sigma < 1 \hmpc$ are not used (the number of
these pairs is negligibly small in our sample). This results in a total
of 128 separation bins in $\sigma$ and $\pi$.
The weighted average values of $(\pi,\sigma)$ of all the contributing
pixel-DLA pairs to every bin are stored, with the same weights as in
equation (\ref{eq:w_delta} (these are generally close but not
exactly equal to the central values of each bin). The measurement is
generally done using a single bin of the mean redshift $z$ of the \lya
forest pixel and the DLA, which is required to be in the range
$2.0 < z < 3.5$, although we also present some results in the next
section using three redshift bins.

%If the bins were infinitely thin (so $\mathbf{r_i} = \mathbf{r_A}$ for 
%all pixels in the bin) it is easy to show that the estimate is unbiased,
%\begin{equation}
% \left< \hat{\xi_{A}} \right> = \frac{\sum_{i \in A}  w_i 
%      ~ \left< \delta_{Fi} \right>_g} {\sum_{i \in A}  w_i} 
%      = \xi_{DF}(\mathbf{r_A}) ~.
%\end{equation}

The covariance of the cross-correlation measured with equation
(\ref{eq:xiA}) in two bins $A$ and $B$ is equal to
\begin{equation}
 \tilde C_{AB} \equiv \left< \hat\xi_{A} \hat\xi_{B} \right> -
 \left< \hat\xi_{A} \right> \left< \hat\xi_{B} \right> 
%  = \frac{\sum_{i \in A} \sum_{j \in B}  w_i ~ w_j ~
%       \left< \delta_{Fi} ~ \delta_{Fj} \right>}
%         {\sum_{i \in A} w_i ~ \sum_{j \in B}  w_j}
   = \frac{\sum_{i \in A} \sum_{j \in B}  w_i ~ w_j ~ C_{ij}}
          {\sum_{i \in A} w_i ~ \sum_{j \in B}  w_j} ~,
 %- \left< \hat\xi_{A} \right> \left< \hat\xi_{B} \right> ~,
 \label{eq:covar}
\end{equation}
where $C_{ij} = \left< \delta_{Fi}\delta_{Fj}\right>$ is the correlation
of the measured values of $\delta_F$
in pixels $i$ and $j$, separated in redshift space by $\mathbf{r}_{ij}$.
Note that the correlation $C_{ij}$, with two
%covariance matrix of the cross-correlation,
subindexes for the two correlated pixels, should not be confused with
the quasar continuum which always has one subindex, and the covariance
matrix of the cross-correlation, $\tilde C$, is a different matrix with
indexes referring to bins in $(\pi,\sigma)$.
There are three main contributions to the correlation $C_{ij}$: 
the intrinsic autocorrelation of the \lya forest at a given separation 
$\xi_F(\mathbf{r}_{ij})$,
the noise term $\left< N_i^2\right>/(C_i^2 \bar F_i^2)$ that we assume to be 
uncorrelated among different pixels, and
continuum fitting errors with a correlation $\xi_c(r_{ij})$ that affect only 
pairs of pixels in the same spectrum,
\begin{equation}
  C_{ij} = \xi_F(\mathbf{r}_{ij}) 
           + \frac{\left< N_i^2\right>}
			{\left(C_i \, \bar F_i \right)^2} \, \delta^K_{ij}
           + \xi_c(r_{ij})\, \delta^D(\sigma_{ij}) ~ .
 \label{eq:cij}
\end{equation}
Here, $\sigma_{ij}$ is the perpendicular component of $\mathbf{r}_{ij}$,
and the Dirac delta function $\delta^D$ indicates that the last term is
non-zero for pixels on the same quasar spectrum only, as opposed to the 
Kronecker function $\delta^K_{ij}$ in the second term that indicates that the 
noise term is only present when both pixels are the same.

%\jr{This is not (technically) a Dirac delta function. To be correct you should 
%give an index for the quasar and use the Kronecker delta.}
%\kg{You did not define $\delta_{ij}^K$}
%\jr{$\langle N_i^2 \rangle$:  is this the same as $\sigma_{Ni}^2$?}
%\af{Actually, the term should be divided by the $C_i^2 \bar F_i^2$. Should fix 
%this.}
%\af{I've improved this text.}

%\jr{The correlations are not just due to ``fitting errors'' but also to the 
%spectrometer PSF and to the mean-flux-subtraction (makes $xi_c$ negative for 
%large separations).}
%\af{Sure, but these should be sub-dominant and we do not include them. 
%We could add a comment on this.}

  In general, a joint analysis of the \lya auto-correlation and a
cross-correlation may be done, and the measured correlation $\xi_F$ can
then be used to compute the cross-correlation covariance. Here, we use
instead a theoretical correlation function computed as the Fourier transform
of the \lya power spectrum measured in \cite{MCDON03} from numerical 
simulations, with modified bias parameters in agreement with the measurement 
of the correlation function measured with the first year of BOSS data in 
\cite{2011JCAP...09..001S}.
%\jr{I guess you ignore terms due to correlations of the DLAs.  
%  Might explain this  (they are widely separated?)}
%\af{Yes, we should add a line on this.}

%the theoretical linear correlation function of the \lya forest
%in redshift space, using the equations of \cite{HAMIL92} for the
%$\Lambda$CDM model with $\Omega_m=0.281$ and $\sigma_8 = 0.8$, 
%\jm{Is this good now?}
%\af{Yes!}

We use a \lya density bias parameter $b_F=-0.168$ and a redshift distortion
parameter $\beta_F=1$ at $z=2.25$, and we include the non-linear term
$D(k,\mu_k)$ with the parameter values of the fiducial model in the
first row of Table 2 of \cite{MCDON03}. These values are
consistent with the results of \cite{2011JCAP...09..001S}. The
amplitude of the correlation function is assumed to be proportional to
$(1+z)^\alpha$, with $\alpha=3.8$, as found from the 1D power spectrum
measurement in \cite{2006ApJS..163...80M}. We expect most of the
correlated errors in the continuum fitting to be removed by the MTC
procedure, so we generally do not include the term $\xi_c$ in equation
(\ref{eq:cij}) except when the MTC is omitted. In the latter case,
a constant contribution $\xi_c = 0.03$, equal to the measured variance
of the value of $\delta_F$ averaged over one spectrum, is added to
$C_{ij}$ for all pixel pairs on a common line of sight.

  The calculation of the cross-correlation covariance using equation
(\ref{eq:covar}) can be time consuming because of the need to evaluate
$\xi_F$ for all pixel pairs with separations within the $A$ and $B$ bins
from every pair of DLAs. Each sum in the numerator of equation
(\ref{eq:covar}) is over every DLA, and over every pixel within the
separation bin. To increase the speed of the calculation, we include only 
pixel pairs with a perpendicular separation smaller than $\sigma_{ij}=20 \hmpc$,
a similar approximation to the one used in \cite{2011JCAP...09..001S} for
computing the covariance matrix. This approach ensures the inclusion of the most
important correlations of our \lya spectra in the covariance matrix.

%\jr{The change from talking about the covariance matrix to talking about the 
%cross-correlation is a bit abrupt. Perhaps you should begin a subsection here.}
%\af{I moved these paragraphs before the covariance.}

%This can perhaps go to section 4:
%Measuring any variation of the cross-correlation amplitude with redshift
%is difficult because our DLAs and \lya forest pixels are distributed
%over a narrow range of redshift, peaking near $z=2.3$. The
%cross-correlation variation is also not expected to be large: the \lya
%density bias evolves as $(1+z)^{\alpha/2+1} = (1+z)^{2.9}$, and the
%linear growth factor is approximately proportional to $(1+z)^{-1}$, so
%for a constant DLA bias the cross-correlation amplitude would evolve as
%$(1+z)^{\alpha/2-1} = (1+z)^{0.9}$.

\subsection{Sub-samples and bootstrap errors}
\label{ss:boot}

  The errors obtained on the measured cross-correlation using
the covariance matrix computed with equations (\ref{eq:covar}) and
(\ref{eq:cij}) rely on the accuracy of this calculation, which may be
affected by the assumed \lya autocorrelation and the neglect of
continuum fitting errors and other possible systematics (such as
correlated sky subtraction errors). It is therefore useful to compute
an alternative set of bootstrap errors by dividing our quasar sample
into a number of sub-samples and testing the variation of the obtained
cross-correlations among the sub-samples.

\begin{figure}[h!]
 \begin{center}
  \includegraphics[scale=0.4, angle=-90]{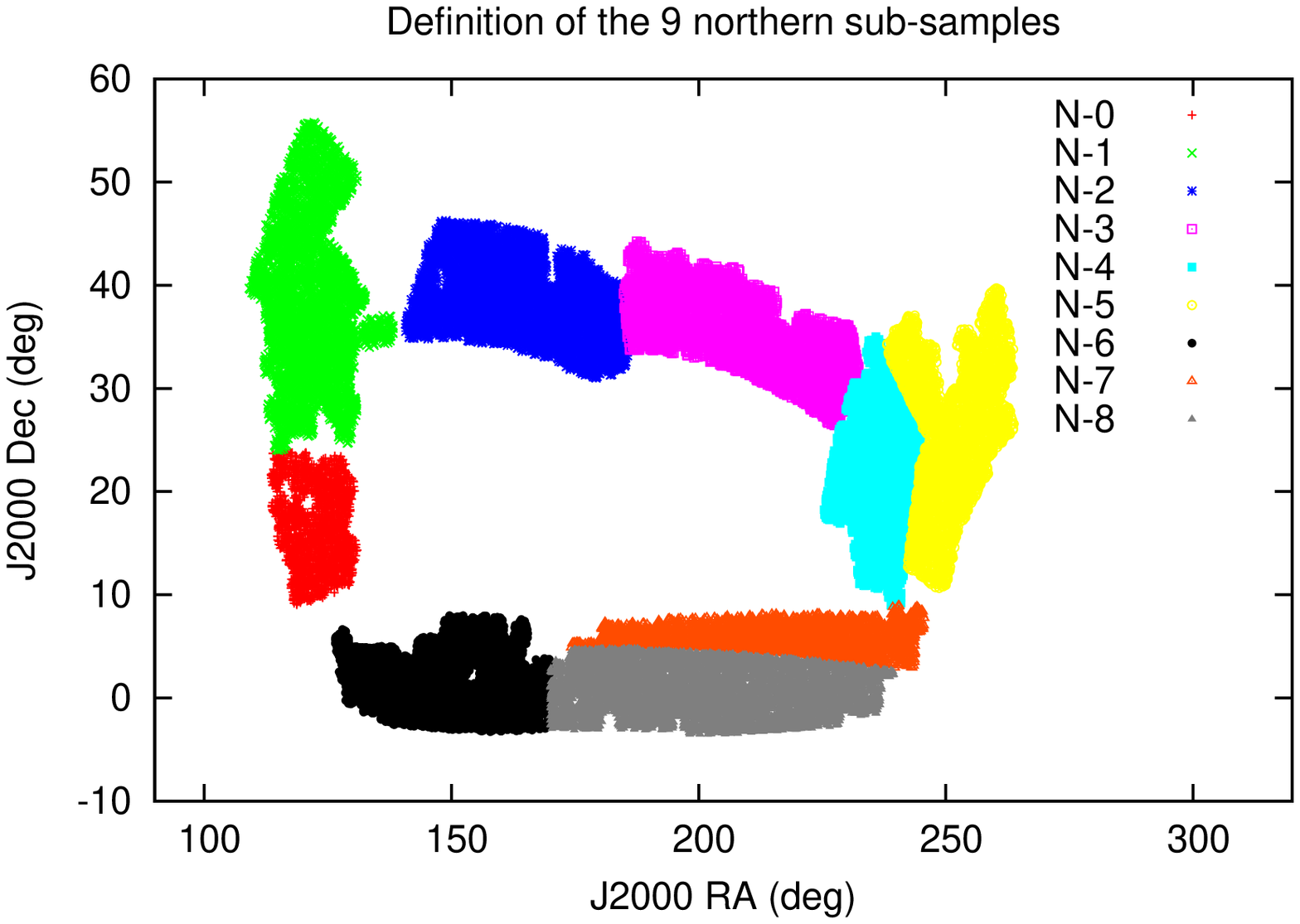}
  \includegraphics[scale=0.4, angle=-90]{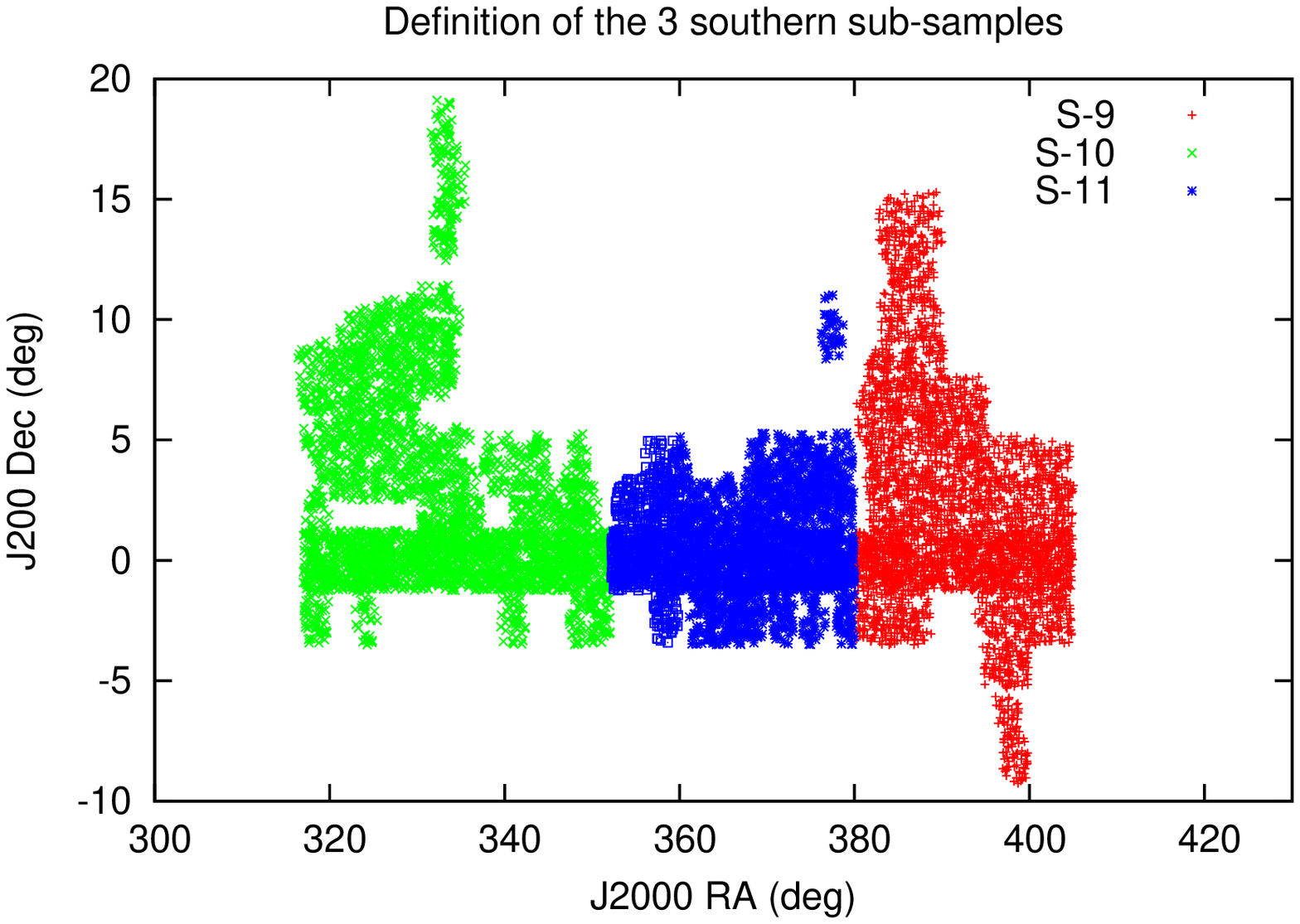}
 \end{center}
 \caption{The 12 sub-samples in which we split the DR9 quasar sample,
  $9$ in the northern Galactic hemisphere (left) and $3$ in the southern one
 (right).}
 \label{fig:chunks}
\end{figure}

  We divide the area covered by DR9 into 12 sub-samples as shown in Figure
\ref{fig:chunks}. Table \ref{tab:chunks} provides the number of quasars
and the number of DLAs present in each sub-sample. The DLA-\lya
cross-correlation and its covariance matrix are separately computed in
each sub-sample. The typical sub-sample size is $\sim 10 \deg $, or
$\sim 700 \hmpc$ 
%\kg{Specify that you're talking about linear scale on the sky,
%although it would also be good to also mention the typical area.}
, which is much larger than the maximum scale of our
analysis, $\sigma < 60 \hmpc$. We are therefore not concerned about the
loss of a small fraction of DLA-\lya pairs that are near the border of
two neighboring sub-samples. Because of the geometry of the observed sky area 
in DR9, most of the sub-samples have short borders with their neighbors.

\begin{table}[h!]
  \centering
  \begin{tabular}{c|cc}
    Sub-sample & \# quasars & \# DLA \\
    N-0 & 2290 & 303 \\
    N-1 & 5176 & 647 \\
    N-2 & 5264 & 670 \\
    N-3 & 5246 & 758 \\
    N-4 & 4010 & 537 \\
    N-5 & 4056 & 545 \\
    N-6 & 4823 & 658 \\
    N-7 & 3255 & 406 \\ 
    N-8 & 6322 & 819 \\
    S-9 & 4339 & 543 \\
    S-10 & 4308 & 503 \\
    S-11 & 3360 & 391 \\
\\
\hline
  \end{tabular}
  \caption{Number of quasars and DLAs included in our analysis in each of the
  12 sub-samples.}
  \label{tab:chunks}
\end{table}

The cross-correlations and their covariance matrices are obtained in each
sub-sample, $\mathbf{\hat\xi_\alpha}$ and $\tilde C_{\alpha}$, and are
combined using the equations
\begin{equation}
 \tilde C^{-1} = \sum_\alpha \tilde C^{-1}_\alpha ~, 
 \label{eq:total_C}
\end{equation}
\begin{equation}
 \mathbf{\hat\xi} = \tilde C \sum_\alpha \tilde C^{-1}_\alpha
 \mathbf{\hat\xi_\alpha} ~. 
 \label{eq:total_xi}
\end{equation}

%The uncertainties computed with the MCMC method rely on having a covariance
%matrix that accurately describes the data. Even though we do find a good fit
%to the theory, with reasonable values of $\chi^2$, we compute the errorbars
%from an alternative method that does not rely on the accuracy of the 
%covariance matrix.

Bootstrap errors are calculated from these 12 sub-samples by generating
$N=100$ random combinations of the 12 sub-samples with repetitions,
and recomputing $\tilde C$ and $\hat \xi$ for each of these 100
combinations \cite{Efron83}.
The dispersion among these combinations of the values of any model parameter 
that is fitted to the result yields the bootstrap error.
%The dispersion among these combinations of the values of $\hat \xi$ or of any 
%model parameter that is fitted to the result yields the bootstrap error.
 
%\af{No, we don't compute bootstrap errors on the correlation function, only
%on the bias parameters. I have changed this text.}

%\begin{equation}
% C^{-1} = \sum_\alpha n_\alpha C^{-1}_\alpha ~, 
%\end{equation}
%\begin{equation}
% \mathbf{\xi} = C \sum_\alpha n_\alpha C^{-1}_\alpha \mathbf{\xi_\alpha} ~,
%\end{equation}
%with $\sum_\alpha n_\alpha = 12$.

\subsection{Fitting the DLA bias}
\label{ss:bias}

In the limit of large-scales, linear theory predicts the precise form
of all the correlations among any pair of tracers of the large-scale
structure in redshift space. The \lya forest probes material at low
density and high redshift, which is less affected by non-linear
evolution and random peculiar velocities than
a set of galaxies moving on orbits within gravitationally collapsed
halos. Therefore, linear theory remains a good approximation down to
smaller scales for the \lya forest than for other tracers.
In linear theory, redshift distortions cause the amplitude of each
Fourier mode in a biased tracer field to be enhanced by the factor
$b(1+\beta \mu_k^2)$, where $b$ is the density bias factor, $\beta$ is
the redshift distortion parameter, and $\mu_k$ is the cosine of the
angle between the Fourier mode and the line of sight
\cite{KAISE87}. The linear cross-power spectrum of the DLAs
and the \lya forest is therefore equal to
\begin{equation}
 P_{DF}(\vk,z) = b_D(z) \left[1+\beta_D(z) \mu_k^2 \right] 
           b_F(z) \left[1+\beta_F(z) \mu_k^2 \right] \, P_L(k,z) ~.
 \label{eq:Px}
\end{equation}
where $b_D$ and $b_F$ are the DLA and \lya density bias factors,
$\beta_D$ and $\beta_F$ are their redshift space distortion parameters,
and $P_L(k,z)$ is the linear matter power spectrum. The
cross-correlation function is the Fourier transform of $P_{DF}$, and
can be computed using the following equations (derived as in \cite{HAMIL92})
\begin{equation}
\xi(\mathbf{r}) = \xi_0(r) P_0(\mu) + \xi_2(r) P_2(\mu) +
  \xi_4(r) P_4(\mu) ~,
 \label{eq:xiP}
\end{equation}
where $\mu = \pi/r$ is the angle cosine in redshift space, $P_0$, $P_2$
and $P_4$ are the Legendre polynomials, and the functions $\xi_0$,
$\xi_2$ and $\xi_4$ are
\begin{equation}
  \xi_0(r) = b_D b_F \left[1 + (\beta_D + \beta_F)/3 + \beta_D\beta_F/5 \right]
  \zeta(r) ~,
\end{equation}
\begin{equation}
  \xi_2(r) = b_D b_F \left[2/3(\beta_D + \beta_F) + 4/7\beta_D\beta_F \right]
  \left[\zeta(r) - \bar\zeta(r)\right] ~,
\end{equation}
\begin{equation}
  \xi_4(r) = 8/35\, b_D b_F  \beta_D\beta_F
  \left[\zeta(r) + 5/2 \bar\zeta(r) - 7/2 \bar{\bar\zeta}(r) \right] ~.
\end{equation}
The function $\zeta(r)$ is the linear correlation function in real
space, and $\bar\zeta(r)$ and $\bar{\bar\zeta}(r)$ are averages of
$\zeta(r)$ within $r$ given in equation (9) of \cite{HAMIL92}. 
We compute $\zeta(r)$ using the $\Lambda$CDM model with the
parameters given at the end of the Introduction.

  We assume the DLA bias $b_D$ to be constant with redshift,
%\jm{is there any case where we do anything different?}
%\af{no, but we measure the bias in 3 redshift bins to check if it evolves}
while for the \lya forest, we assume the bias evolves as
$b_F(z)/b_F(z=2.25) = [(1+z)/3.25]^{\alpha/2+1}$,
where $\alpha=3.8$ describes the evolution of the amplitude of the 
line-of-sight power spectrum measured in \cite{2006ApJS..163...80M}.
The rapid evolution of the amplitude of $\delta_F$ in the \lya forest is
a consequence of the rapid change in the mean transmitted fraction with
redshift.
%\jr{For the beginner, you might say a word on why the Lya forest bias has 
%stronger z-dependence than the DLA's.} \af{done}.

  For the redshift distortion of the DLAs, we assume $\beta_D =
f(\Omega) / b_D \approx b_D^{-1}$, where $f(\Omega)$ is the logarithmic
derivative of the linear growth factor \cite{KAISE87}, which is
close to unity at $z>2$ in the $\Lambda$CDM model. This assumption may not be
exactly correct for DLAs if the probability of detection of a DLA
depends on the large-scale peculiar velocity gradient because of the
\lya forest absorption that is superposed on the damped wings. This
effect would generate a peculiar velocity bias analogously to the
case of the \lya forest, but we expect this effect to be small.

  For the \lya forest, $b_F$ and $\beta_F$ are independent because the
$\delta_F$ tracer is subject to a strong bias by the large-scale
peculiar velocity gradient \cite{MCDON03}. We use the results for the
\lya autocorrelation of \cite{2011JCAP...09..001S} to fix
$b_F (1+\beta_F) = -0.336$. Most of our results will be presented
assuming $\beta_F=1.0$, in which case the only free parameter is $b_D$. 
Some results will also be obtained when $\beta_F$, which was poorly determined 
by \cite{2011JCAP...09..001S}, is also treated as a free parameter.
Note that the \lya forest has a negative bias, since denser regions of the
universe have a lower transmitted flux.
This implies that the cross-correlation is also negative.
%A new study of the \lya clustering from the BOSS DR9 will be presented 
%elsewhere, with updated values for the bias parameters. We present here the 
%results for different values of the \lya bias parameters, to allow better
%constraints on the DLA bias whenever we obtain new results on the \lya biases.

  The model of the linear theory in equation (\ref{eq:xiP}) is generally
modified to account for the effect of the MTC, which removes the mean
value of $\delta_F$ in each quasar spectrum, therefore distorting the
DLA-\lya cross-correlation. The way this correction is modeled and
calculated is explained in Appendix \ref{app:correc}.
Instead of evaluating the theoretical model at the center of every bin
in the $\sigma, \pi$ components of the DLA-\lya separation, we evaluate
it at the weighted average values of $\sigma, \pi$ for all the DLA-\lya
pairs that contribute to the bin.
To minimize possible non-linear effects on the correlation function, all
the fits are done using only bins with $ r > 5 \hmpc$; this eliminates
the two bins with smallest $\sigma$ and $\pi$, reducing the number of
bins to 126.

For each theoretical model that predicts the cross-correlation at each
bin $A$ of $(\pi,\sigma)$, $\xi_{t,A}$, we compute its $\chi^2$ as
\begin{equation}
 \chi^2 = (\hat\xi_A - \xi_{t,A}) \tilde C^{-1}_{AB} (\hat\xi_B - \xi_{t,B}) ~, 
\end{equation}
where the indexes $A$ and $B$ are summed over all bins. The likelihood
function is
\begin{equation}
 L = \frac{\exp \left[ -0.5 \, \chi^2 \right] }{(2\pi)^{N/2} \, |\tilde C|} ~,
\end{equation}
where $|\tilde C|$ is the determinant of the covariance matrix of the
cross-correlation (computed using 
equations \ref{eq:covar} and \ref{eq:total_C}), $N=126$ is the number of 
bins, and we have assumed that the distribution of errors in the 
cross-correlation is Gaussian.

Our best estimate of the DLA bias is the one that maximizes the likelihood,
and its uncertainty is evaluated using two different approaches.
First we estimate the errors on the fitted parameters with the Monte Carlo 
Markov Chain technique (MCMC errors), where we sample the parameter space 
using the likelihood function described above and compute the dispersion
around the best fit value.
Secondly, the fits are repeated with the bootstrap realizations
of the 12 sub-samples randomly selected with repetitions as described
previously, obtaining the bootstrap error (BS error) from the dispersion in the
fitted parameters among the bootstrap realizations. Both the
MCMC and BS errors will be presented.

\section{Results}
\label{sec:results}

In this section the results of the cross-correlation of DLAs
and the \lya absorption in redshift space are presented. After obtaining
a fit to our fiducial model with only one free parameter, the bias
factor of DLAs, we analyze the consistency of the covariance matrix and
the bootstrap errors, the evidence for the
presence of redshift distortions, and the dependence of our best fit
model on the range of separations that are used. 
%\jr{Not sure that this is what you want to say.  Can't the bootstrap be used 
%to generate a covariance matrix? So you mean ``consistency of bootstrap errors 
%and errors calculated from equation 3.12.''}
%\af{you could do that if you had few bins and a low of chunks, but we have 12 
%chunks and 128 bins, so we compare the errors in the bias from a likelihood 
%analysis using a covariance matrix, and the error in the bias from bootstrap.}
We then investigate a variety of possible dependences of the DLA bias factor 
on several variables: redshift, column density, and the \lya forest redshift
distortion parameter. In general the results presented here have been
corrected with the MTC, and the fitted models are also corrected as
described in Appendix \ref{app:correc}, except when we test for the
effect of the MTC.

%\af{Nathalie and Christophe have found a bug on KG's files (the noise correction
%is wrong), so we might want to redo everything with the fixed version of the 
%files. It would only effect the weighting, but in the case that Pasquier does
%significant changes in his DLA catalogue it might be worth doing it. Will also
%run with smaller sigma of 2 instead of 0.1}

\subsection{Measured cross-correlation and bias parameter}

\begin{figure}[h!]
 \begin{center}
  \begin{tabular}{cc} 
   \includegraphics[scale=0.5, angle=-90]{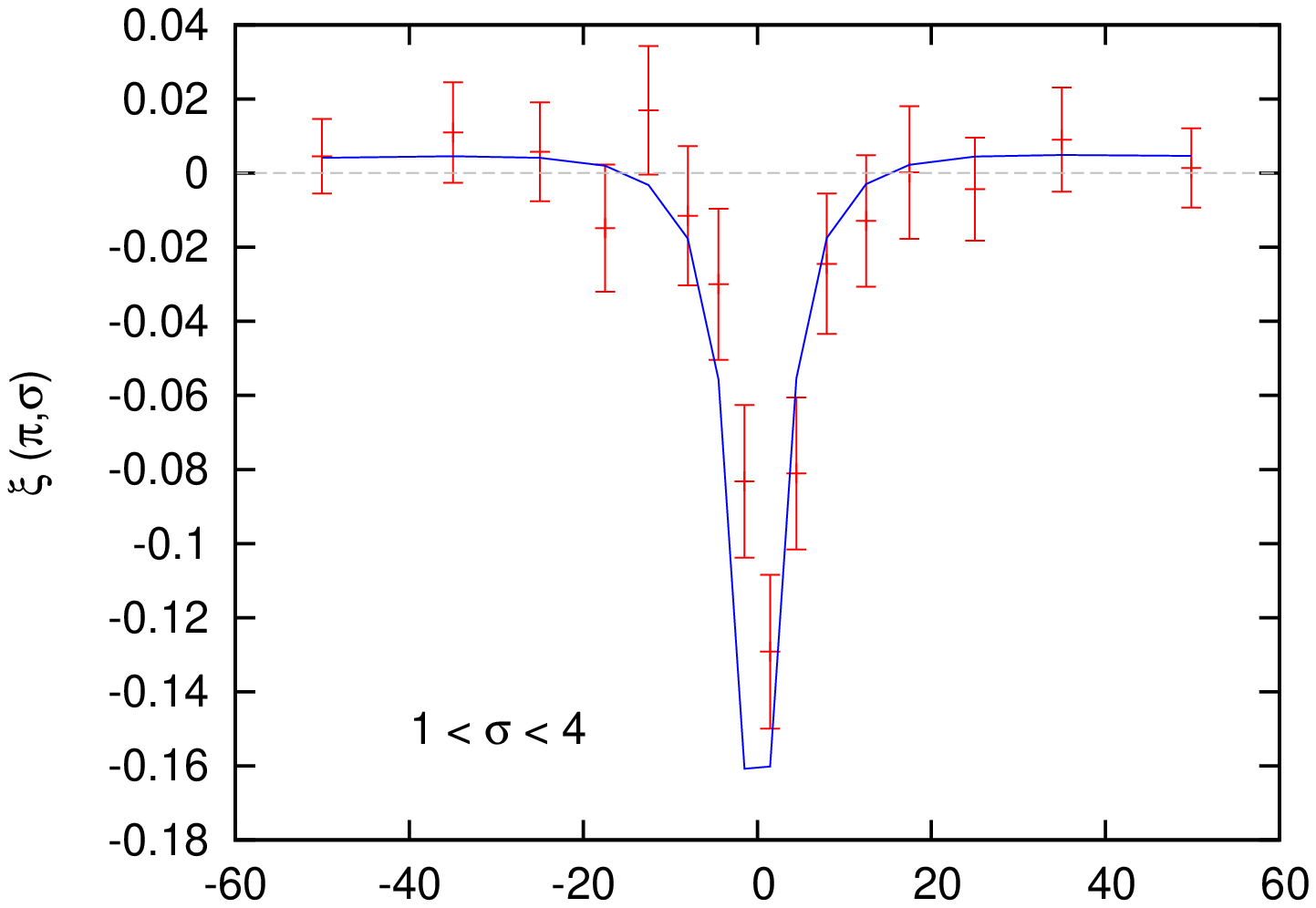} &
   \includegraphics[scale=0.5, angle=-90]{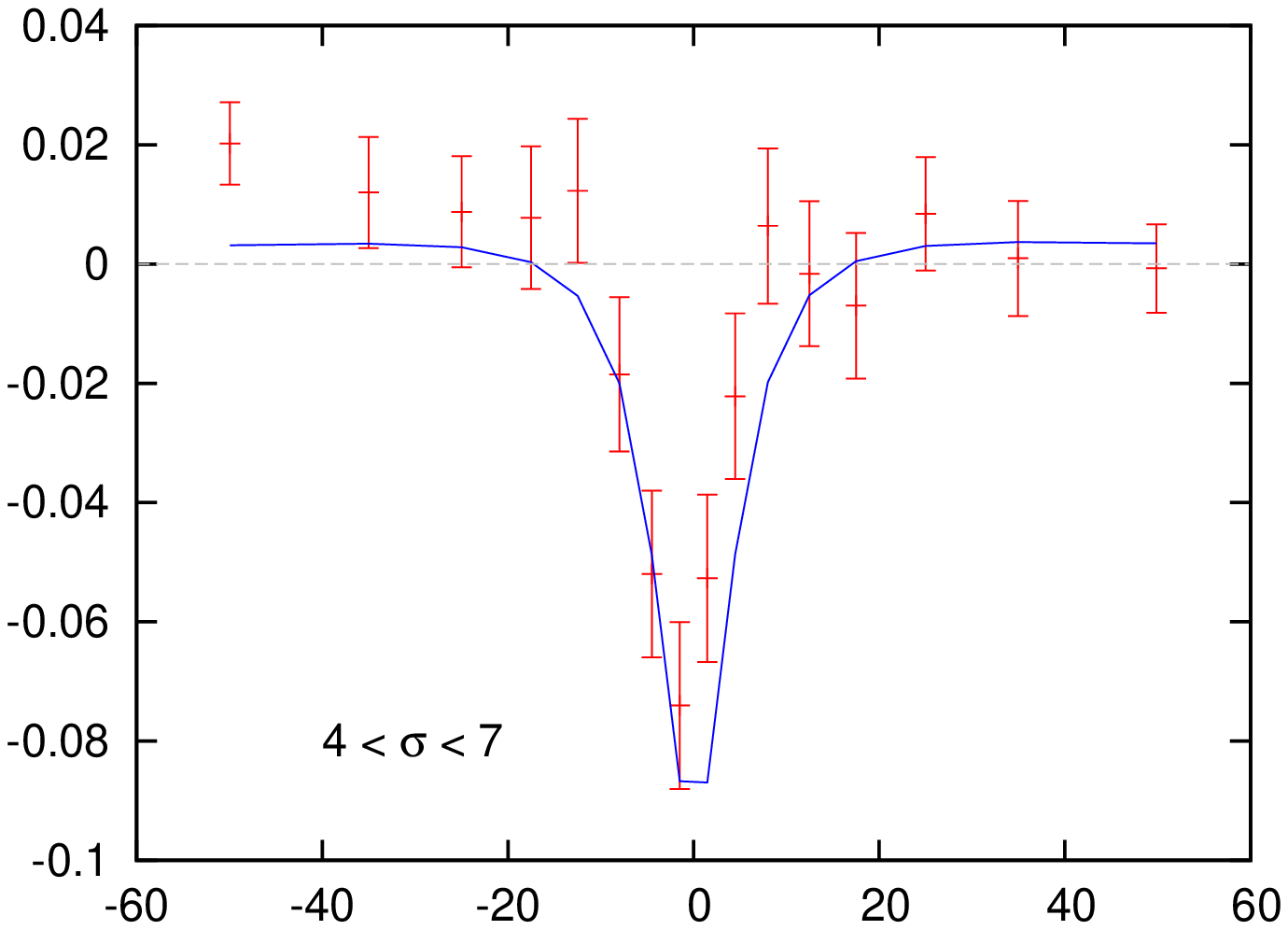} \\
   \includegraphics[scale=0.5, angle=-90]{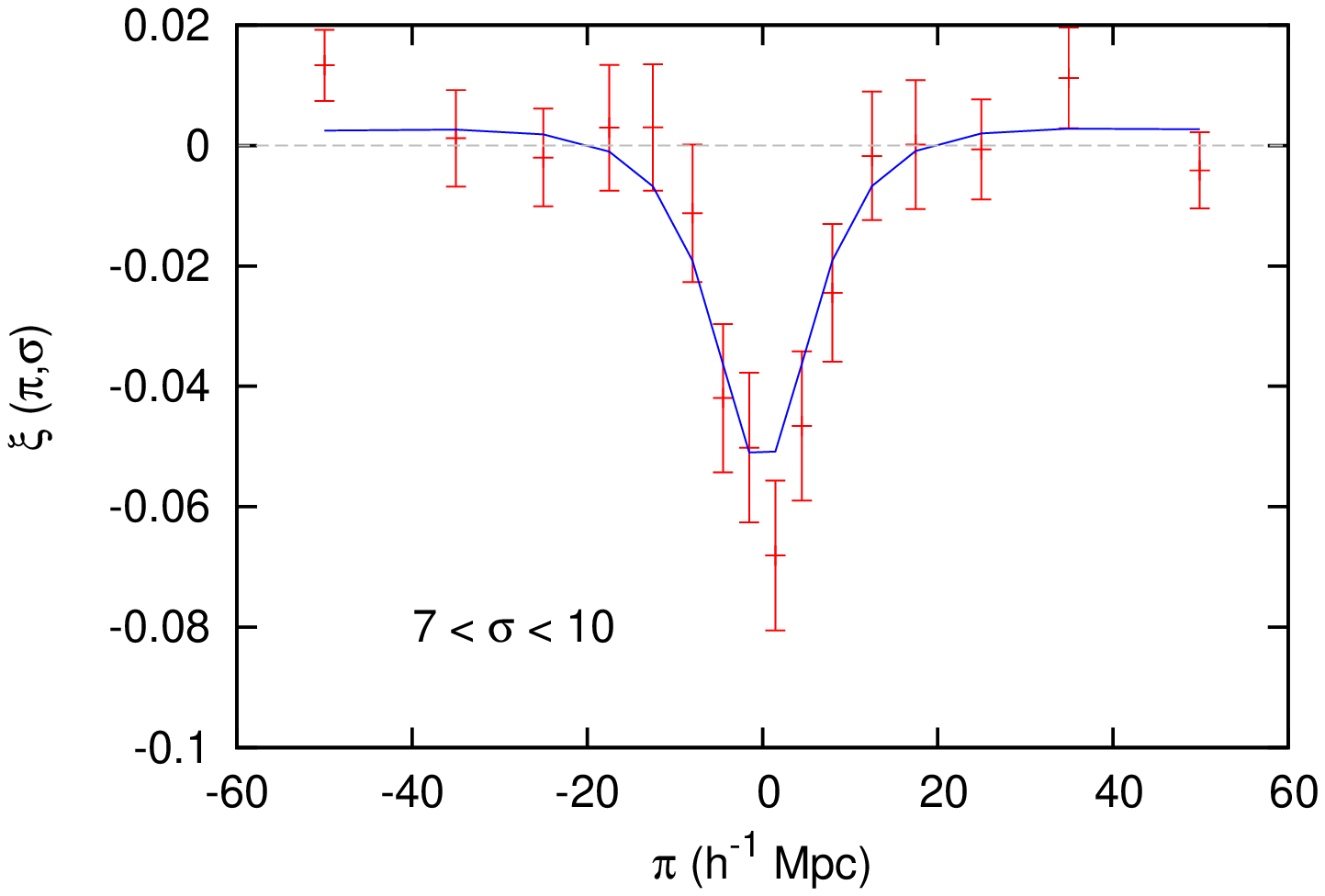} & 
   \includegraphics[scale=0.5, angle=-90]{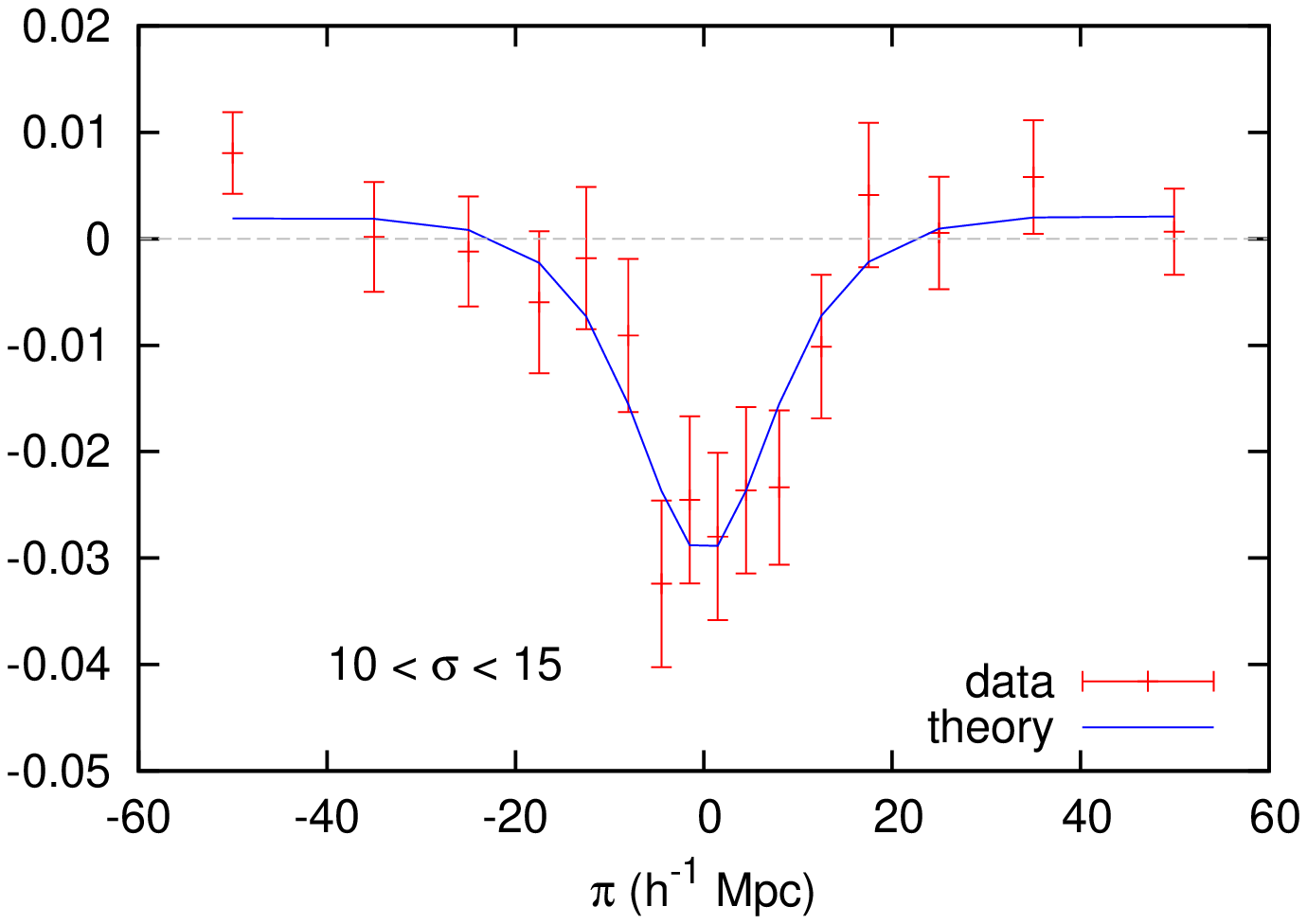} \\
  \end{tabular}
 \end{center}
 \caption{Measured cross-correlation in different bins of $\sigma$:
  from top-left to bottom-right,
    $1 < \sigma < 4 \hmpc$,
    $4 < \sigma < 7 \hmpc$,
    $7 < \sigma < 10 \hmpc$, and 
    $10 < \sigma < 15 \hmpc$.
  Solid lines show the best fit model for the fiducial analysis, including 
  the correction derived in appendix \ref{app:correc}.
 }
 \label{fig:cross_1}
\end{figure}

\begin{figure}[h!]
 \begin{center}
  \begin{tabular}{cc} 
   \includegraphics[scale=0.5, angle=-90]{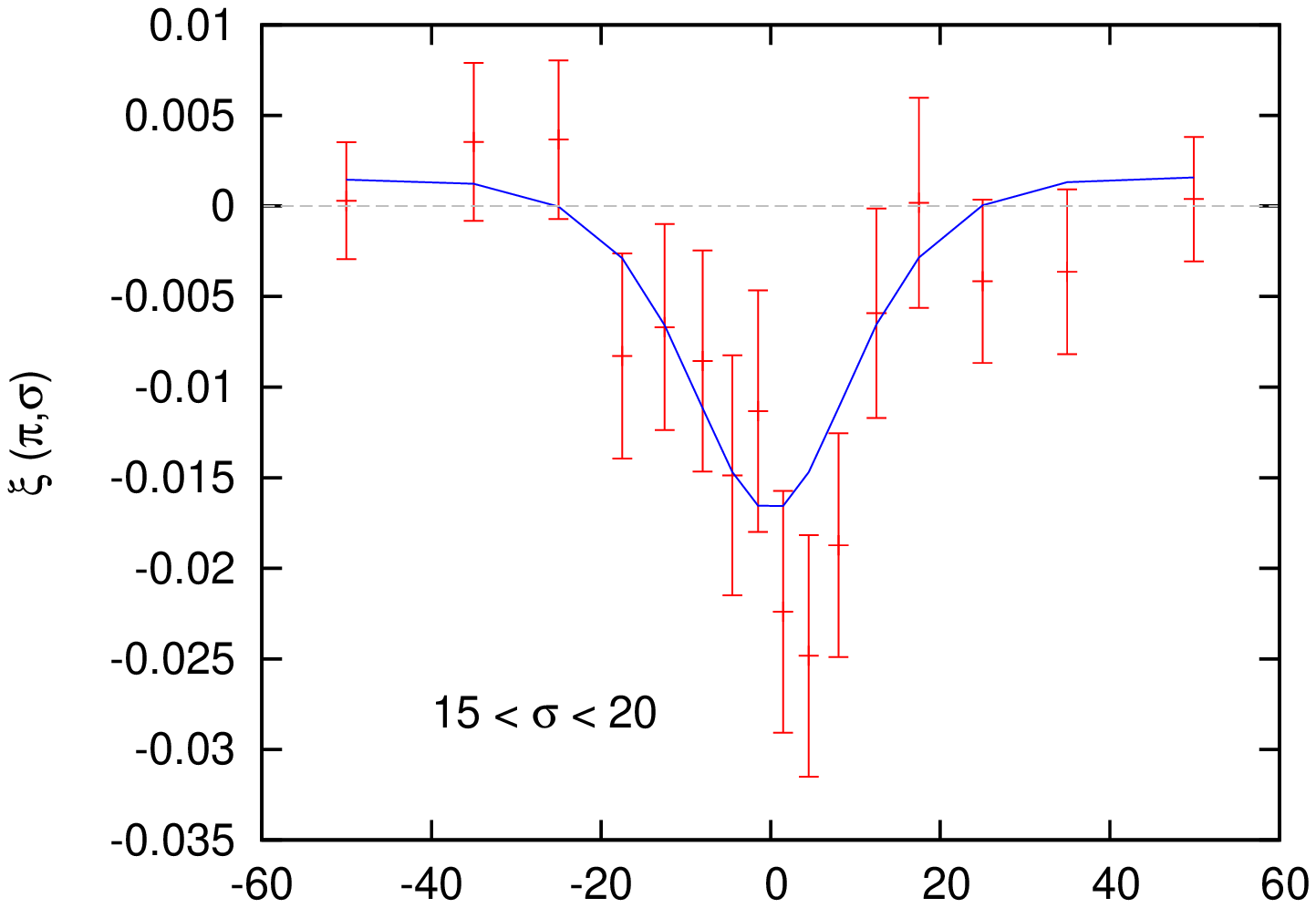} &
   \includegraphics[scale=0.5, angle=-90]{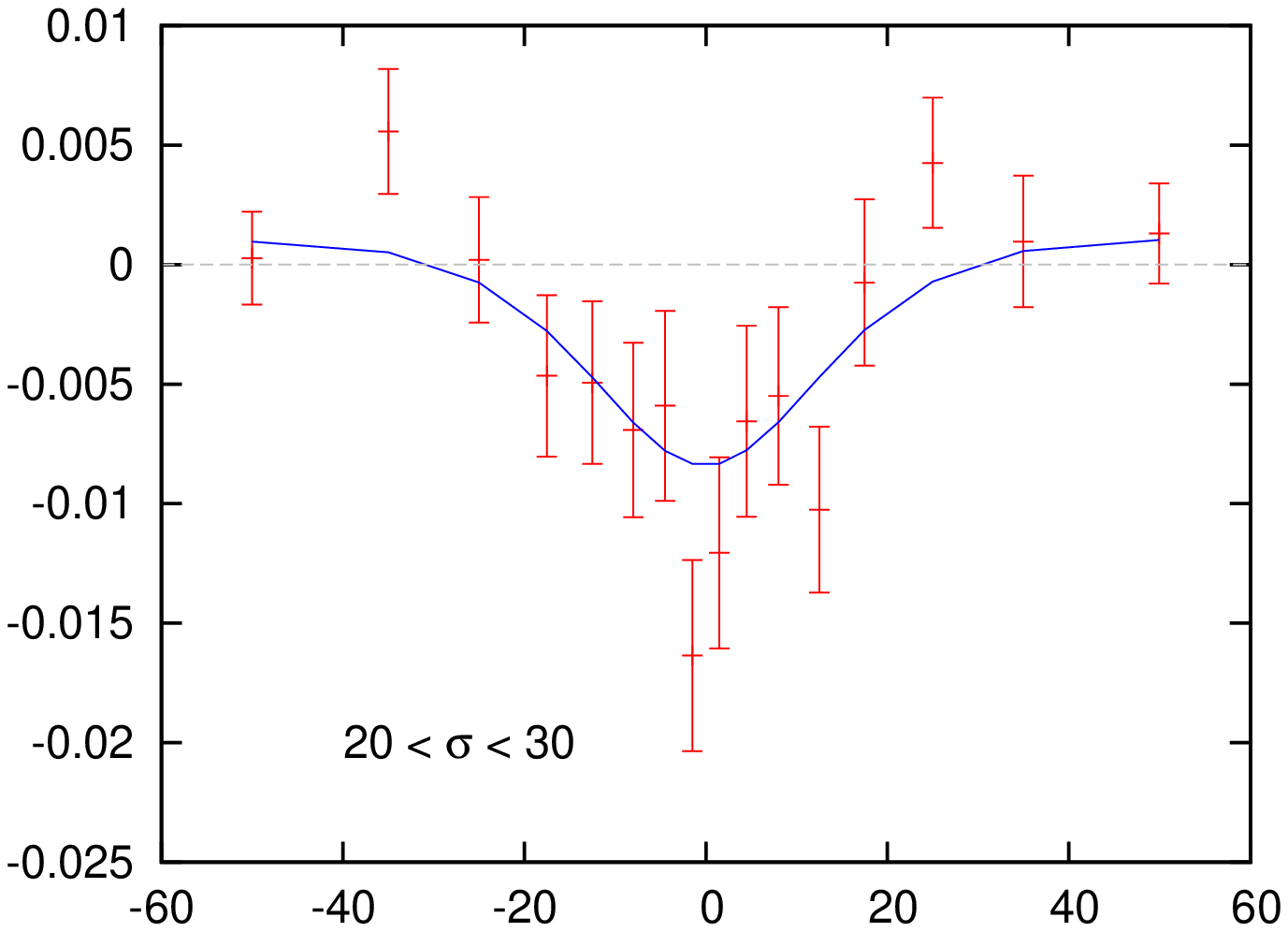} \\
   \includegraphics[scale=0.5, angle=-90]{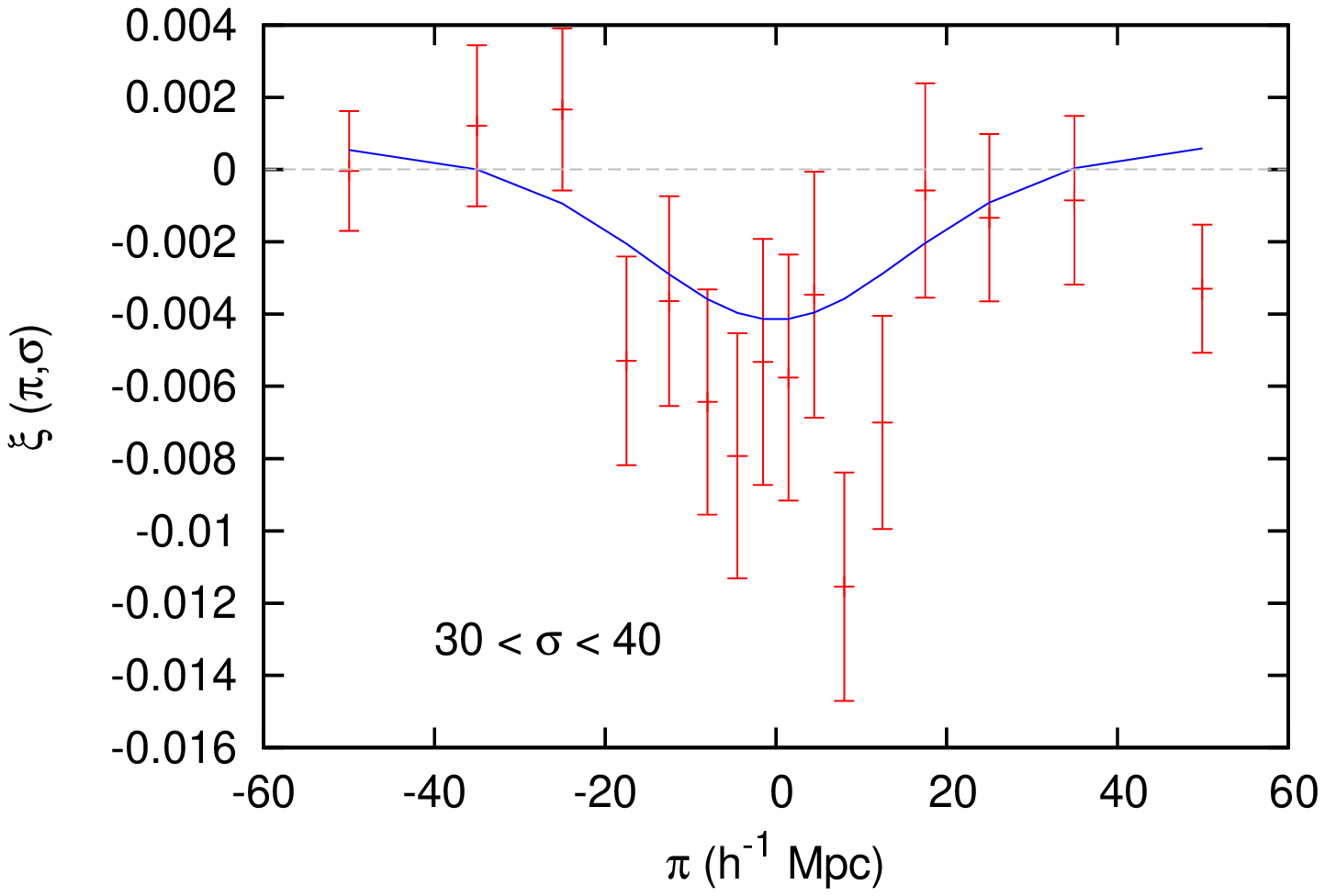} & 
   \includegraphics[scale=0.5, angle=-90]{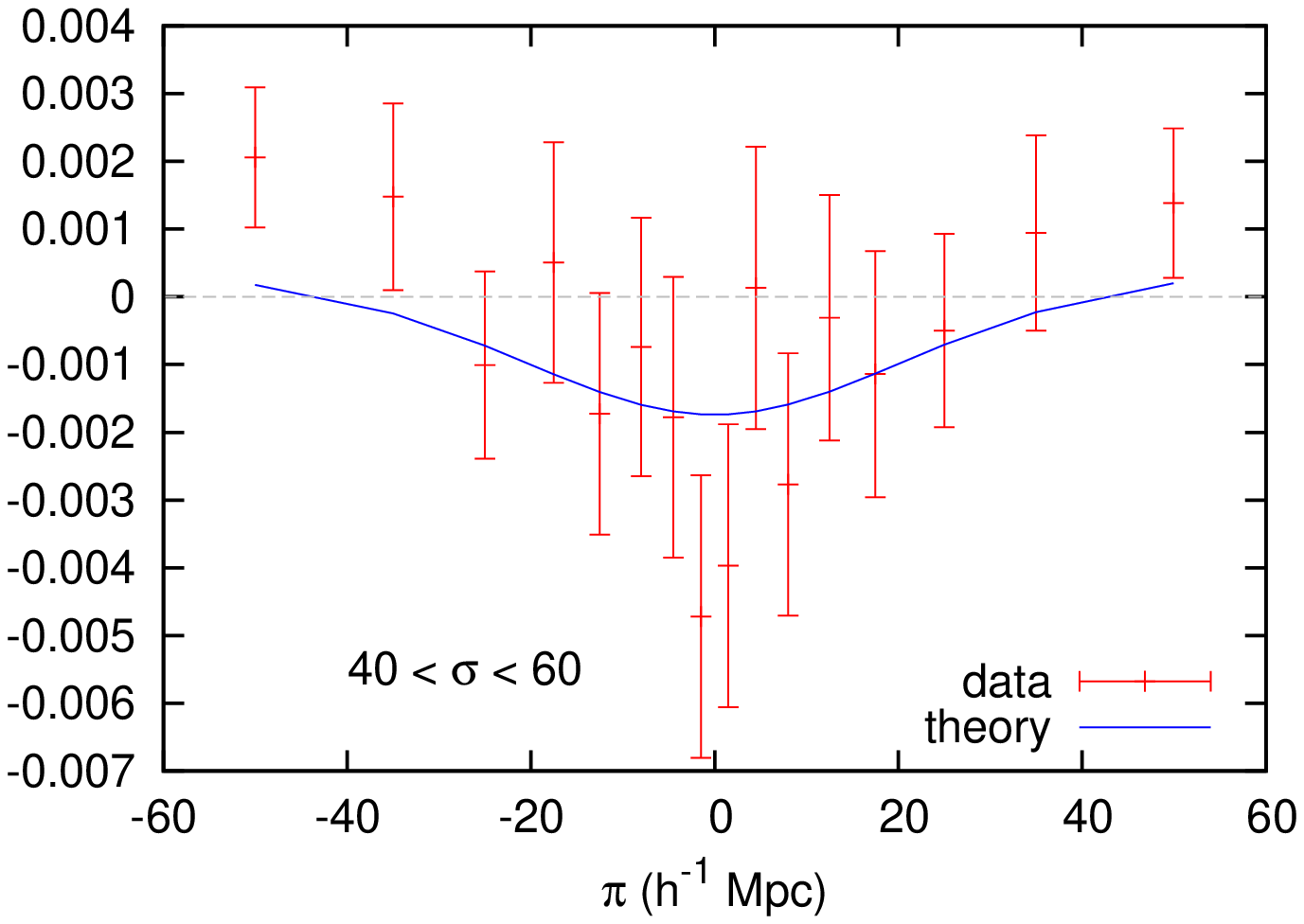} \\
  \end{tabular}
 \end{center}
 \caption{Same as figure \ref{fig:cross_1}, for the $\sigma$ bins,
   from top-left to bottom-right,
    $15 < \sigma < 20 \hmpc$,
    $20 < \sigma < 30 \hmpc$,
    $30 < \sigma < 40 \hmpc$, and
    $40 < \sigma < 60 \hmpc$.
 }
 \label{fig:cross_2}
\end{figure}

\begin{figure}[h!]
 \begin{center}
  \includegraphics[scale=0.35]{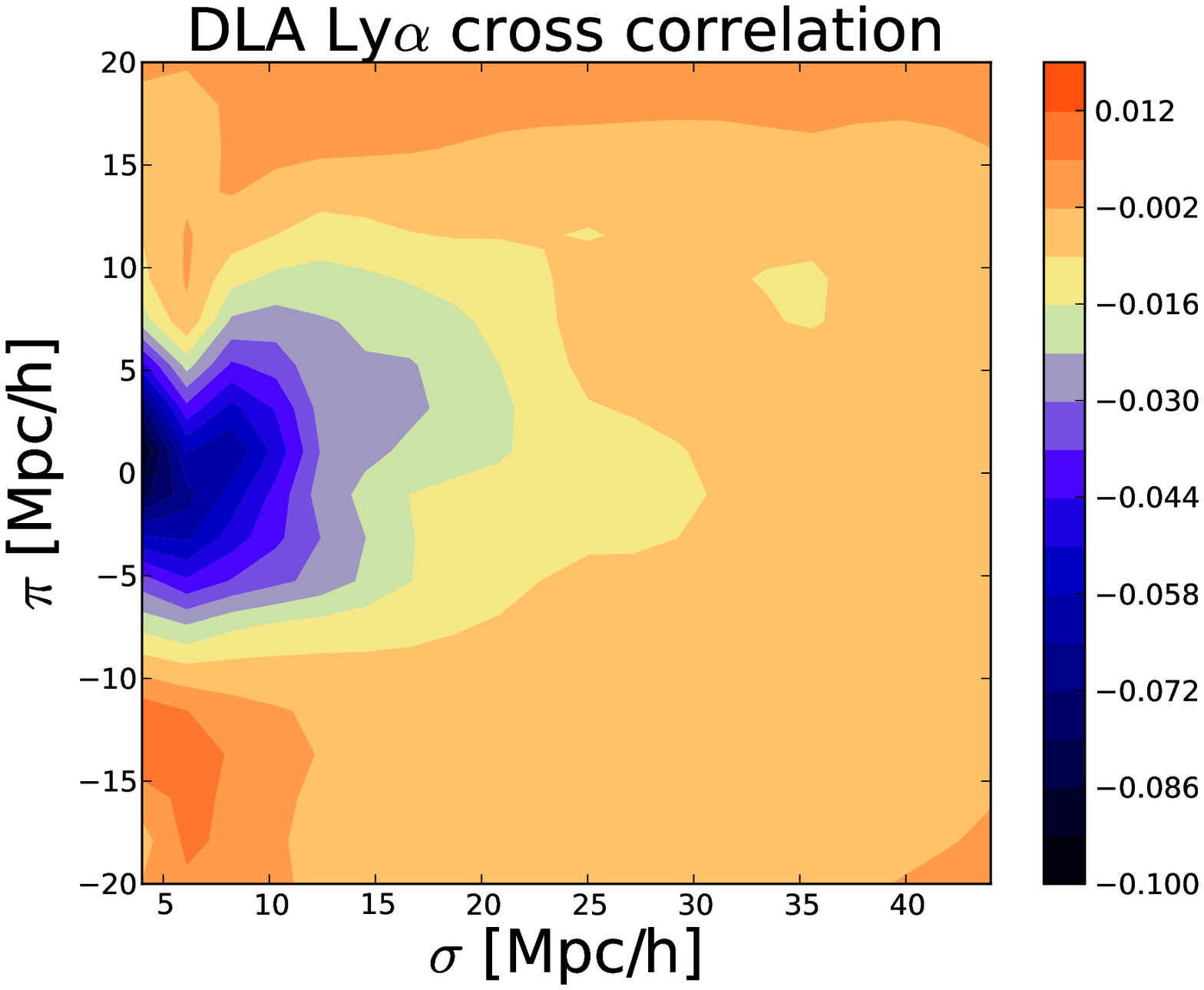} 
  \includegraphics[scale=0.35]{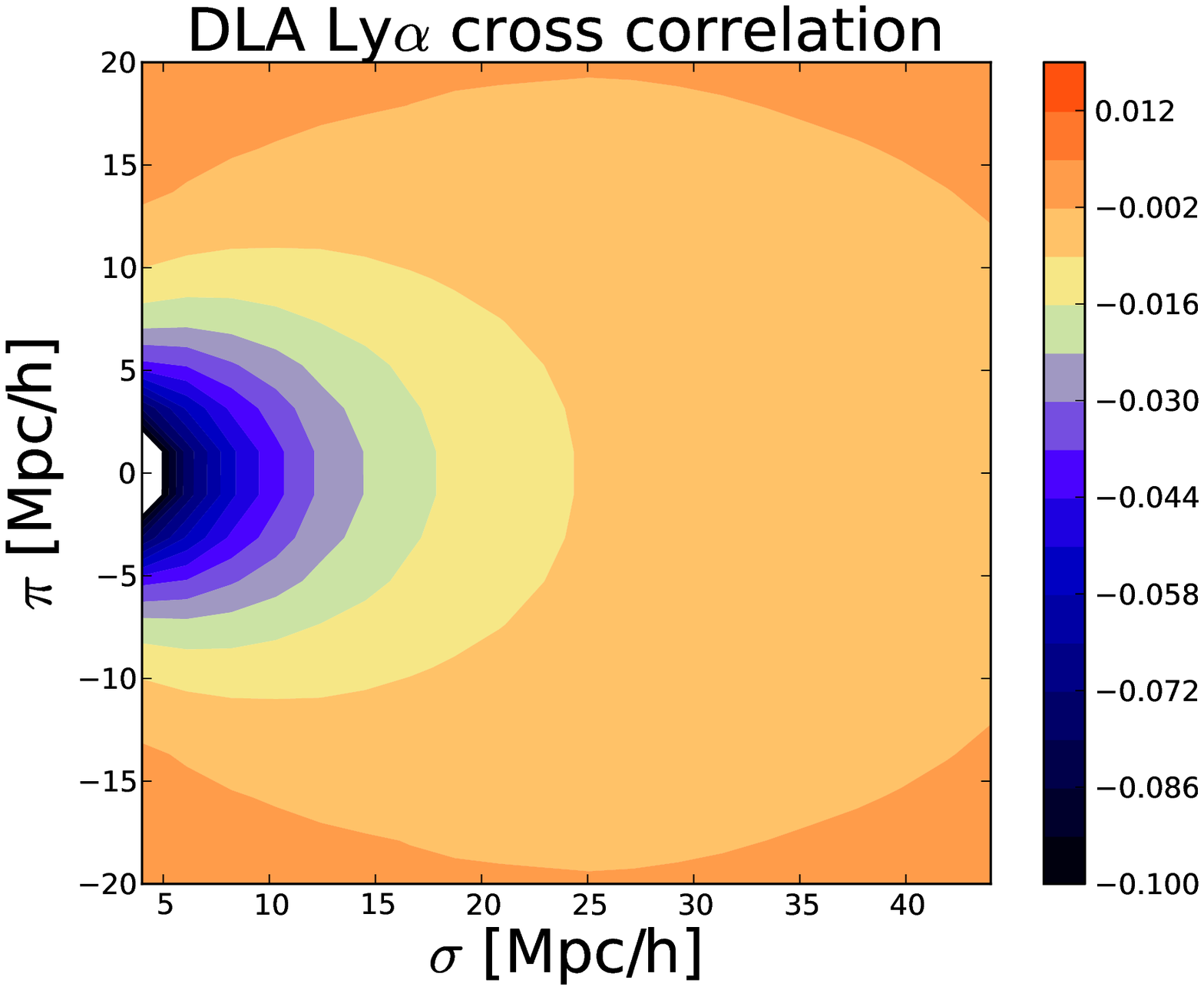} 
 \end{center}
 \caption{Contour plot of the measured redshift space cross-correlation for 
  our fiducial analysis (left) and our best fit theoretical model that includes
  the MTC correction (right).
 }
 \label{fig:kaiser}  
\end{figure}

The results of the cross-correlation as a function of $\pi$ for our
fiducial model are shown as red points with errorbars in figures
\ref{fig:cross_1} and \ref{fig:cross_2} in eight panels, corresponding
to our eight bins in $\sigma$. The errorbars are the square root of the
diagonal elements of the covariance matrix $\tilde C$ computed as
described in section \ref{ss:cov} (equation \ref{eq:total_C}). The blue
solid curves are our best fit fiducial model (all of our fits
exclude bins at $r = (\sigma^2 + \pi^2)^{1/2} < 5\hmpc$, which are only
the two bins at the lowest $\sigma$ and $\pi$, even though
they are shown in the figures), which fixes $\beta_F=1$ and fits only
one free parameter, the DLA bias. The result obtained
from our covariance matrix is $b_D=2.17 \pm 0.20$, with a $\chi^2$
of 106 for 125 degrees of freedom, and the bootstrap analysis gives the same
errorbar. Note that we are measuring only the  
amplitude of the cross-correlation, which is proportional to
$b_D b_F \sigma_8^2$ and depends on $\beta_F$, and the error on $b_D$
here reflects only that in our measurement of the amplitude. The results
are also shown in a contour plot in the left panel of figure
\ref{fig:kaiser}, and the best fit fiducial model is shown in the right
panel with the same contour levels.

  As seen in the figures and indicated by the $\chi^2$ value, the fit of the 
standard $\Lambda$CDM model to the DLA-\lya forest cross-correlation is 
excellent. The measurements are fully consistent with the radial dependence and
anisotropy that is expected in linear theory for the model that agrees
with all other cosmological determinations of the large-scale power spectrum.

%In Figure \ref{fig:kaiser} we show the cross-correlation contours in the 
%$\pi - \sigma$ plane, comparing the measurement (left panel) to the best-fit 
%theory (right panel). The lack of rotational symmetry of the contours are 
%caused by the redshift space distorsions, that are detected at high confidency 
%as shown in the next subsection.

%If we compute the $\chi^2$ for the null hypothesis, $b_D=0$, we get a value of
%$\chi^2 = 331$, compared to our best-fit value of $\chi2 = 123$. 
%We could then claim that our detection has a significance of $14\sigma$.
%If we re-do the analysis without the "mean flux regulation", we obtain 
%$\chi^2 = 303$, compared to $\chi^2 = 114$, i.e., also a $14\sigma$ 
%detection.

\subsection{Effect of the MTC}

All our results are generally obtained including the MTC, and the
fitted models include the correction described in Appendix \ref{app:correc}.
We now check the effect of the MTC by recomputing the cross-correlation
without including the correction, i.e., by using directly the PCA-only
continuum. In this case, we add to our covariance matrix the additional
term $\xi_c=0.03$ in equation (\ref{eq:cij}) due to the
error of the continuum fitting for pixels on a common line of sight.
%\jm{There is no explanation here on how you calculate this term,
%I actually have no idea how you do it!}
%\af{what we add is this 0.025 extra variance to every pair of pixels that 
%belong to the same line of sight.}

\begin{figure}[h!]
 \begin{center}
  \begin{tabular}{cc}
   \includegraphics[scale=0.5, angle=-90]{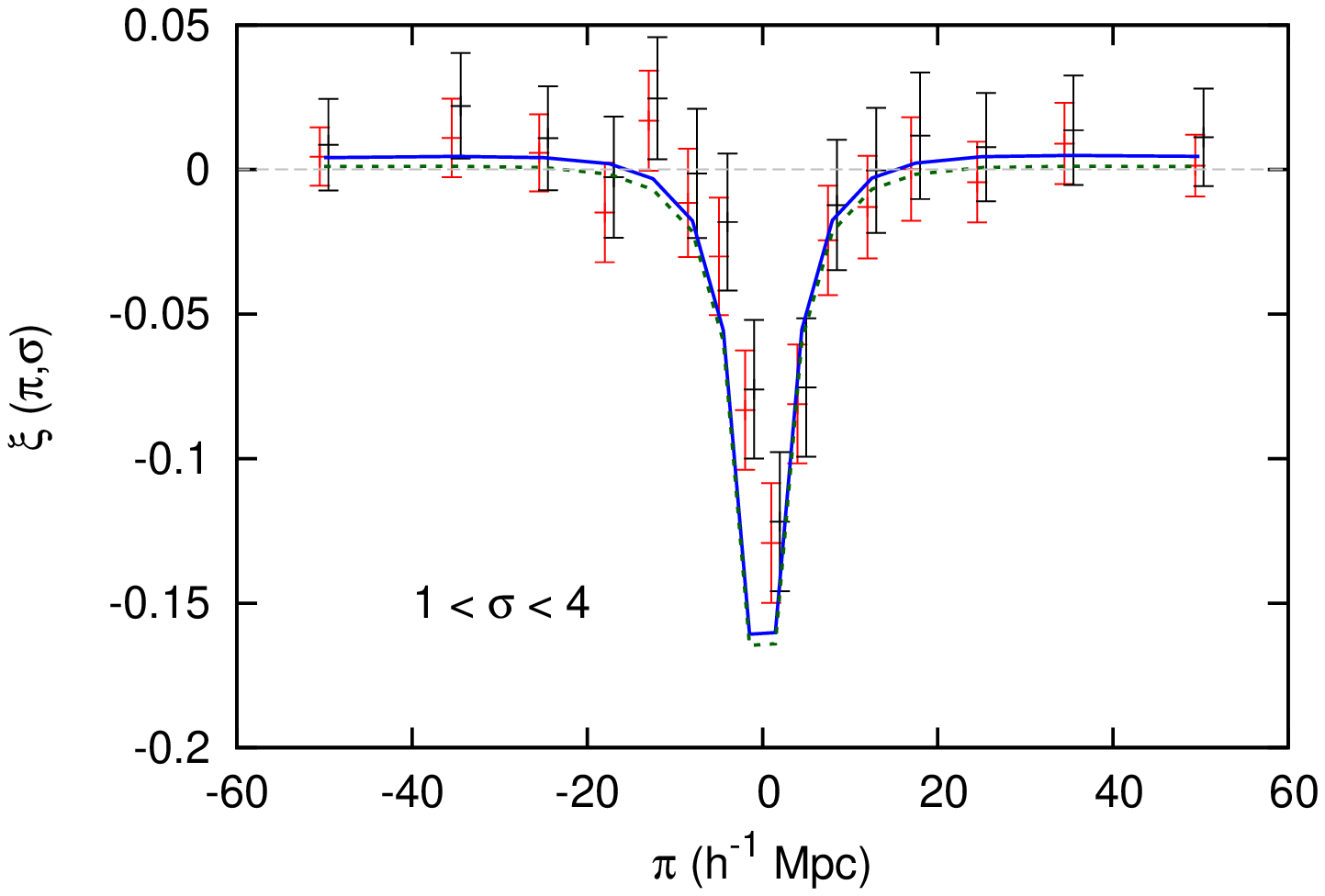} &
   \includegraphics[scale=0.5, angle=-90]{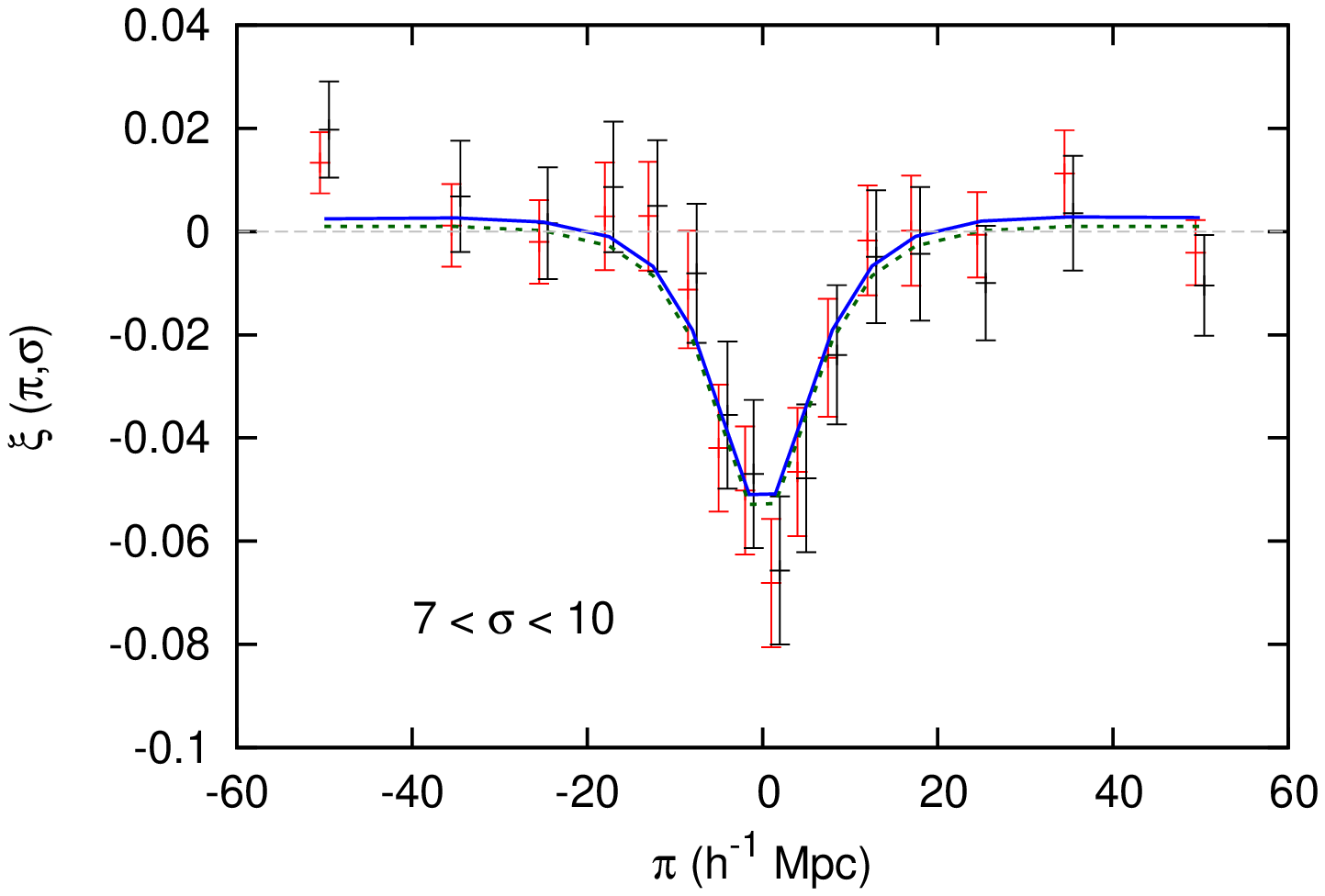} \\
   \includegraphics[scale=0.5, angle=-90]{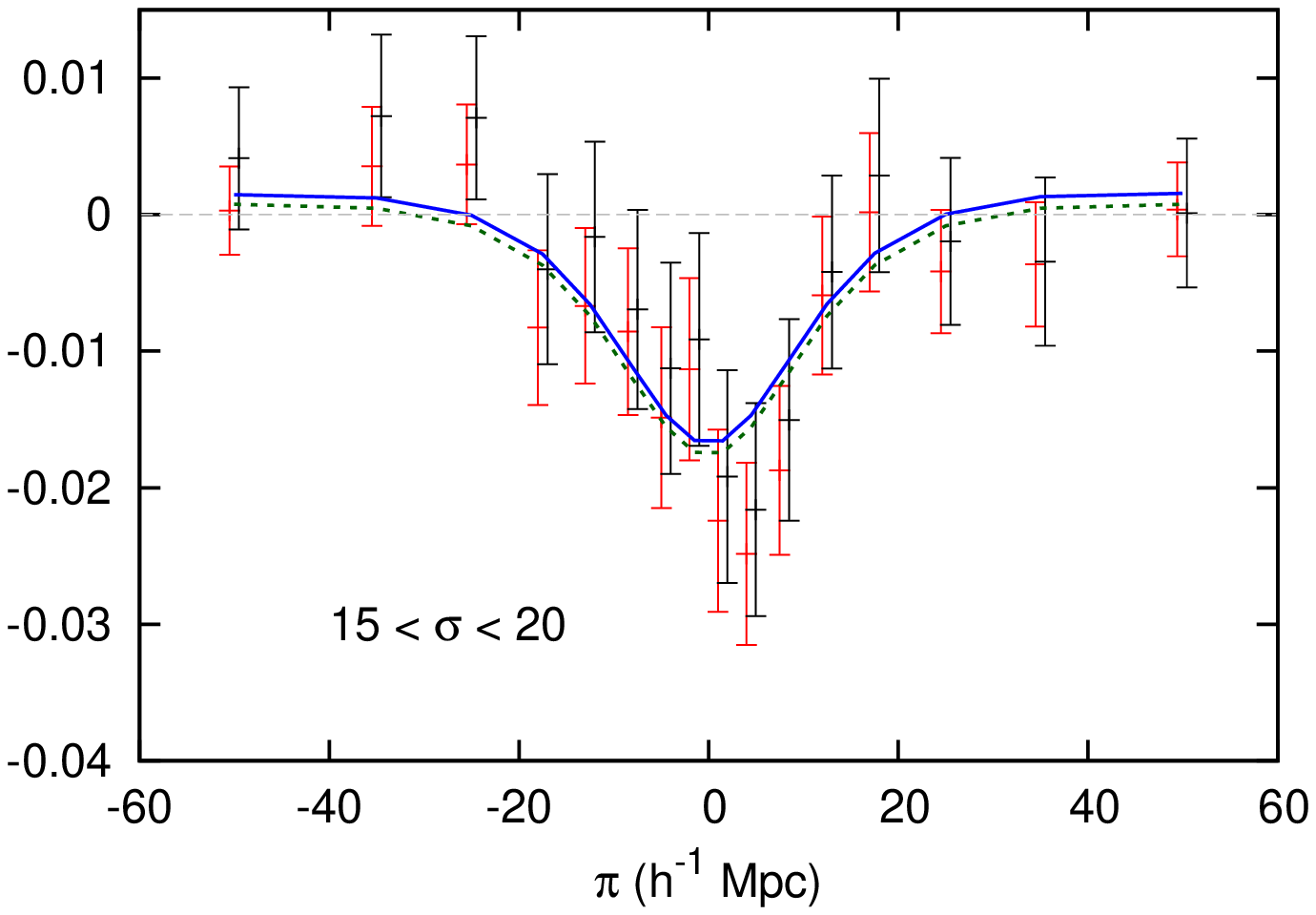} &
   \includegraphics[scale=0.5, angle=-90]{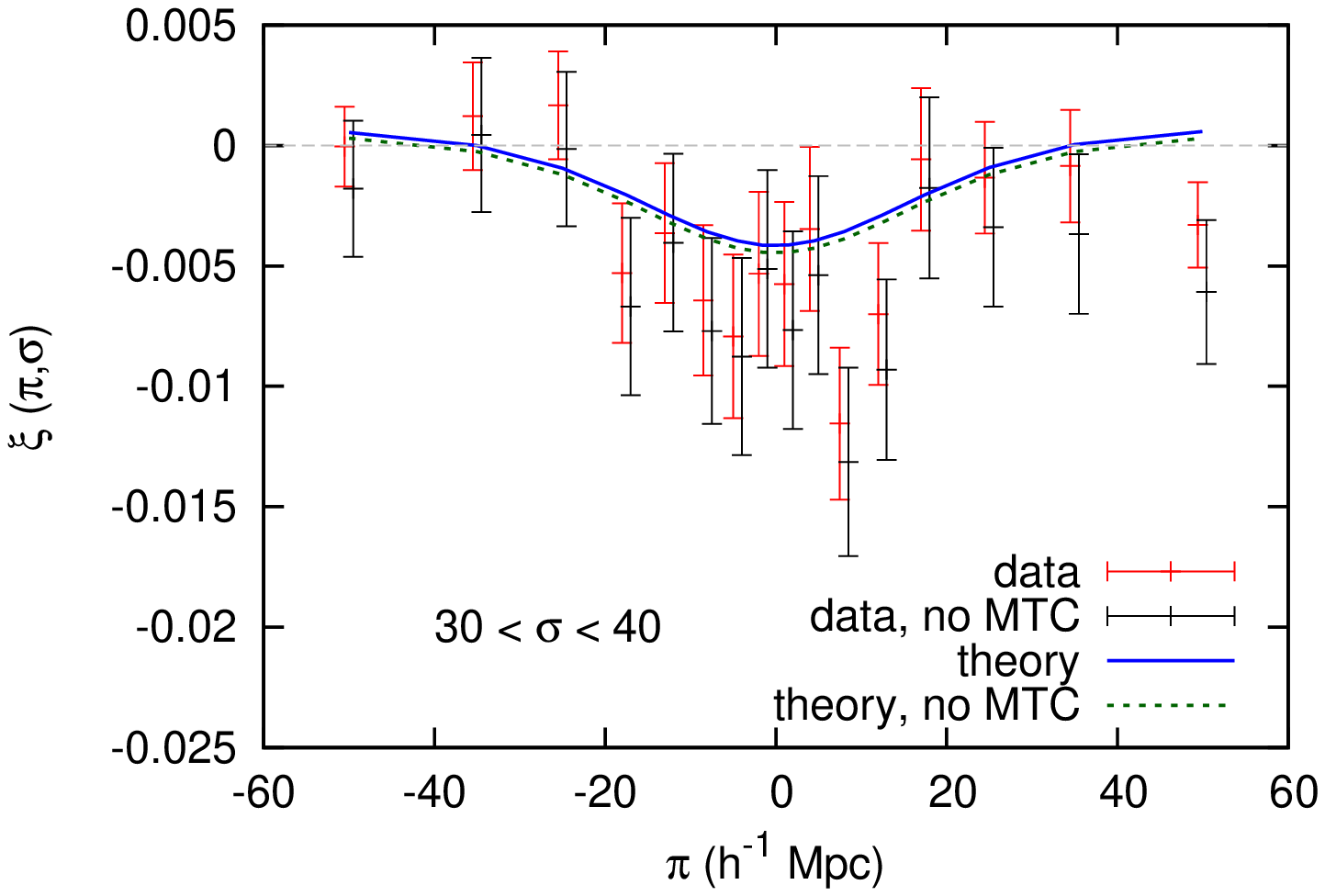} \\
  \end{tabular}
 \end{center}
 \caption{Effect of the MTC on the measured cross-correlation in the
four selected bins of $\sigma$ that are indicated.
%, from top-left to bottom-right,
%    $1 < \sigma < 4 \hmpc$,
%    $7 < \sigma < 10 \hmpc$,
%    $15 < \sigma < 20 \hmpc$, and
%    $30 < \sigma < 40 \hmpc$.
    The fiducial measurement is plotted with red errorbars, while the 
    measurement without the MTC is shown in black (with a small horizontal 
    shift for better visualization). The blue solid line is 
    the fiducial theory including the MTC from appendix \ref{app:correc},
    and the dotted green line without the MTC.
  }
 \label{fig:MFC}
\end{figure}

  The measurements with and without the MTC are compared in figure
\ref{fig:MFC} (red and black errorbars), for four selected $\sigma$
bins. The best fit model is also shown with and without including the
correction of appendix \ref{app:correc} (blue solid and dotted green
line, respectively). Table \ref{tab:bias} shows the result for the bias
for the NOMTC case (uncorrected data and uncorrected fit model), and
for the case where we apply the MTC to the data but we leave the fitted
model uncorrected (NOCOR case).

\begin{table}
 \centering
  \begin{tabular}{c|cccc}
    	       & $b_D$ & BS errors & MCMC errors & $\chi^2$ (d.o.f) \\
\hline
    FIDUCIAL   &  2.17 &      0.20 &      0.20 &   106 (125) \\     
    NOMTC      &  2.11 &      0.21 &      0.22 &   109 (125) \\     
    NOCOR      &  2.00 &      0.19 &      0.20 &   111 (125) \\     
    NODLA      &  2.25 &      0.22 &      0.21 &   109 (125) \\     
\hline
    LOWZ ($2<z<2.25$)   &  2.18 &     0.41 &      0.33 &   116 (125) \\     
    MIDZ ($2.25<z<2.5$) &  2.16 &     0.32 &      0.34 &   109 (125) \\     
    HIGHZ ($2.5<z<3.5$) &  1.88 &     0.57 &      0.37 &    92 (125) \\     
\hline
    LOWNHI ($\log(N_{HI})<20.4$)  &  2.27 &   0.30 &   0.29 &  133 (125) \\     
    HIGHNHI ($\log(N_{HI})>20.4$) &  1.89 &   0.26 &   0.30 &  110 (125) \\     
  \end{tabular}
  \caption{Best fit value of the DLA bias with its bootstrap errors, 
   MCMC errors and $\chi^2$ value of the fit, for various analyses:
    FIDUCIAL (with the MTC and the theory  
     corrected with the expression derived in appendix \ref{app:correc}),
    NOMTC (PCA-only continuum fitting, uncorrected theory),
    NOCOR (MTC, uncorrected theory),
    NODLA (spectra containing DLAs are rejected),
    data split in redshift bins (LOWZ, MIDZ, HIGHZ) and finally the DLA
    sample split in two bins of column density (LOWNHI, HIGHNHI).
  }
  \label{tab:bias}
\end{table}

  The effect of the MTC is clearly a minor one; it has little effect on the 
result on the DLA bias, and the errorbars are only increased by $\sim$ 10\% 
when the continuum fitting errors are included.
%. The covariance error is increased
%by less than 10\% when the continuum fitting errors are included, and the
%$\chi^2$ and the bootstrap error actually decline slightly in the NOMTC
%case. In section \ref{ss:scale} we shall show that the effect of the MTC 
%becomes more important when fitting the bias on the largest separations. 
%This may indicate that the continuum errors are not well corrected by the MTC.
 The effect of the correction of appendix \ref{app:correc} is small compared 
to the present errorbars of the measurements, but our model characterization 
of the effect of the MTC is
supported by the larger value of $\chi^2$ in the NOCOR case.
%\dps{suggests to not split the names, NOCORR and so on}
%\jm{We figured from previous paper that JCAP style is to split these names}
We keep the correction despite the fact that with the present level of
noise its effect is not significant, because it will be important for
future studies with smaller errorbars and larger separations.
%\jm{The smaller errorbars when MTC is included are just a consequence
%of adding continuum errors to the covariance matrix in the NO MTC case,
%right? So that is no proof that the MTC is adequate...}
%\af{Sure. The only place where the MTC step helps is in the determination 
%of $\beta_F$, and I guess also on the errorbars at very large scales as seen
%in figure 10.}

\subsection{Tests of the covariance matrix}

%  We have seen that the bootstrap error on the fitted bias parameter of
%the DLAs is generally slightly larger ($\sim 20 \%$) than the one computed 
%from our covariance matrix (see table \ref{tab:bias}). 
  We have seen that the bootstrap error on the fitted bias parameter of the 
DLAs is generally in good agreement with the one computed from our covariance
matrix using the MCMC technique (see table \ref{tab:bias}).
%As discussed in section \ref{ss:cov}, several approximations are made in
%our calculation of the covariance matrix that could introduce 
%probably explain this
%small difference. 
As a further test of the accuracy of our errors, we now compare the scatter 
between the values of the measured cross-correlation in bins, using 
the 12 sub-samples defined in section \ref{ss:boot}, with the predicted 
errors from the diagonal elements of the covariance matrix.

  For the purpose of this comparison, instead of averaging the
cross-correlation values from the sub-samples weighted with their
inverse covariance matrices, as in equation \ref{eq:total_xi}, we 
average by weighting each sub-sample with the sum of the weights
of all the pixels that contributed to each bin $A$, $\tilde w_A^\alpha$, 
where the index $\alpha$ labels each sub-sample. This estimate of the
cross-correlation in bin A is
\begin{equation}
 \hat{\xi_A} = \frac{\sum_\alpha \hat\xi_A^\alpha \tilde w_A^\alpha}
		{\sum_\alpha \tilde w_A^\alpha} ~.
\end{equation}
and the uncertainty $\sigma_A$ of this estimate is
\begin{equation}
 \sigma_A^2 = \frac{1}{\left( \sum_\alpha \tilde w_A^\alpha \right)^2} 
        \sum_\alpha (\tilde w_A^\alpha)^2  
		\left[ (\hat\xi_A^\alpha)^2 - (\hat{\xi_A})^2\right] ~.
 \label{eq:rms}
\end{equation}

\begin{figure}[h!]
 \begin{center}
  \begin{tabular}{cc}
   \includegraphics[scale=0.5, angle=-90]{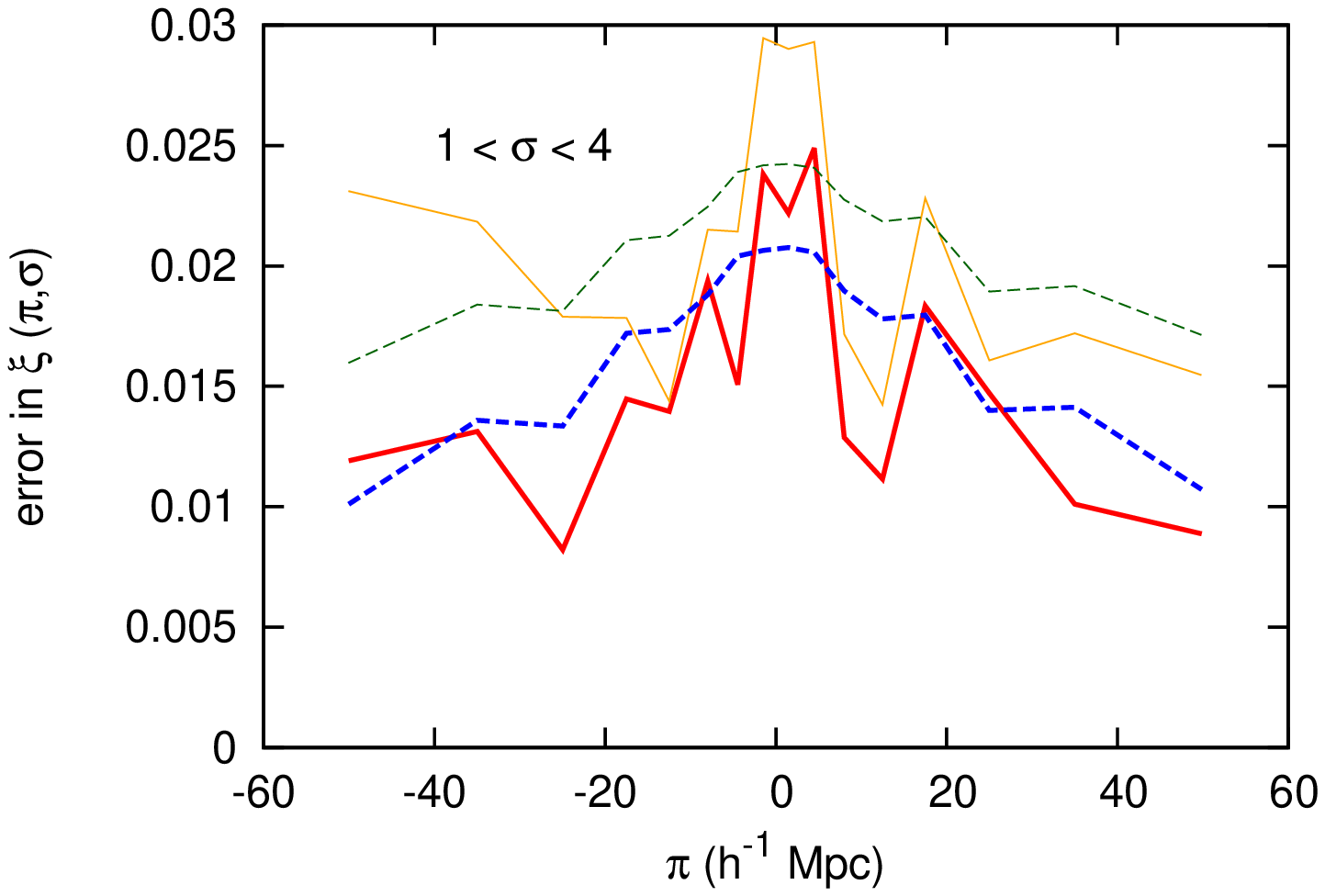} &
   \includegraphics[scale=0.5, angle=-90]{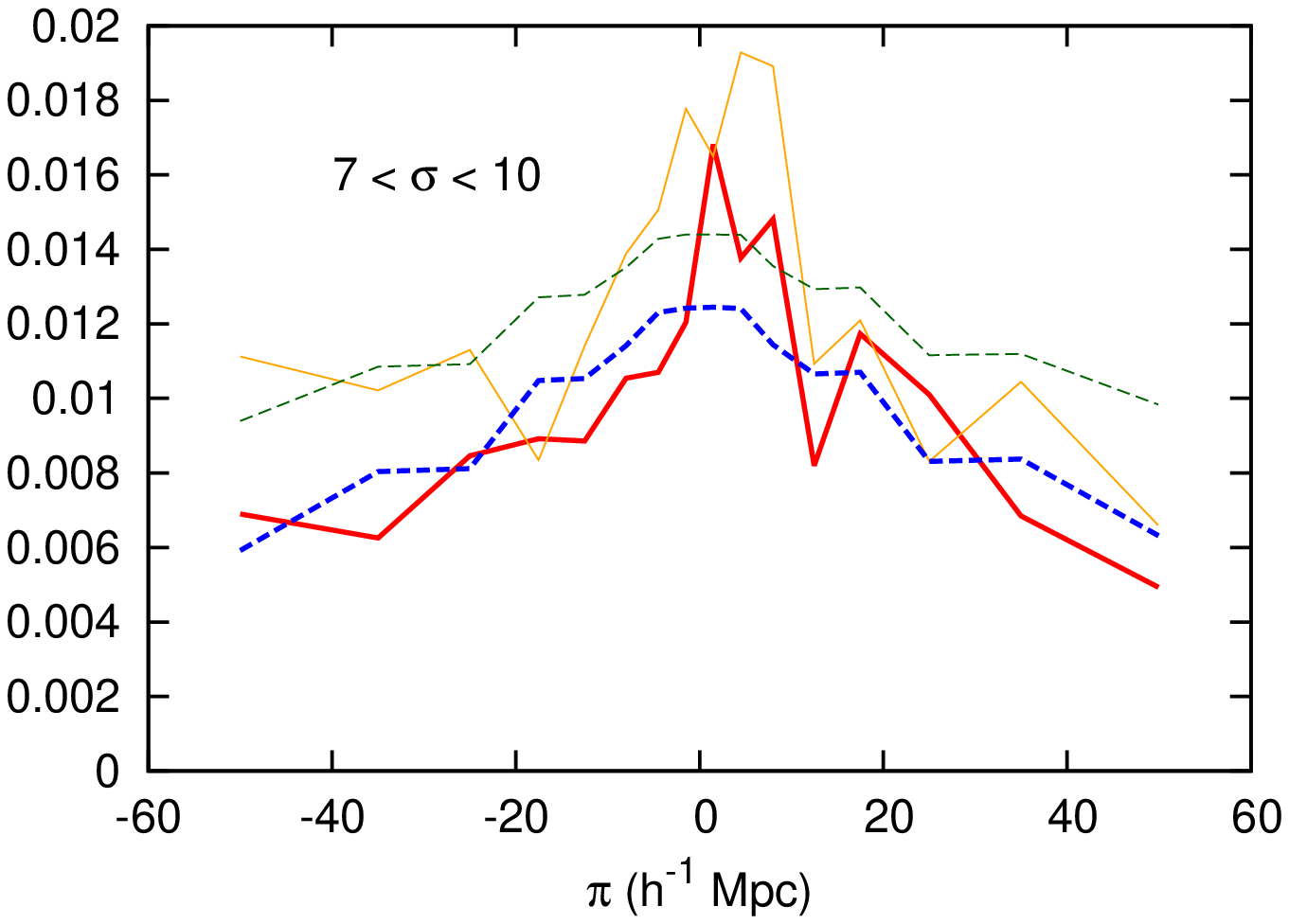} \\
   \includegraphics[scale=0.5, angle=-90]{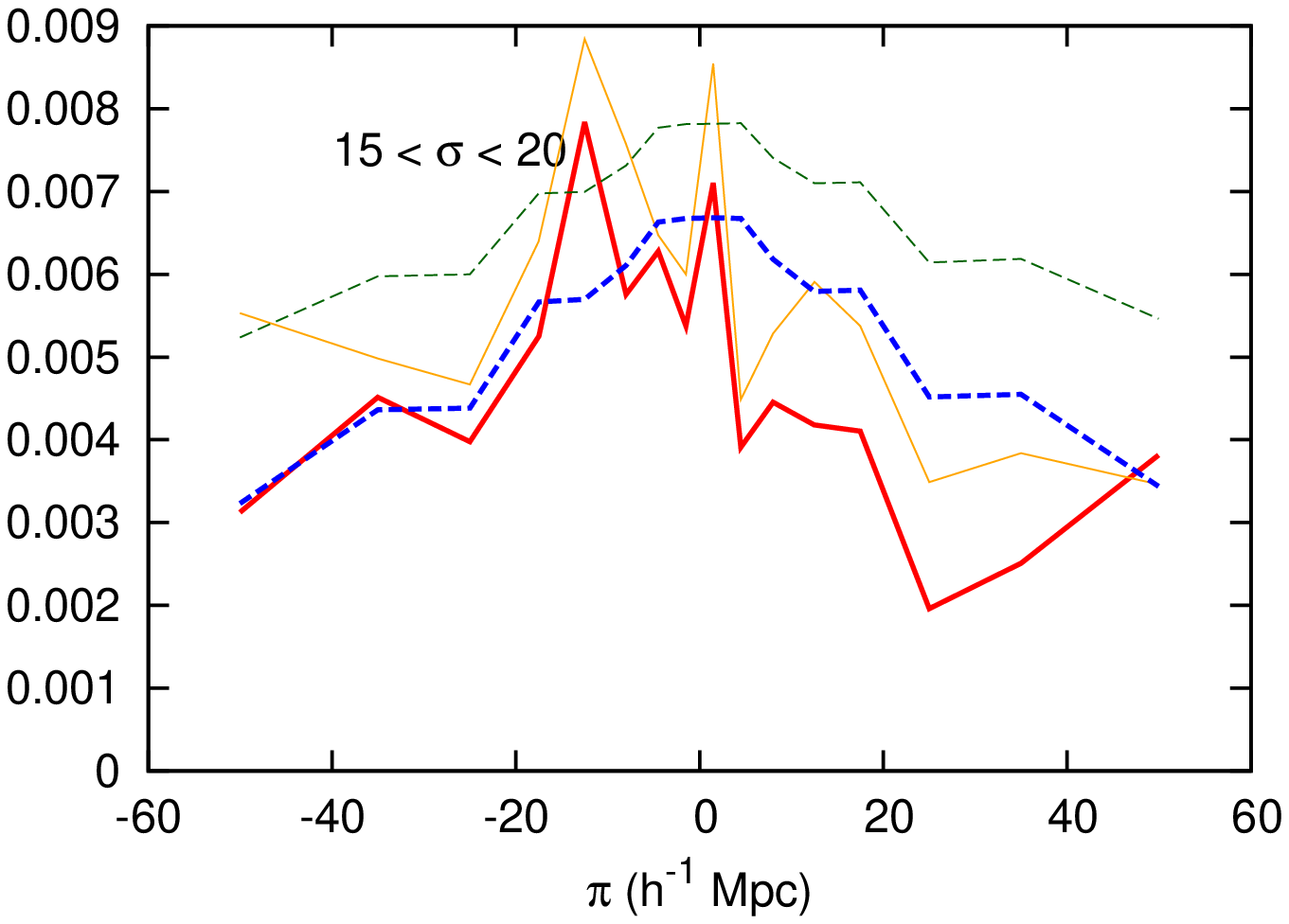} &
   \includegraphics[scale=0.5, angle=-90]{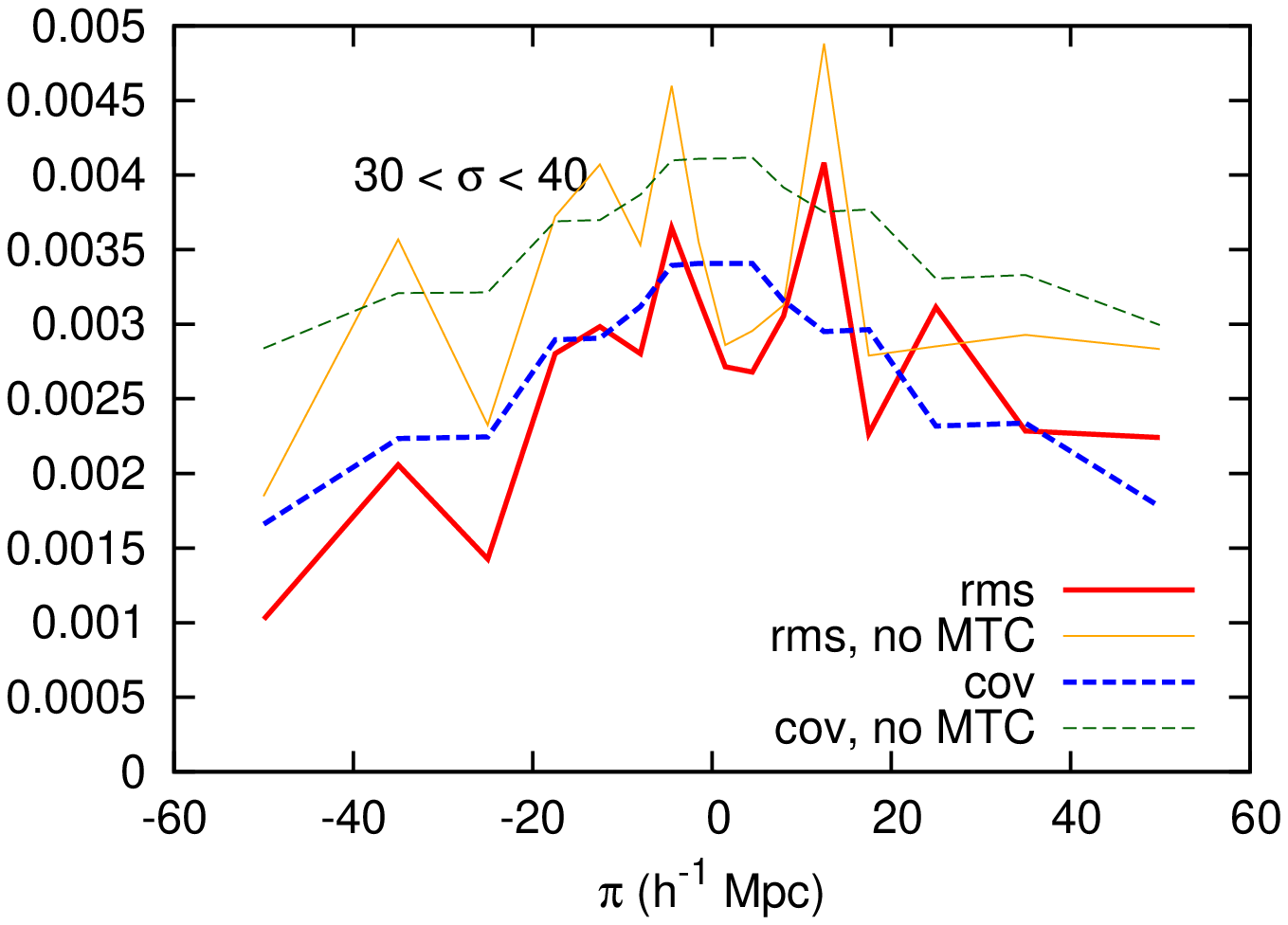} \\
  \end{tabular}
 \end{center}
 \caption{Errors of the cross-correlation in four selected bins of
$\sigma$, as indicated in each panel.
%from top-left to bottom-right,
%   $1 < \sigma < 4 \hmpc$,
%   $7 < \sigma < 10 \hmpc$,
%   $15 < \sigma < 20 \hmpc$, and
%   $30 < \sigma < 40 \hmpc$.
   The solid lines show the estimate from the scatter between sub-samples 
   (equation \ref{eq:rms}), while the dashed lines show the estimate from 
   the diagonal elements of the covariance matrix. 
 }
 \label{fig:errors}
\end{figure}

%\dps{Compress panels and add $\hmpc$ in the legend}

In figure \ref{fig:errors} the errors computed from the 
scattering among the sub-samples and from the diagonal elements of the
covariance matrix are compared. The errors from the scatter among
sub-samples (solid lines) are noisy because of the small number of
sub-samples, but in general they agree remarkably well with the 
ones obtained from the covariance matrix (dashed lines).
The thick lines are for the fiducial case, and the thin lines for the
NOMTC case. As before, the NOMTC case incorporates continuum fitting
errors that account for the larger covariance errors by $\sim$ 10\%.
%Generally the NO MTC case has larger bootstrap errors
%than the MTC case, as we would expect if the MTC actually helps
%reducing the uncertainty in the cross-correlation measurement,
%although the value of the DLA bias parameter increases its bootstrap
%when the MTC is applied (table \ref{tab:bias})
Generally the NOMTC case has larger scatter than the MTC case, as we would
expect if the MTC actually helps reducing the uncertainty in the 
cross-correlation measurement. 
%\jm{Do we have any idea why that is??? Perhaps the correlated errors
%have some funny property that is not seen in these plots?}
%\af{It's just fluctuations. The important thing is that the error on the bias
% gets larger when using only large separations (figure 11)}

%\kg{The beginning of the section says you're trying to figure out why the 
%bootstrap and likelihood errors are discrepant, but you don't actually 
%explain the difference explicitly anywhere in the section.}

\subsection{Evidence for the redshift distortions}
\label{ss:reddist}

  The gravitational evolution origin of the DLA-\lya absorption
cross-correlation we detect is tested by the presence of the predicted
redshift distortions in linear theory. The expected elongation of 
contours perpendicular to the line of sight is apparent in figure
\ref{fig:kaiser}. By fitting only the DLA bias factor and making the
reasonable choice $\beta_F=1$ based on previous observations of the
\lya autocorrelation in \cite{2011JCAP...09..001S}, the observed
anisotropy is consistently matched by our linear model. The detection
of redshift distortions can be quantified by comparing with a fit that
forces isotropy requiring $\beta_F = \beta_D = 0$. This isotropic fit
yields $\chi^2 = 150$, implying a detection of more than $6\sigma$ when
compared to $\chi^2=106$ (125 d.o.f.) for our best fit model with 
$\beta_F=1$ and
$\beta_D=f(\Omega)/b_D$ (points at $r< 5 \hmpc$ are not used in the
fits in both cases). 
%Even though the significance is slightly
%reduced if we take into account the larger size of our bootstrap errors
%compared to those derived from the covariance matrix, the detection
%of redshift distortions is still extremely robust. 
This result provides strong
support for the interpretation that the measured cross-correlation is
induced by the gravitational growth of large-scale structure that is
traced by the DLAs and the \lya forest.
%Without the "mean flux regulation", we obtain
%$\chi^2 = 153$, i.e., also a $6\sigma$ detection.

\subsection{Scale dependence of the DLA bias}
\label{ss:scale}

\begin{figure}[h!]
  \begin{center}
   \includegraphics[scale=0.70, angle=-90]{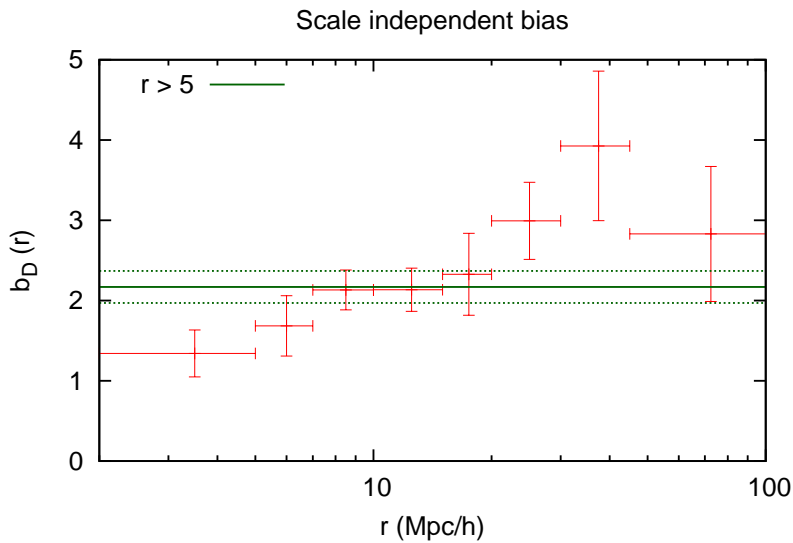}
  \end{center}
  \caption{Fitted DLA bias in several bins of the separation $r$ for our
   fiducial analysis. % (left) and for the NOMTC case (right).
   The green lines show the values obtained when combining all bins above
   $r > 5 \hmpc$.
  }
 \label{fig:bias_r}
\end{figure}

In addition to the redshift distortions, the predicted dependence of
the cross-correlation on the separation $r$ in the $\Lambda$CDM model is
another important test that our measurement agrees with the theoretical
expectation. This can be rephrased as testing that the inferred
DLA bias factor is constant in different intervals of $r$.
The results of this exercise are shown in figure \ref{fig:bias_r}.
% for the fiducial model  and for the analysis without the MTC step. 
%Even though the uncertainty on the global measurement of the DLA bias is 
%similar in both analysis, in this figure we can see that on large separations 
%the MTC step allows us to get smaller errorbars.

%\af{I thought it could be good to add also the noMTC model since its on the 
%largest scales that one can see how it helps}

  The value of the bias is consistent with being constant for
$r > 5 \hmpc$. At smaller scales we expect non-linear effects to be
important, so it is not surprising that the first point has a discrepant
value of the bias. The first bin that we use in our fits,
$5 \hmpc < r < 7 \hmpc$, also has a lower bias than the ones at larger
radii by $1.3\sigma$, and it may already be affected by non-linear
effects to some degree. In addition to the physical
non-linearities in the distribution of gas and DLAs in redshift space,
the transformation from optical depth to the transmitted fraction
introduces an additional non-linearity that in this case may be the
dominant effect.

  The presence of non-linearities may be accounted for in the future
by using predictions from hydrodynamic numerical simulations of
structure formation. When the full BOSS dataset is available, the reduced 
error bars will allow for a better measurement of these non-linear effects.
%\jm{It might be interesting to apply the simple correction assuming
%a log-normal distribution of optical depths, and that the optical
%depths are all multiplied by the 1+xi factor, to obtain a correction
%for the transmitted fraction. It probably raises the point at 5-7 Mpc
%by something like 5\% }
%\af{I think this is beyond the scope of this project. May be for QSO-LyaF.}
%\jr{comments on $\pi$ symmetry.}
%\af{Will add here notes fits on positive and negative $\pi$ only, they agree.}
  We have also checked the symmetry of the cross-correlation under a
sign change of the $\pi$ coordinate. There is no statistically
significant difference between
the fitted fiducial model when only the bins with $\pi>0$ or those
with $\pi < 0$ are used. %, and the values of $\chi^2$ are good in both
%cases.
%\dps{What do you mean by "good"? consistent fits?}

%$$b_D = 2.11 \pm 0.34 (BS) \pm 0.25 (MC) ~ , \quad \chi^2 = 63 (63 d.o.f.)
%  \qquad , \qquad \pi < 0$$

%$$b_D = 2.18 \pm 0.25 (BS) \pm 0.25 (MC) ~ , \quad \chi^2 = 62 (63 d.o.f.)
%  \qquad , \qquad \pi > 0$$

%\af{should think carefully about the minimum r to use from a theoretical 
%point of view. At what scale does the linear bias model break down at this 
%redshift for these halo masses?}

\subsection{Effect of the DLAs on the \lya forest}
\label{ss:noDLA}

  In our fiducial analysis we include in the \lya sample 3047 lines of sight
that contain at least one DLA. As explained in section \ref{sec:data},
we apply a mask on the central region of the absorption profiles and correct 
for the damped wings by multiplying the continuum estimate with a Voigt profile.
To test the effect of this correction we measure the 
cross-correlation using only the remaining 49402 lines of sight. The results
can be seen in table \ref{tab:bias} under the name NODLA, and are clearly 
consistent with our fiducial analysis.

\subsection{Dependence on the \lya bias}
\label{ss:beta}

  Our fits for the DLA bias are based on assuming fixed values for
the \lya forest bias parameters: the well constrained combination
$b_F (1+\beta_F) = -0.336$ from \cite{2011JCAP...09..001S}, and
$\beta_F=1$. The value of $\beta_F$ is still highly uncertain, but
improved measurements are expected in the near future from BOSS.
We therefore examine the dependence of our result for $b_D$ on the
assumed value of $\beta_F$.

\begin{figure}[h!]
  \begin{center}
   \includegraphics[scale=0.7, angle=-90]{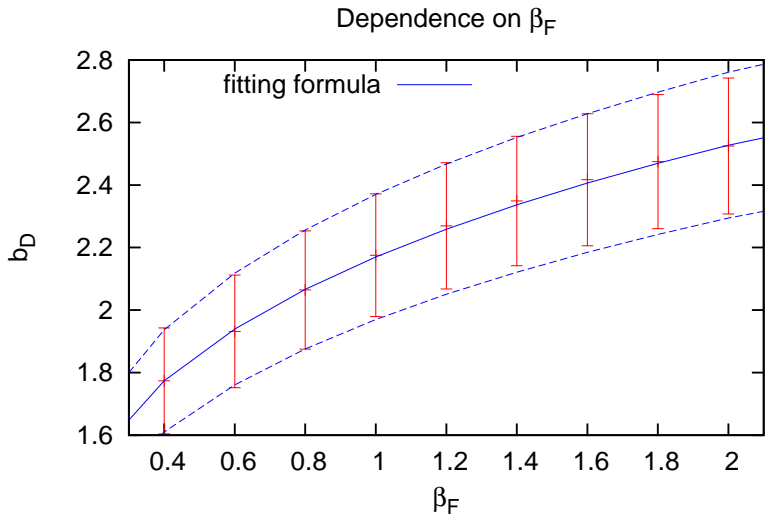}
  \end{center}
  \caption{Best fit value of the DLA bias for different values of $\beta_F$, 
   for our fiducial analysis. The dependence is well described with the 
   fitting formula $b_D = (2.17 \pm 0.2) \beta_F^{0.22}$ (blue line). 
  }
 \label{fig:fits_beta} 
\end{figure}

In figure \ref{fig:fits_beta} we plot the results assuming different values
of $\beta_F$. We show that for the most plausible range $0.4 < \beta_F < 2.0$,
the dependence is well fitted by a simple function
\begin{equation}
  b_D = \left(2.17  \pm 0.20 \right) \beta_F^{0.22} ~.
 \label{eq:fitted}
\end{equation}

\begin{figure}[h!]
  \begin{center}
   \includegraphics[scale=0.55]{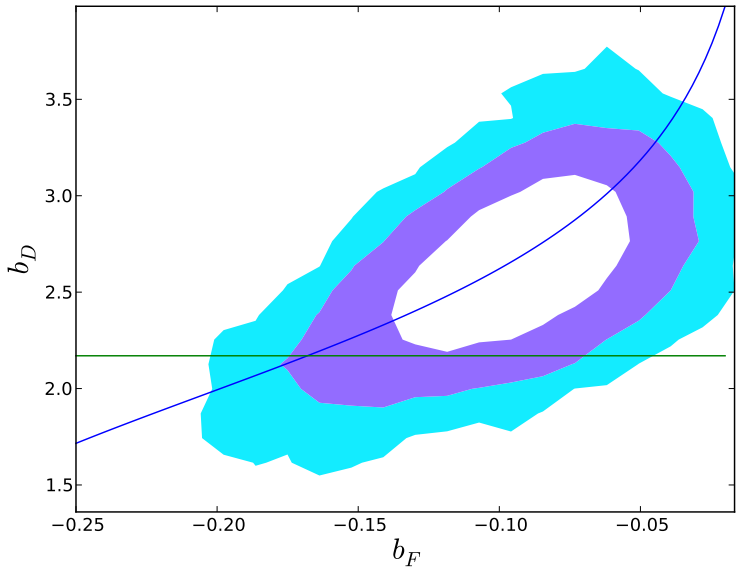}
  \end{center}
  \caption{Degeneracy between the value of $b_D$ and $b_F$ when fitting both
   parameters at the same time in our fiducial analysis. Since we keep fixed 
   the well constrained quantity $b_F (1 + \beta_F) = - 0.336$, 
   each value of $b_F$ is related to a value of $\beta_F$.
   The horizontal green line shows the best fit value when fixing $\beta_F=1$
   ($b_F=-0.168$), $b_D=2.17$. The blue line shows the fitted function in 
   equation \ref{eq:fitted}.
  }
 \label{fig:joint} 
\end{figure}

Figure \ref{fig:joint} shows the likelihood contours for the
two-parameter fit when $b_F$ is allowed to vary, keeping fixed the
combination $b_F (1 + \beta_F) = -0.336$, for our fiducial analysis.
The data clearly prefer larger values of $\beta_F$ (smaller values
of $b_F$) than the one we have assumed so far ($\beta_F=1$, $b_F=-0.168$). 
However, the best fit value of $\beta_F$ is quite sensitive to the applied 
correction, and so we prefer to be conservative by assuming a fixed, low
value of $\beta_F$, minimizing the required value of the DLA bias.

%\af{I suggest only ploting bias-bias plot for the fiducial case.}

\subsection{Redshift evolution}
\label{ss:zevol}

Until now we have assumed that the value of the DLA bias is constant with 
redshift. We now evaluate the cross-correlation in three different redshift 
bins, 
$2.0 < z < 2.25$ (LOWZ), $2.25 < z < 2.5$ (MIDZ) and $2.5 < z < 3.5$ (HIGHZ). 
The third bin is wider because the \lya forest data available
for cross-correlation is much more sparse at high redshift in BOSS.
We assume the redshift evolution of the \lya forest bias factor to be
$b_F \propto (1+z)^{2.9}$, based on the evolution of the line of sight
power spectrum measured in \cite{2006ApJS..163...80M}.
The best fit values for the three redshift bins, shown in table
\ref{tab:bias}, provide some evidence that the DLA bias decreases
with redshift. This evolution is opposite to that expected if the DLA host 
halos are approximately at constant velocity dispersion: in this case the DLA
bias should increase with redshift as the host halos become
increasingly rare fluctuations. The observed trend is, however, of
low statistical significance.

\subsection{Dependence on column density}

  Finally, we split our sample of DLAs into two data sets as a function of 
column density, separated at $N_{HI} = 10^{20.4} \cm^{-2}$. If systems with 
higher column density resided in halos of higher mass, one would expect the 
bias of the high column density sub-sample (HIGHNHI) to be larger than that of 
the low column density (LOWNHI). The results shown in table \ref{tab:bias} are 
consistent with a bias that is independent of column density, showing, if 
anything, an opposite trend that is not statistically significant.
%The $\chi^2$ of the LOWNHI sample is marginally discrepant; this is influenced 
%by two bins that appear to contain real statistical fluctuations and is not 
%indicative of any actual deviation from the model, together with the effect of 
%a slight underestimation of the elements of our covariance matrix.
%\jr{What do you mean? How does something ``appear'' to be statistical?}

  A possible systematic error that may affect our DLA
bias is the inclusion in our DLA catalogue of systems that are not
actual DLAs, but regions of absorption with lower column density than
our threshold which pass our column density cut because
of a combination of spectral noise and the \lya forest superposed on
the damped wings. At the same time, DLAs that are just above our
column density cut may more easily be attributed a column density
below the threshold (and be therefore omitted from the catalogue) when
the superposed \lya forest is particularly weak. If this error is important, 
the \lya forest superposed on the wings of DLAs may
introduce a correlation with the \lya forest in nearby lines of sight
that would systematically enhance our measured $b_D$. We do not believe
this effect is important because preliminary tests of the DLA
selection with mocks show that there is a low rate of miss-identifications
(\cite{2012A&A...547L...1N}), and because of the consistency of the radial
dependence of the cross-correlation and the anisotropy with theoretical
expectations found in sections \ref{ss:reddist} and \ref{ss:scale},
but we plan to examine this selection effect
more carefully in the future using the techniques described in
\cite{2012JCAP...07..028F}.

%\begin{itemize}
% \item red vs blue forest
% \item $\pi>0$ vs $\pi<0$
% \item bright / faint spectra
% \item different continua (I tried pipeline continua, no difference)
%\end{itemize}

\subsection{Result for the sample affected by BAL Contamination}

As seen in figure \ref{fig:lr}, there is a significant excess of systems 
detected in the window    
$1005 {\, \rm\AA} \le \lambda_r \le 1037 {\, \rm\AA}$, probably due to 
contamination by BALs with small Balnicity index $BI$. The systems detected 
in this window were rejected in the fiducial analysis.
We measured the cross-correlation using only the 989 systems that fall 
into this range of restframe wavelength, including those detected in lines
of sight with a Balnicity index in the range $0 < BI < 1000 \kms$.
The measured bias is $b_D=0.68 \pm 0.50$ (BS errors), showing that at
least half of these systems are indeed contaminants, and confirming that
the measured cross-correlation is present only for physically
real DLAs.

\section{Discussion}
\label{sec:disc}

%\af{add comments on recent models suggested by Nic Ross: 
% \cite{2010arXiv1010.5014C}, \cite{2009MNRAS.397..411T} and 
% \cite{2011MNRAS.418.1796F}
%}
%\af{Add also comments on recent paper suggested by Joe Hennawi: 
% \cite{2012arXiv1201.3653E} 
%}
%\af{Patrick Petitjean suggested this recent paper by Matteo Viel on 
%galactic winds: \cite{2012arXiv1207.6567V}
%}

  The measurement of the cross-correlation of DLAs and the \lya forest
absorption, presented for the first time in this paper, provides a new
observational constraint on the nature of DLAs: the bias factor of their
host halos. We have measured this bias factor to be
$b_D = (2.17 \pm 0.20)\beta_F^{0.22}$, at a mean redshift of $z=2.3$.
The analysis of numerical simulations of the \lya forest predict a value
of $\beta_F\simeq 1.5$ (\cite{MCDON03}), and even though the first
measurement of $\beta_F$ from the \lya forest autocorrelation
\cite{2011JCAP...09..001S} allowed for a wide range of $\beta_F$ around
unity, a more recent analysis also suggests a value larger than one
(Slosar, private communication).

  As discussed in the introduction, the bias factor constrains the mass
distribution of the host halos of DLAs. A small bias factor implies low
halo masses and an association of most DLAs with dwarf galaxies, while
a large bias factor means that most DLAs are extended clouds in
massive halos. The observational determination of the mean bias factor
is therefore a new constraint that models of DLAs must satisfy.

  Theoretical models of DLAs need to provide the average cross section of a
halo of mass $M$ for producing an absorption system with $N_{HI} >
10^{20.3}\cm^{-2}$. The predicted rate of incidence is given by equation
(\ref{eq:rate}), and the new observable of the mean bias factor to be
matched by models is obtained with equation (\ref{eq:biash}).
Numerical simulations of galaxy formation following the collapse and
radiative cooling of gas in halos have found that the rate of incidence
predicted for DLAs is affected by two uncertain factors: the limited
resolution of the simulations and the impact of galactic winds
(\cite{1996ApJ...457L..57K}, \cite{1997ApJ...484...31G}, 
\cite{2001ApJ...559..131G}). As the
resolution is improved and halos are well resolved down to the lowest
masses for which photoionized gas can collapse, the rate of incidence
is increased; at the same time, when a prescription for supernova-driven
galactic winds is included leading to the ejection of gas from low-mass
halos when star formation occurs, the gas mass and the DLA cross
sections in low-mass halos decrease as the energy that is deposited in
winds is
increased (\cite{2004MNRAS.348..421N},\cite{2007ApJ...660..945N},
\cite{2008ApJ...683..149R}, \cite{2008MNRAS.390.1349P}).

  These models based on cosmological simulations do not at present
provide fundamental physical predictions for the properties of DLAs,
because the way that the gas distribution in halos is affected by
cooling and fragmentation, radiative transfer, star formation and
galactic winds depends on extremely complex physical processes that can
only be roughly approximated through simple parameterized recipes that
are incorporated into the equations being solved in the simulations.
Nevertheless, they can provide a basic guide as to the general
characteristics of star formation rates and wind strengths that are
needed to satisfy the observational constraints.

\subsection{Constraints on the halo mass distribution from the bias factor}

  We now examine the expected bias factors for simple relations of the
DLA cross section and the halo mass that are useful fits to the results
found in the numerical simulations of DLAs. The DLA systems are likely
distributed over a wide range of halo mass, but they are not expected in
halos that are too shallow to hold photoionized gas and allow it to
radiatively cool. This condition corresponds to a halo circular
velocity of $v_c\sim 20 \kms$, or a halo mass at $z=2.3$ $M_h \sim
 v_c^3/(3\pi HG)\sim 10^9 \msun$. 
Many numerical models of DLAs, starting with
\cite{1997ApJ...484...31G}, 
have obtained fits
to simulation results with a power-law relation for the cross section,
\begin{equation}
 \Sigma(M)=\Sigma_0 (M/M_{min})^\alpha \qquad\qquad ( M > M_{min} ) ~.
\label{eq:plcs}
\end{equation}
The distribution of halo masses and the bias factor of halos has been
thoroughly examined in analytic models and numerical simulations and
are robustly predicted in the $\Lambda$CDM model 
(e.g., \cite{1999MNRAS.308..119S}; \cite{2010ApJ...724..878T}). 
Here we use the mass distribution and halo-bias relation given by 
\cite{2010ApJ...724..878T}.

\begin{figure}[h!]
 \begin{center}
  \includegraphics[angle=-90,scale=0.5]{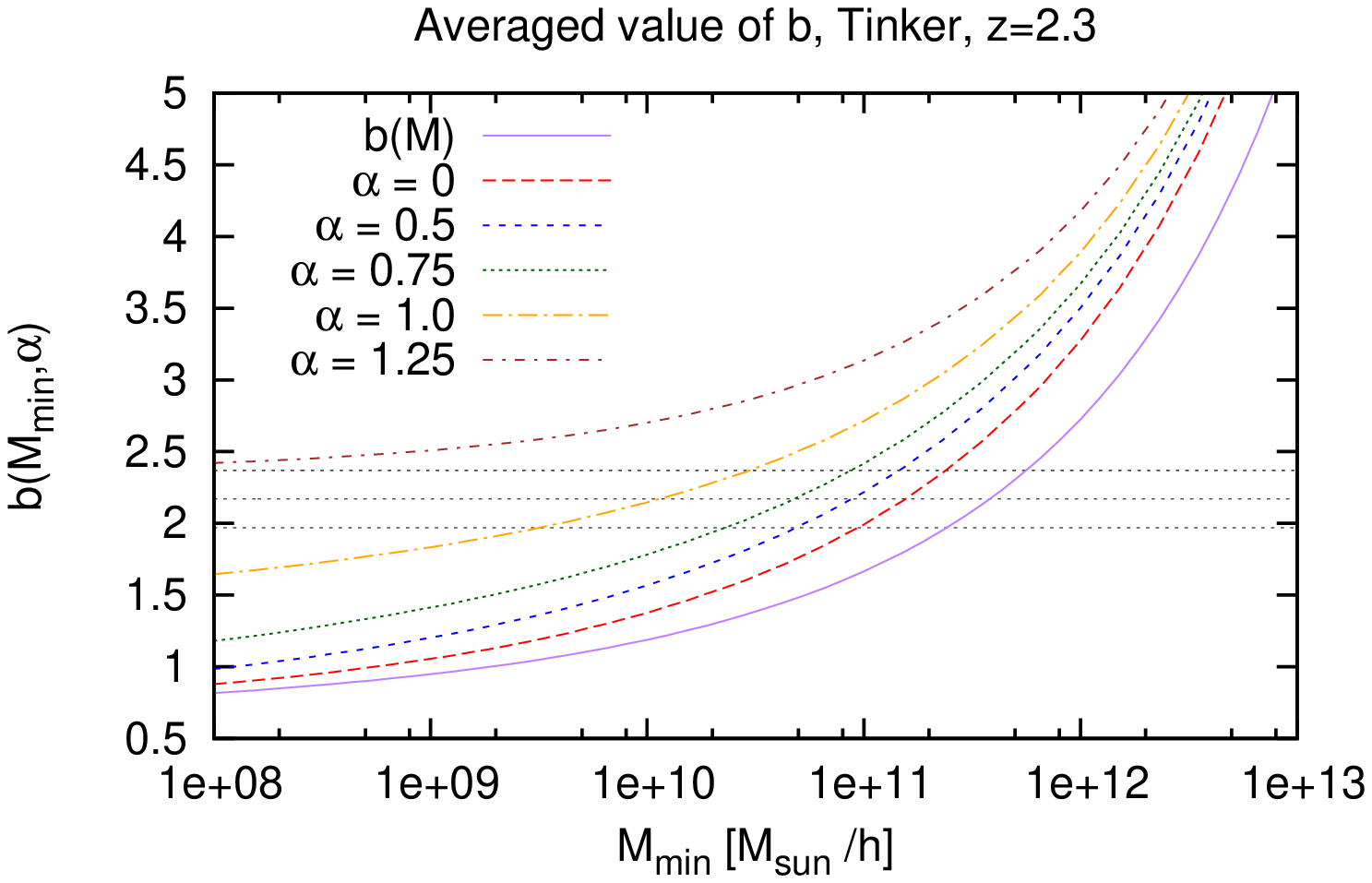}
  \includegraphics[angle=-90,scale=0.5]{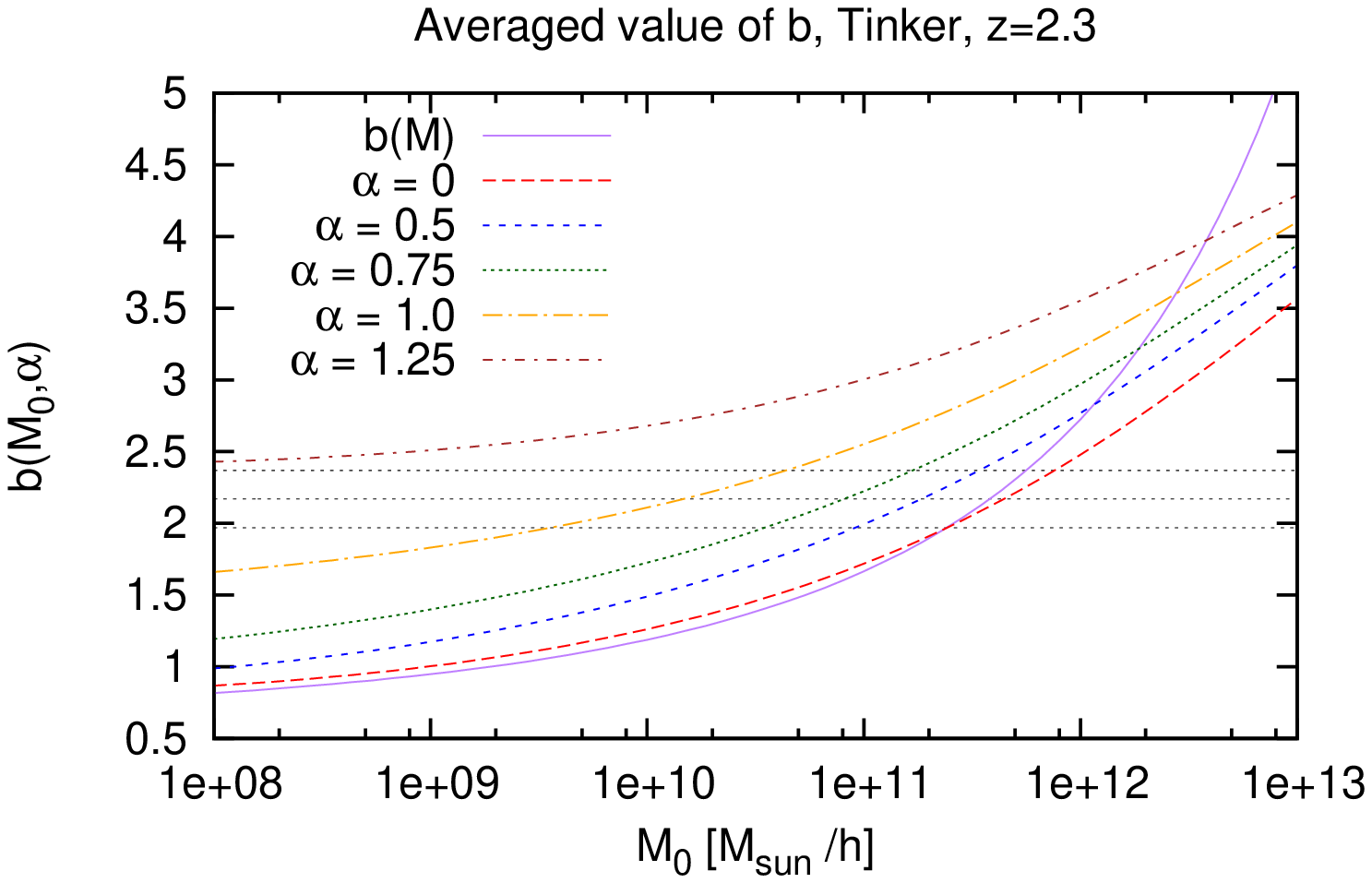}
 \end{center}
 \caption{Average bias as a function of $M_{min}$ when the cross
section depends on halo mass as $M_h^\alpha$ with a sharp lower limit
at $M_{min}$ (left), and for the model of equation (\ref{eq:pontzcs})
with a smooth suppression at masses below $M_0$, for several values
of $\alpha$, computed at $z=2.3$ using \cite{2010ApJ...724..878T}. The 
horizontal dotted lines show the result of this paper for the mean DLA
bias with 1$\sigma$ errorbars.
}
 \label{fig:bias_alpha}
\end{figure}

%\begin{figure}[h!]
% \begin{center}
%  \includegraphics[angle=-90,scale=0.5]{bias_alpha_z3}
%  \includegraphics[angle=-90,scale=0.5]{bias_alpha_pontzen_z3}
% \end{center}
% \caption{Same figure, evaluated at z=3.}
% \label{fig:bias_alpha_z3}
%\end{figure}

  In the left panel of figure \ref{fig:bias_alpha}, the halo mass-bias
relation of \cite{2010ApJ...724..878T} is shown as the solid purple line.
The other lines show the average value of the bias weighted according
to the power-law cross section of equation (\ref{eq:plcs}), for several
values of $\alpha$, as a function of $M_{min}$. All the relations are
computed at $z=2.3$, the mean redshift of our DLA bias measurement.
Halos having the value of the mean bias we have measured for DLAs have
masses of $\sim 6\times 10^{11}\msun$ (we use $h=0.71$). Realistically,
the host halo masses should vary over a broad range, but this mass
should be a typical one. Our measurement of the bias factor clearly requires 
DLAs to be mostly associated with host halos of relatively massive galaxies.

  Fixing the lower limit in equation (\ref{eq:plcs}) at
$M_{min}\simeq 10^9 \msun$, the best value of $\alpha$ that fits our
measured bias for $\beta_F=1$ is $\alpha=1.1 \pm 0.1$.
To better interpret the implication of this result, consider a simple
model where the DLAs are clouds found within a fixed fraction $f_a$ of
the virial radius of the halo, $r_{vir}\propto M^{1/3}$, with a
covering factor $f_c$. Then the cross section of a halo is $\Sigma(M)
\propto f_a^2 f_c M^{2/3}$. In other words, $\alpha=2/3$ if DLAs are
found within a fixed fraction of the virial radius and have a constant
covering factor. This value of $\alpha$ is ruled out by our result
at the $4\sigma$ level for $M_{min}=10^9 \msun$. Therefore, DLAs must
cover an increasing fraction of the projected area of halos with
increasing mass, up to masses larger than $\sim 10^{12} \msun$.
For $\beta_F=0.6$, $\alpha=2/3$ is still ruled out at the $3\sigma$
level.

  Most of the models from cosmological simulations predict shallower
slopes than are implied by our result of the DLA bias. Models without
strong effects from galactic winds predict a slope close to $\alpha=2/3$
(see, e.g., \cite{2007ApJ...660..945N}, \cite{2008MNRAS.390.1349P},
\cite{2009MNRAS.397..411T}, \cite{2012arXiv1201.3653E}, 
\cite{2012arXiv1209.2118B} for 
recent work on the effects of galactic winds and differences among
hydrodynamic codes). Models with the strongest winds are barely
consistent with our measured bias factor: model Q5 of
\cite{2004MNRAS.348..421N},\cite{2007ApJ...660..945N}, with $\alpha=1$ 
and $M_{min}\simeq 10^9 \msun$, yields a bias about $1.5\sigma$ below our 
measurement. The model of momentum-driven winds of 
\cite{2009MNRAS.397..411T}, with $\alpha=0.92$ at $z=2.25$, is also
barely consistent.

  The right panel in figure \ref{fig:bias_alpha} shows the same
relation of the mean bias for a model where the sharp cutoff of the
cross section at $M_{min}$ is replaced by a smooth function:
\begin{equation}
  \Sigma(M) = \Sigma_0 \left( M\over M_0 \right)^2 \, 
   \left(1 + {M \over M_0} \right)^{\alpha-2} ~.
\label{eq:pontzcs}
\end{equation}
We find that the numerical results of \cite{2008MNRAS.390.1349P} (see their
figure 4) are well matched by this formula with $\alpha=0.5$,
$M_0=10^{9.5} \msun$, and $\Sigma_0 = 40 \kpc^2$. Their results are
at $z=3$; to shift them to $z=2.3$ we assume that the properties
of DLAs stay constant in halos of fixed circular velocity, which
shifts their characteristic halo mass to $M_0 = 10^{9.7} \msun$.
Their model predicts a mean bias factor $b_D\simeq 1.35$, which is
ruled out by our measurement at more than the $3\sigma$ level if
$\beta_F > 1$. Even if $\beta_F=0.6$ their model is ruled out at
a confidence level of $2.5\sigma$. 

%\cite{2009MNRAS.397..411T} found a mild evolution of $\alpha$ with 
%redshift in their simulation with momentum-driven winds. If confirmed,
%this could reduce the tension between our measurement at $z=2.3$ and 
%the $z=3$ simulations of \cite{2008MNRAS.390.1349P}.

  Note that the models need to predict only the average DLA cross
section as a function of the halo mass, because the mean DLA bias does
not depend on the scatter of the cross section for a fixed halo mass.

%\af{Referee Report Comment \# 2: Regarding the alpha parameter that 
%parameterizes the cross-section: the simulations predict a relatively large 
%redshift evolution for alpha. This is not discussed in the paper (the bias 
%evolution is discussed). It would be nice to briefly discuss this, in order 
%to offer a more comprehensive way of comparison to the community of simulators.
%Does the tension with simulations become less evident at z=3? }
%
%\af{I don't know what simulation predicts this, I couldn't find any paper with
%this statement (actually, Tescari++ 2009).
% However, it is true that if one computes what is the effective
%bias from Pontzen++ at z=3 it is closer to the measurement (I have added a 
%figure for discussion). Since we argue that the DLA bias does not evolve 
%strongly with redshift, it might be possible that the tension is smaller at
%z=3}

%\af{We should add a comment on the fact that we compare z=3 simulations 
%to z=2.3 measurement assuming $\alpha$ is constant with redshift, and that 
%might not be the case (as found in simulations by Tescari et al. 2009).}

\subsection{Constraints on the cross sections from the rate of incidence}

  The  most recent determination of the rate of DLA incidence using the
BOSS 9th Data Release shows that a fraction $0.226 \pm 0.005$ of the sky
is covered by a DLA with $N_{HI} > 10^{20.3}\cm^{-2}$ in the redshift
range $2 < z < 3$, as derived from the column density distribution in
table 1 of \cite{2012A&A...547L...1N} (the error quoted here is only
the statistical one obtained from the same table, and does not include
the systematic error due to corrections for incompleteness and impurity).
This corresponds to a rate of incidence per unit of absorption pathlength,
$d\chi = (1+z)^2 H_0/H(z)\, dz$, equal to $0.066 \pm 0.002$, which we
assume to be constant with redshift
This observed rate of incidence fixes a required normalization of the 
relation of the DLA cross section and halo mass for any model.
%This is much larger than the fraction covered by observed
%galaxies over the same redshift range.

  We consider here a few examples
and examine the value of the physical cross section $\Sigma$ at the
characteristic halo mass $M_h=10^{12}\msun$ to illustrate the typical
implications of the observed rate of incidence and bias factor of DLAs:
\begin{enumerate}
 \item A model with $\alpha=1$ and $M_{min}=10^9 \msun$, which produces
a bias $b_D = 1.8$, requires
$\Sigma(M=10^{12} \msun) = 1400 \kpc^2$ to match the observed incidence
rate. This is a factor of two larger than the cross section in the
Q5 model of \cite{2004MNRAS.348..421N}, and a factor $1.5$ larger than
the prediction of the momentum-driven wind model of
\cite{2009MNRAS.397..411T}.

 \item A model with equation (\ref{eq:pontzcs}), with $\alpha=1$ and
$M_0=10^{10}\msun$, matches our observed bias for 
$\Sigma(M=10^{12} \msun) = 2400 \kpc^2$.
 \item A model with equation (\ref{eq:pontzcs}), with $\alpha=0.5$,
needs an increased $M_0 = 3\times 10^{11} \msun$ to match the observed
bias, and then requires $\Sigma(M=10^{12} \msun) = 4400 \kpc^2$.
\end{enumerate}
This illustrative example shows that the measured DLA bias factor can be
matched either by an extended DLA host halo distribution with
$\alpha > 1$ and a low mass cutoff, or with a narrower range of halos
around the characteristic mass of $10^{12} \msun$.
A more extended distribution of halo mass reduces the required
cross section because the rate of incidence is accounted for by a wider
range of halo masses.

\subsection{Consequences for the nature of DLAs}

  Generally, the models of DLAs discussed in the literature based on
hydrodynamic simulations of structure formation fail to correctly
predict the properties of DLAs for a common reason: the cross sections
they predict for massive halos ($M\simeq 10^{12}\msun$) are too
small. The models that come closest to matching the observations are the
Q5 model of strong winds in \cite{2004MNRAS.348..421N}, which still
predicts a rate of incidence and a baryonic content of DLAs that is a
factor of 2 below the observed ones, and the momentum-driven wind model
of \cite{2009MNRAS.397..411T}, which with $\alpha=0.92$ would predict
a DLA bias that is still $2\sigma$ below our best value for $\beta_F=1$,
and cross sections that are too small by a factor $\simeq 1.5$.
%basically be in agreement with observations if the area covered by DLAs
%and their baryonic content in the simulation could be increased by a
%factor of two in all halos.

  Our observational result therefore implies that some neglected
physical mechanism keeps the atomic gas spread out over an area larger
than predicted by present models in host halos of massive galaxies. The
natural expectation is that galactic winds that are even stronger than
in the Q5 model of \cite{2004MNRAS.348..421N} and momentum-driven wind
model of \cite{2009MNRAS.397..411T}, and that have an
increased impact on the halo gas distribution, are required.
We emphasize, however, that DLAs in low-mass halos do not need to be
further suppressed: in fact, these objects also help account for the
total rate of incidence. The requirement for stronger winds is to allow
atomic gas to stay at large radius over several orbits in massive halos.

  To better understand the consequences of our result, we consider the
Q5 model of \cite{2004MNRAS.348..421N} with all the cross sections
increased by a factor two, to match both the bias factor and rate of
incidence of DLAs. Then, the mean cross section in a $10^{12} \msun$
halo with circular velocity $v_c\simeq 200 \kms$ at $z=2.3$ is $\Sigma =
1400 \kpc^2$. The corresponding proper radius containing the DLAs for
a circular cross section is $\sim 20 f_c^{-1/2} \kpc$, or about 20\% of
the virial radius of the halo if the covering factor $f_c$ is near
unity, and the orbital period is $\sim 5\times 10^8 f_c^{-1/2}$ years.
These clouds can therefore complete no more than a few orbits at this
radius over the age of the universe at $z=2.3$, and would be forming a
halo system that is dynamically supported by random motions, because the
large radius implies that rotation may account for only a small fraction
of the velocity dispersion.

  The massive nature of most DLA host halos helps account for the large
velocity widths observed for the metal lines associated with
DLAs (\cite{1997ApJ...487...73P}, \cite{1998ApJ...495..647H}), as noted
also by \cite{2009MNRAS.397..511B}.
At the same time, the problem of the rate of energy dissipation in DLAs 
(\cite{1999ApJ...519..486M}, \cite{2008ApJ...683..149R}) is solved,
because the long orbital periods imply long collision times among the
DLA clouds, and the available gravitational energy that can be
dissipated in cloud collisions is increased in massive halos.

  Once the cross sections are known, we can infer the fraction of
baryons present in DLA clouds if we make the simple assumption that the
distribution of column densities is independent of halo mass, which is
supported by the lack of dependence of our measured bias factor on
column density. From table 1 in \cite{2012A&A...547L...1N}, 
we infer that the mean column density of all DLAs with
$N_{HI} > 2 \times 10^{20} \cm^{-2}$ is $\bar N_{HI} = 7.8\times
10^{20} \cm^{-2}$. In a halo of $10^{12} \msun$ of total mass and
cross section $\Sigma = 1400 \kpc^2$ (and using a hydrogen abundance
by mass $X=0.76)$, this implies that a fraction of 8\% of all the
baryons in the halo are in the form of atomic gas in the DLA clouds.
Models with a narrower halo mass range as mentioned above, with higher
required cross sections, would also require a larger baryon fraction
in DLAs.

  Finally, there may be a possible conflict with observations that
limit the luminosity of galaxies associated with DLAs. In general,
highly luminous galaxies have not been found in the proximity of DLAs,
although there are cases where they are present: for example, a galaxy
with a star formation rate of $25 \msun/{\rm yr}$ was found
coincident with a DLA of very high column density by
\cite{2012A&A...540A..63N}. This star formation rate is just as expected
for a $10^{12}\msun$ halo dominated by a central galaxy, if as much
as half of the baryons in the halo are turning to stars in a central
galaxy over the age of the universe at $z=2.3$. The lack of luminous
galaxies in most other cases may have various explanations:
the star formation rate in a massive halo may be distributed over
several satellite galaxies, or maybe underestimated observationally
because of dust absorption. Associated galaxies in massive halos may
often be at rather large impact parameters and may sometimes not be
recognized for this reason. It is also likely that there is a large
scatter in the relations of galaxy luminosity and DLA cross section to
halo mass, and halos with luminous central galaxies may not be the same
ones as halos with large DLA cross sections. We therefore do not think
that our inferred large host halo mass for DLAs is necessarily in
conflict with the rarity of luminous associated galaxies found so far.
Precise observational upper limits on the mean luminosity of associated
galaxies within at least 10 arc seconds of DLAs (corresponding to the
expected halo virial radii) would be useful.

%\pp{
%      If I follow what you say, we should find a massive galaxy close to
%      most DLAs at z=2.5. The distance should be something like 25 kpc
%      therefore about 3 arcsec (far enough from the QSOs).
%      Not only this galaxy should be massive but also should have intense
%      SF. I don't think we can say this from observations. You should
%      therefore be more careful in your writting.
%      Note also that paper nb 58 (Haehnelt et al. 1998) indeed conclude
%      on high mass because of velocities. However Ledoux et al. (1998,
%      A\&A, 337, 51)
%      contradicted these conclusions from detailed analysis of
%      absorption profiles.
%}

\section{Conclusions}

  The first detection of the cross-correlation of DLAs with the \lya
forest absorption on nearby lines of sight is presented in this paper.
The cross-correlation is well matched by linear theory predictions
and shows strong evidence for the presence of redshift space distortions.
We therefore believe that the excess \lya absorption
detected around DLAs is explained by the average large-scale
gravitational mass inflow around the host halos of DLAs that is
predicted when large-scale structure grows by gravitational
evolution. We have used the amplitude of the cross-correlation to infer
the bias factor of DLAs, related to the mass distribution of the host
halos.

  The new observational constraint of the bias factor of DLAs, together
with the increasingly accurate determinations of the DLA rate of
incidence, lead us to the conclusion that a majority of DLAs at
$z \sim 2$ to 3 arise in giant halos of atomic gas clouds around typical
galaxies in dark matter halos of masses $\sim 10^{12} \msun$. A simple
model that
works well is that DLAs are present in all halos able to accrete
photoionized gas ($M > 10^9 \msun$), with a cross section $\Sigma(M)
\propto M^{\alpha}$, with $\alpha \simeq 1.1$ when $\beta_F=1$, and
slightly steeper for larger values of $\beta_F$. If the mean DLA column
density is independent of halo mass, the value $\alpha=1$ corresponds to
the case where a fixed fraction of the halo mass is present in the form
of gas in the DLAs, so the observations do not require a
large increase of the fraction of mass in DLAs with halo mass. However,
the observations clearly imply that this gas in DLAs must be enormously
extended in massive halos, with characteristic proper radii
$\sim 20 f_c^{-1/2} \kpc (M/10^{12} \msun)^{\alpha/2}$
which are an increasing fraction of the virial radius as the halo mass
increases.

  This leads to a very different picture of high-redshift galaxies
compared to their local counterparts: typical galaxies at $z>2$ are
surrounded by atomic gas in the halo with column densities comparable to
those in present galactic disks, but covering a much larger area and
containing $\sim 10\%$ of all the baryons in the halo, with kinematics
that can only be dominated by random motions instead of rotation. This
is clearly a fundamental fact that must be accounted for in any models
of galaxy formation.

  The cross-correlation with the \lya forest can be used for other
objects, such as quasars and metal-line systems, to study their relation
with their large-scale environment and infer their bias factor. In the
case of DLAs, we hope that their bias factor can be measured as a
function of metallicity as soon as a rough measure of metal abundance
can be obtained, which is difficult because of the low signal-to-noise
of most of the BOSS spectra, the saturation of metal
lines, and the ionization correction uncertainties. This would set
constraints on the halo mass - metallicity relation and the
correspondence to the mass-metallicity relation for galaxies.

%\jh{I have only one substantive comment about your conclusions about the mass 
%scale of DLAs. It seems to me the halo mass scale that you derive is too high, 
%and inconsistent with studies that have tried to identify DLAs in emission. 
%I would be happy to briefly chat with you on the phone about this. I'd like to 
%understand how robust your measurement of the mass-scale is, although I 
%realize a full parameter study is beyond the scope of this paper. 
%But it may be worth emphasizing in your conclusions the tension with other 
%observational studies which deduce that DLAs tend to have small cross-sections
%and faint counterparts.}

\begin{acknowledgments}

We would like to thank Joseph Hennawi, Patrick McDonald, Matt McQuinn, 
Uros Seljak, An\v{z}e Slosar and Matteo Viel for many illuminating discussions.

This research used resources of the National Energy Research Scientific 
Computing Center (NERSC), which is supported by the Office of Science of 
the U.S. Department of Energy under Contract No. DE-AC02-05CH11231, and
was also supported by Spanish grants AYA2009-09745 and CSD2007-00060.
 
Funding for SDSS-III has been provided by the Alfred P. Sloan
Foundation, the Participating Institutions, the National Science
Foundation, and the U.S. Department of Energy Office of Science.
The SDSS-III web site is http://www.sdss3.org/.

SDSS-III is managed by the Astrophysical Research Consortium for the
Participating Institutions of the SDSS-III Collaboration including the
University of Arizona,
the Brazilian Participation Group,
Brookhaven National Laboratory,
University of Cambridge,
Carnegie Mellon University,
University of Florida,
the French Participation Group,
the German Participation Group,
Harvard University,
the Instituto de Astrofisica de Canarias,
the Michigan State/Notre Dame/JINA Participation Group,
Johns Hopkins University,
Lawrence Berkeley National Laboratory,
Max Planck Institute for Astrophysics,
Max Planck Institute for Extraterrestrial Physics,
New Mexico State University,
New York University,
Ohio State University,
Pennsylvania State University,
University of Portsmouth,
Princeton University,
the Spanish Participation Group,
University of Tokyo,
University of Utah,
Vanderbilt University,
University of Virginia,
University of Washington,
and Yale University.

\end{acknowledgments}

\bibliography{cosmo,cosmo_preprints}
\bibliographystyle{revtex}

\appendix
\section{Effect of the Mean Transmission Correction on the cross-correlation}
\label{app:correc}

In section \ref{sec:mfc} the Mean Transmission Correction (MTC) has been
introduced to reduce the variance in the total absorption of each
individual spectrum. The MTC consists of renormalizing the transmitted
fractions according to equations (\ref{eq:aqfac}) and (\ref{eq:fimtc}).
This correction is nearly equivalent to requiring the mean perturbation
$\delta_{Fi}$ to be zero when averaged over each individual spectrum,
except for the fact that the values of $\bar F$ after the MTC are not
exactly equal to $\bar F_e$, as shown in figure \ref{fig:mf}.
%we show that the continuum fitted spectra have an
%extra source of variance that could be explained by an error in the amplitude
%of the continua, with a r.m.s. of $\sim 16\%$, probably due to 
%spectro-photometric errors in the BOSS spectra.

%For this reason we multiply each continuum by a constant factor to force 
%each spectrum to satisfy $\left< \delta_F \right> = 0$ when averaged over 
%the whole forest.

The correction was also applied for the study of the \lya 
correlation function in \cite{2011JCAP...09..001S} to reduce the
variance arising from spectrophotometric errors. As noted there,
the correction eliminates correlated absorption at all scales,
systematically biasing the measured correlation function. A method
to correct for this systematic bias when fitting a theoretical model
to the observed correlation function was described in Appendix A of
\cite{2011JCAP...09..001S}, which was tested with a set of mocks of \lya
spectra.
Here, we derive an equivalent correction for the case of the
cross-correlation of the \lya forest with DLAs, or with any
other objects with measured redshifts. For the MTC correction used
in this paper,
we consider only the simple case where all pixels have the same
weight and instrumental noise is ignored, although the correction can
be generalized to other cases. We also assume that the average of the
corrected $\delta_{Fi}$ in each spectrum is exactly zero.

  Let $\xi_{DF}(\sigma,\pi)$ be the true cross-correlation as a function
of the comoving separation between the DLA and the \lya pixel in the
transverse and parallel directions in redshift space, $\sigma$ and
$\pi$. We consider a specific \lya forest spectrum that extends from
$\pi_1$ to $\pi_2$ in the parallel separation from the DLA. The
corrected cross-correlation $\xi^{(q)}_{DF}(\sigma,\pi)$, measured
after the average of $\delta_F$ has been subtracted from the \lya
spectrum, is obtained by simply subtracting its value averaged over the
range from $\pi_1$ to $\pi_2$,
\begin{equation}
 \xi_{DF}^{(q)}(\sigma,\pi) = 
   \xi_{DF} (\sigma,\pi) - \frac{1}{\pi_2-\pi_1} \int_{\pi_1}^{\pi_2}
 d\pi^\prime \, \xi_{DF} (\sigma,\pi^\prime) ~.
 \label{eq:corr}
\end{equation}
The superindex $(q)$ on the corrected cross-correlation indicates that
this function depends on the specific quasar spectrum and DLA, through
the variables $\pi_1$ and $\pi_2$.

%  A \lya absorption spectrum measured in a quasar at redshift $z_q$ 
%starts and ends at observed wavelengths $\lambda_1(1+z_q)$ and
%$\lambda_2(1+z_q)$, with $\lambda_1 = 1041 \, {\rm \AA}$ and 
%$\lambda_2 = 1185 \, {\rm \AA}$ for the analysis presented in this
%paper. We neglect the fact that quasars with $z < 2.4$ have a \lya
%spectrum starting at a longer wavelength than $\lambda_1(1+z_q)$ because
%of the wavelength range of the BOSS spectrograph. For a DLA at redshift
%$z_D$, with its absorption line centered at wavelength
%$\lambda_\alpha (1+z_D)$, the comoving parallel separations $\pi_1$ and
%$\pi_2$ can be expressed, to first order in the wavelength differences
%from the DLA, as
%\begin{equation}
% \pi_1 = {c\over H(z_D)}\, {\lambda_1(1+z_q) - \lambda_\alpha (1+z_D)
%           \over \lambda_\alpha } ~, \qquad
% \pi_2 = {c\over H(z_D)}\, {\lambda_2(1+z_q) - \lambda_\alpha (1+z_D)
%           \over \lambda_\alpha } ~.
%\end{equation}
%Defining $R \equiv c(z_q-z_D)/H(z_D)$ (which is approximately the
%comoving separation between the quasar and the DLA), we obtain
%\begin{equation}
% \pi_1 = {R \lambda_1\over \lambda_\alpha} - \frac{c(1+z_D)}{H(z_D)} 
%		\frac{\lambda_\alpha-\lambda_1}{\lambda_\alpha} ~, \qquad
% \pi_2 = {R \lambda_2 \over \lambda_\alpha} - \frac{c(1+z_D)}{H(z_D)} 
%		\frac{\lambda_\alpha-\lambda_2}{\lambda_\alpha} ~.
% \label{eq:pi12}
%\end{equation}

  A \lya absorption spectrum measured in a quasar at redshift $z_q$ 
starts and ends at absorption redshifts 
$(1+z_1) = \lambda_1(1+z_q)/\lambda_\alpha$ and
$(1+z_2) = \lambda_2(1+z_q)/\lambda_\alpha$, with 
$\lambda_1 = 1041 \, {\rm \AA}$ and $\lambda_2 = 1185 \, {\rm \AA}$ for 
the analysis presented in this paper. 
We neglect the fact that quasars with $z < 2.4$ have a \lya
spectrum starting at a higher redshift than $z_1$ because
of the wavelength range of the BOSS spectrograph. 
For a DLA at redshift $z_D$, the values of $\pi_1$ and $\pi_2$
for this particular spectrum are the comoving separation from $z_D$ to
$z_1$ and $z_2$.
The correction to apply (second term on the right hand side of equation 
\ref{eq:corr}) to a given pair DLA-spectrum is then a function 
of $z_D$ and $z_q$.
Defining $R$ as the comoving separation between $z_D$ and $z_q$, we can 
express $\pi_1$ and $\pi_2$ as a function of $R$ and $z_D$.
%Assuming that all the systems are in our central redshift $z_D=2.3$ allows 
%us to approximate the correction as a function of $R$ only.

In practice we calculate the correction as a function of $R$ only, by 
considering that all our systems are at redshift $z_D=2.3$, 
which is the median redshift of the \lya forest pixels
weighted by their contribution to our cross-correlation measurement,
and we neglect the changes with redshift except when we measure the
cross-correlation at different redshift intervals in section \ref{ss:zevol}.
In order to compute the integral in the right hand side of equation 
\ref{eq:corr} we use the linear theory described in section \ref{sec:method}.

\begin{figure}[h!]
  \begin{center}
   \includegraphics[scale=0.6, angle=-90]{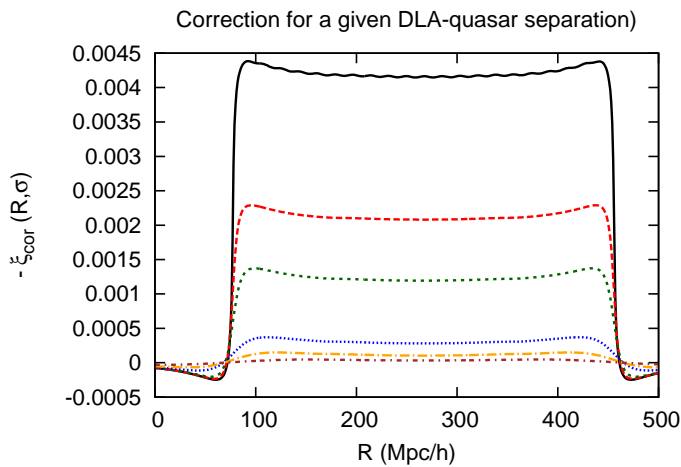}
  \end{center}
  \caption{The correction $\xi_{cor}(\sigma, R)$ as a function of DLA-quasar 
    radial separation $R$, for different values of $\sigma$ 
    (from top to bottom): ($1, 5, 10, 30, 50, 100$) comoving $\hmpc$.
  }
  \label{fig:correct} 
\end{figure}

  The correction to the cross-correlation function is defined as
%\kg{in the integral of the right-hand side, did you have to assume a linear 
%theory model for $\xi_{DF}$? 
%If so, you should state it explicitly in the text.}
%\jm{Figure 16 seems to be plotting Delta xi instead of - Delta xi?
%Changing to the new notation I think the vertical axis should just be
%xicor(sigma,R), but please check that the signs are right...}
\begin{equation}
  \xi_{cor}(\sigma,R)\equiv \xi_{DF}(\sigma,\pi) -
 \xi_{DF}^{(q)}(\sigma,\pi) =
   \frac{1}{\pi_2-\pi_1} \int_{\pi_1}^{\pi_2}
 d\pi^\prime \, \xi_{DF} (\sigma,\pi^\prime) ~.
\label{eq:xicor}
\end{equation}
%using equations (\ref{eq:pi12}). 
This correction depends on $R$, and on
the fixed wavelengths $\lambda_1$ and $\lambda_2$, but not on $\pi$.
The function $\xi_{cor}$ is
shown in figure \ref{fig:correct} as a function of $R$ for several
values of $\sigma$, evaluated at $z_D=2.3$
%\jm{Is this the redshift at which you evaluated the expressions you
%had as a function of z? I have these new terms in equations
%(\ref{eq:pi12}) of the ratio of lambdas which I suspect are small and
%will not change the results very much, but I think these are more
%consistent, referring everything to the frame of the DLA}.

  The correction is symmetric with respect to a sign change of $R-R_c$,
where $R_c \simeq 265 \hmpc$ is the value of $R$ when the
DLA redshift is at the center of the forest, i.e., $\pi_1 = - \pi_2$.
The absolute value of the correction drops sharply for
$R > 455 \hmpc$ or $R < 75 \hmpc$ (or equivalently, $\pi_1 > 0$ or
$\pi_2 < 0$, respectively), when the DLA redshift is outside the
\lya forest region of the quasar and the strongest contribution to
$\xi_{cor}$ near $\pi=0$ is outside the range of the integral in
equation (\ref{eq:xicor}).

  To obtain the average effect on the cross-correlation when measured
over our entire sample of DLAs and quasar spectra, we define the
function $p(R | \pi)$ to be the conditional probability of $R$, given
that a \lya forest pixel is present at a comoving parallel separation
$\pi$ from a DLA. This conditional probability depends in a complex
way on the redshift distribution of DLAs and quasars in our sample,
but can easily be computed. The final correction to the
cross-correlation is
\begin{equation}
 \Delta \xi (\sigma, \pi) = \int dR ~ p(R | \pi) \,
			\xi_{cor}(\sigma,R) ~ .
\end{equation}

\begin{figure}[h!]
  \begin{center}
   \includegraphics[scale=0.6, angle=-90]{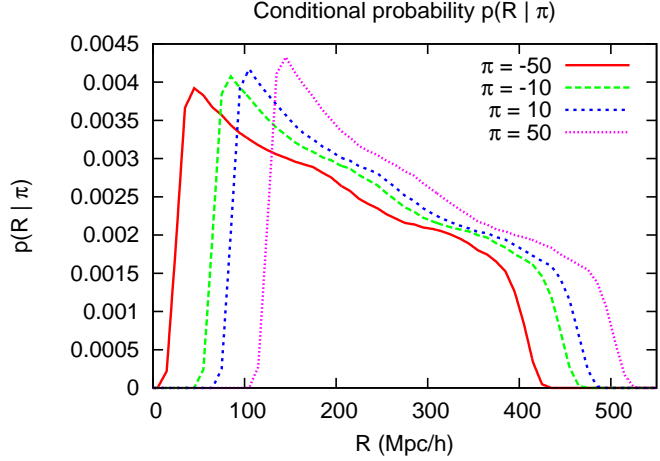}
  \end{center}
  \caption{Conditional probability of $R$ (DLA-quasar radial separation) 
    given $\pi$ (radial separation between the \lya pixel and the DLA).
    $p(R | \pi)$ is shown as a function of $R$ for different values of 
    $\pi$, computed for our sample of DLAs and quasar spectra.
  }
  \label{fig:prob} 
\end{figure}

\begin{figure}[h!]
  \begin{center}
   \includegraphics[scale=0.6, angle=-90]{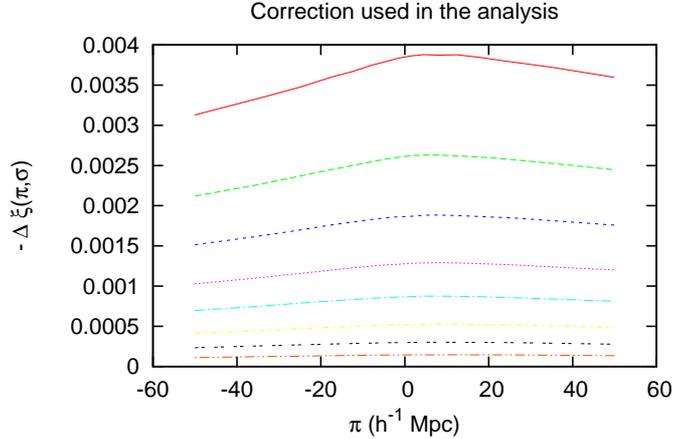}
  \end{center}
  \caption{Final correction applied to the theoretical cross-correlation and 
   used when fitting the DLA bias. The correction is shown as a function of 
   radial separation $\pi$ for different bins of $\sigma$ (from top to bottom):
    $1 < \sigma < 4 \hmpc$,
    $4 < \sigma < 7 \hmpc$,
    $7 < \sigma < 10 \hmpc$,
    $10 < \sigma < 15 \hmpc$,
    $15 < \sigma < 20 \hmpc$,
    $20 < \sigma < 30 \hmpc$,
    $30 < \sigma < 40 \hmpc$, and
    $40 < \sigma < 60 \hmpc$.
  }
  \label{fig:correction} 
\end{figure}

In figure \ref{fig:prob} we plot the conditional probability $p(R | \pi)$
measured for the distribution of DLA and quasar redshifts in our
sample, for four different values of $\pi$. This conditional probability
is approximately invariant if $R-\pi$ is fixed, but not exactly so. The 
quantity $p(R | \pi)$ is not symmetric under a change of sign of $\pi$, so
the correction 
$\Delta \xi (\sigma,\pi)$ is also asymmetric. This effect is seen in 
figure \ref{fig:correction} which shows the correction we apply as a function
of $\pi$, for the different bins of $\sigma$ used in our analysis
(increasing from top to bottom; see section \ref{ss:cov}).

In figure \ref{fig:MFC} (section \ref{sec:results}) we plot the measured 
cross-correlation function, together with the best fit model, with and 
without the correction. The effect of this correction is clearly not
crucial for our investigation, but will be more important for 
studies aiming to measure the cross-correlation on larger scales. 
We expect this to be more important when measuring the cross-correlation of
quasars and the \lya absorption, because of the larger number of quasars
and their higher bias factor.

\section{Optimal quadratic estimator}
\label{app:estim}

In section \ref{sec:method} our simple method to estimate the 
cross-correlation of \lya absorption and the density of DLAs is described.
In this appendix the precise optimal quadratic estimator for this
cross-correlation is derived, assuming that the fields are Gaussian,
and the assumptions required to obtain the simple estimator of equation
(\ref{eq:xiA}) are described.

\subsection{Definitions and notation}

The estimator of the fluctuation $\delta_{Fi}$ at each pixel $i$ in
the whole set of quasar spectra is
\begin{equation}
 d_i^F \equiv \frac{f_i}{\bar f_i } - 1 = \delta_{Fi} + \frac{N_i}{\bar f_i } ~,
\end{equation}
where $f_i$ is the measured physical flux, $\bar f_i = C_i \bar F_i$ is
the averaged value of $f_i$ at redshift $z_i$, 
and $N_i$ is the noise, which is assumed to have a Gaussian distribution
with variance $\left< N_i^2 \right>$. The covariance matrix of this
estimator is, neglecting any errors from the continuum fit,
\begin{equation}
 C^F_{ij} \equiv \left< d_i^F \, d_j^F \right> = \xi_F({\bf r}_{ij})
	+ \frac{\left< N_i^2\right> }{\bar f_i^2} \delta_{ij}^K ~, 
\end{equation}
where $\xi_F({\bf r}_{ij})$ is the correlation function of the \lya
fluctuation in the two pixels separated by ${\bf r}_{ij}$ and $\delta^K$
is the Kronecker delta function.
This matrix is the same as $C_{ij}$ in equation \ref{eq:cij}, except
that the continuum fit errors are not included in the analysis of this
appendix.

  The other ingredient that we need to model is the overdensity of detected 
DLAs. Our discussion is completely general and applicable to any set
of detected objects that are cross-correlated with the \lya forest, and we
refer to these detected objects in a specific survey as systems.
We divide the survey volume in redshift space into $N_c$ small cells of volume
$V_c$, such that the number $h_i$ of systems detected in each cell 
$i$ is either 0 or 1, and the mean number of systems per cell, $\bar h_i$,
is much smaller than one. The overdensity in a cell, $\delta_{hi}$, is 
defined according to
\begin{equation}
 h_i = \bar h_i  \left[ 1 + \delta_{hi} \right] + \epsilon_i ~,
\end{equation}
where $\epsilon_i$ is the shot-noise, with 0 mean and a variance
$\sigma_{\epsilon i}^2 = \bar h_i $.
%The field is 
%defined in the whole volume of the survey, so only a small fraction of the
%cells will have a system, or in other words, $\bar h_i$ will always 
%be extremely small.
The estimator of the density fluctuation from the measured systems in
the survey is
%Assuming that we know the mean number density $\bar h_i$ at each cell, we can 
%compute a noisy estimate of the fluctuations $\delta_{hi}$:
\begin{equation}
 d_i^H \equiv \frac{h_i}{\bar h_i } - 1 = \delta_{hi} 
				+ \frac{\epsilon_i}{\bar h_i} ~,
\end{equation}
and its associated covariance matrix $C^H_{ij}$ is
\begin{equation}
 C^H_{ij} \equiv \left< d_i^H d_j^H \right> = \xi_H({\bf r}_{ij}) 
	+ \frac{1}{\bar h_i } \delta_{ij}^K ~, 
\end{equation}
where $\xi_H({\bf r}_{ij})$ is the correlation function of detected systems. In
general, these detected systems may be associated with dark matter halos,
and then $\xi_H$ is the correlation function of halos weighted with
their probability of yielding a detected system in the survey as a function
of halo mass. This probability is the mean number of objects per halo when
the detected systems are galaxies or quasars, or the cross section for the
case of DLAs or metal absorption lines.

  It is convenient to define a global data vector,
\begin{equation}
 \mathbf d = (\mathbf d^F, \mathbf d^H) ~,
\end{equation}
of dimension $N_F + N_c$, where $N_F$ is the total number of \lya pixels and
$N_c$ the number of cells. The covariance matrix of the global data vector can
be written as
\begin{equation}
 \mathcal C = \left( \begin{array}{cc}
	C^F & C^X \\
	(C^{X})^T & C^H \end{array} \right) ~, 
\end{equation}
where $C^X$ is the covariance of the two fields,
\begin{equation}
 C^X_{ij} = \left< d_i^F d_j^H \right> = \xi_X({\bf r}_{ij})~,
\end{equation}
and $\xi_X({\bf r}_{ij})$ is the quantity we want to obtain, the
cross-correlation function of the \lya absorption and the overdensity
of systems. 
%This covariance matrix can be decomposed as follows:
%\begin{equation}
% \mathcal C = \mathcal C^F + \mathcal C^H + \mathcal C^X ~,
%\end{equation}
%where we have defined the matrices
%\begin{align} 
% \mathcal C^F & \equiv \left( \begin{array}{cc}
%	C^F & 0 \\
%	0 & 0 \end{array} \right) \nonumber \\
% \mathcal C^H & \equiv \left( \begin{array}{cc}
%	0 & 0 \\
%	0 & C^H \end{array} \right) \nonumber \\
% \mathcal C^X & \equiv \left( \begin{array}{cc}
%	0 & C^X \\
%	(C^X)^T & 0 \end{array} \right) ~. 
%\end{align}
In this appendix the calligraphy letters $ \mathcal C$ are used for
matrices with dimension $(N_F+N_c)\times(N_F+N_c)$. Generally, we are
interested in measuring the cross-correlation in some bins in the
separation (both the transverse and parallel components). Designating
the bin number with a subindex $\alpha$, and the model value of the
cross-correlation in this bin as $\xi_{\alpha}$, the matrix $C^X$ can
be expressed as:
\begin{equation}
 C^X_{ij} = M^\alpha_{ij} \, \xi_\alpha ~,
 \label{eq:X}
\end{equation}
where $M_{ij}^\alpha$ is a matrix with non-zero elements only when the 
separation between the pixel $i$ and the cell $j$, $r_{ij}$, lies inside
the bin $r_\alpha$.

  The derivative of the covariance matrix $\mathcal C$ with respect to the 
cross-correlation $\xi_\alpha$ in one of the bins is
\begin{equation}
 \mathcal C_{,\alpha} %= \mathcal M^\alpha 
	= \left( \begin{array}{cc}
	0 & M^\alpha \\
	M_\alpha^T & 0 \end{array} \right) ~,
\end{equation}
and the covariance matrix is decomposed as
\begin{equation}
 \mathcal C = \mathcal C_{,\alpha} \xi_\alpha + \mathcal N  ~, 
\end{equation}
with % $\mathcal N = \mathcal C^F + \mathcal C^H$.
\begin{equation}
 \mathcal N 
        = \left( \begin{array}{cc}
        C^F & 0  \\
        0 & C^H \end{array} \right) ~.
\end{equation}

\subsection{General solution}

An unbiased quadratic estimator can generally be written as 
(\cite{1997PhRvD..55.5895T},\cite{1998ApJ...506...64S}):
\begin{equation}
 \hat \xi_\alpha = \left( A^{-1} \right)_{\alpha \beta} \, E_\beta ~,
\end{equation}
with 
\begin{equation}
 A_{\alpha \beta} = \rm Tr \left[ \mathcal W \, \mathcal C_{,\alpha} \,
	\mathcal W \, \mathcal C_{,\beta} \right] ~,
\end{equation}
and
\begin{equation}
 E_\beta = \rm Tr \left[ \mathcal W \, \mathcal C_{,\beta} \, 
	\mathcal W \left( \mathbf d \mathbf d^T - \mathcal N \right) \right] ~,
\end{equation}
where $\mathcal W$ is any symmetric matrix that we use to weigh our data 
vector.

For a Gaussian data vector, the optimal estimator is
$\mathcal W = \mathcal C^{-1}$, and $A$ is the Fisher 
matrix. Any other weight matrix $\mathcal W$ still yields an unbiased
estimator, but the estimator becomes more optimal as 
$\mathcal W$ approaches $\mathcal C^{-1}$.

The expected value of the estimator is
\begin{equation}
 \left< \hat \xi_\alpha \right> = \left( A \right)^{-1}_{\alpha \beta} 
       \left< E_\beta \right>
    = \left( A \right)^{-1}_{\alpha \beta} 
       \rm Tr \left[ \mathcal W \, \mathcal C_{,\alpha} \, 
	\mathcal W \left( \mathcal C - \mathcal N \right) \right]
    = \left( A \right)^{-1}_{\alpha \beta} \, A_{\beta \gamma} \, \xi_\gamma
    = \xi_\alpha  ~.
\end{equation}

\subsection{Approximations} 

Inverting covariance matrices is a computationally demanding process. Since we
have thousands of spectra with hundreds of pixels, the matrix $C^F$ is in
practice impossible to invert, not to mention the matrix $\mathcal C$.
Here, we explain the different approximations that one can do in order to get
an estimator that is sub-optimal but realistic.

\subsubsection*{Noise-dominated regime}

The first approximation is to consider that the contribution of the 
cross-correlation to the covariance is negligible, i.e., we can approximate
$\mathcal C^{-1} \sim \mathcal N^{-1}$. The weight matrix is then
\begin{equation}
\mathcal W = \mathcal N^{-1} = \left( \begin{array}{cc}
(C^F)^{-1} & 0 \\
0 & (C^H)^{-1} \end{array} \right) ~ .
\end{equation}

The product $\mathcal W \, \mathcal C_{,\alpha}$ is:
\begin{equation}
  \mathcal W \,  \mathcal C_{,_\alpha} = \left( \begin{array}{cc}
	0 & (C^F)^{-1} \, M_\alpha \\
	(C^H)^{-1} \, M_\alpha^T  & 0 \end{array} \right) ~.
\end{equation}

We can now compute $E_\alpha$ and $A_{\alpha \beta}$ for this weighting:
\begin{equation}
 E_\alpha = \rm Tr \left[ \mathcal N^{-1} \, \mathcal C_{,\alpha} \, 
	\mathcal N^{-1} \left( \mathbf d \mathbf d^T 
	- \mathcal N \right) \right] = 
    2 \rm Tr \left[ (C^F)^{-1} \, M_\alpha (C^H)^{-1} 
	\mathbf d_H \mathbf d_F^T \right] ~,
\end{equation}
and
\begin{equation}
 A_{\alpha \beta} = \rm Tr \left[ \mathcal N^{-1} \, \mathcal C_{,\alpha} \, 
	\mathcal N^{-1} \, \mathcal C_{,\beta}\right] = 
    2 \rm Tr \left[ (C^F)^{-1} \, M_\alpha \, (C^H)^{-1} \, M^T_\beta \right]~.
\end{equation}

\subsubsection*{Rare-objects regime}

In this case the covariance matrix of the systems overdensity, $C^H$, will be 
dominated by the shot-noise, and we can do a second approximation for 
part of the weighting matrix:
\begin{equation}
 C^H_{ij} \sim \bar h_i^{-1} \delta_{ij}^K ~. 
\end{equation}

We can now simplify $E_\alpha$ even further:
\begin{align} 
 E_\alpha & = 2 ~ \rm Tr \left[ (C^F)^{-1} \, M_\alpha \, (C^H)^{-1} 
	~ \mathbf d_H ~ \mathbf d_F^T \right]  \\ \nonumber
 & = 2 \sum\limits_{i=0}^{N_F} \sum\limits_{j=0}^{N_F} 
     \sum\limits_{k=0}^{N_c} \sum\limits_{l=0}^{N_c} 
     (C^F)^{-1}_{ij} M^\alpha_{jk} (C^H)^{-1}_{kl} d^H_l d^F_i \\ \nonumber
 & = 2 \sum\limits_{k=0}^{N_H} 
     \sum\limits_{i=0}^{N_F} \sum\limits_{j=0}^{N_F} 
     M^\alpha_{jk} F^{-1}_{ij} d^F_i ~, 
\end{align}
where we have used that 
$(C^H)^{-1}_{ij} \, d_j^H \sim \bar h_i \, d_i^H = (h_i - \bar h_i)$ 
equals to one when there is a system in the cell and zero otherwise, 
because the constant $\bar h_i$
averages to zero when cross-correlated with the \lya field $d_F$.
%(it has 0 mean).
The first summations in $k$ and $l$ are over the $N_c$ cells in 
the volume, while the last summation in $k$ is only over those $N_H$ cells 
with a system on it.

  The matrix $A_{\alpha \beta}$ can also be simplified in a similar manner:
\begin{align} 
A_{\alpha \beta} & = 
 2 ~ {\rm Tr} \left[ (C^F)^{-1} \, M_\alpha \, (C^H)^{-1} \, 
	M_\beta^T \right] \\ \nonumber
 & = 2 \sum\limits_{i=0}^{N_F} \sum\limits_{j=0}^{N_F} 
     \sum\limits_{k=0}^{N_c} \sum\limits_{l=0}^{N_c} 
     (C^F)^{-1}_{ij} \, M^\alpha_{jk} \, (C^H)^{-1}_{kl} \, 
	M^\beta_{il} \\ \nonumber
 & = 2 \sum\limits_{k=0}^{N_H} 
     \sum\limits_{i=0}^{N_F} \sum\limits_{j=0}^{N_F} 
     M^\alpha_{jk} \, (C^F)^{-1}_{ij} \, M^\beta_{ik} ~.
\end{align}

  The last expressions can be rewritten in a more compact form by defining a 
new matrix $M_\alpha^\prime$, which has dimension $N_H \times N_F$ 
(instead of $N_c \times N_F$ for the matrix $M_\alpha$) because it is
defined only in the cells where a system is detected,
\begin{equation}
 E_\alpha = 2 ~ {\rm Tr} \left[ (C^F)^{-1} \, M^\prime_\alpha 
	~ \mathbb{1} ~ \mathbf{1} ~ \mathbf d_F^T \right] ~,
\end{equation}
and
\begin{equation}
 A_{\alpha \beta} = 2 ~ {\rm Tr} \left[ (C^F)^{-1} \, M^\prime_\alpha 
	~ \mathbb{1} \, M_\beta^{\prime T} \right] ~.
\end{equation}
We have included the identity matrix $\mathbb{1}$ and vector $\mathbf{1}$,
both with dimension $N_H$, to ease the reference to the previous expressions. 

\subsubsection*{Independent-spectra regime}

Pixels of two different spectra are only weakly correlated, especially in 
the limit of low quasar density where the typical separation between 
quasar lines of sight is large. In this limit, an additional approximation
can be made for the weighting matrix in which only pixels from the same
spectrum are correlated. Then,
the correlation matrix $C^F$ and its inverse are block 
diagonal, with a block $C^F_q$ for each quasar:
\begin{equation}
 C^F = \left( \begin{array}{cccccc}
	C^F_1 & 0 & ... & 0 & ... & 0 \\
	0 & C^F_2 & ... & 0 & ... & 0 \\
	... & ... & ... & ... & ... & ... \\ 
	0 & 0 & ... & C^F_i & ... & 0 \\
	... & ... & ... & ... & ... & ... \\ 
	0 & 0 & ... & 0 & ... & C^F_{N_q} \end{array} \right) ~.
\end{equation}
Instead of having to invert a matrix of size $N_F \times N_F$ (with
$N_F \sim 5 \times 10^7$) we need only $N_q$ matrices of size 
$N_p \times N_p$, where $N_q \sim 10^5$ is the number of spectra and 
$N_p \sim 500$ is the number of pixels in a typical spectrum.

\subsubsection*{Independent-pixels regime}

Our final approximation is that each pixel in the spectrum is independent. 
In this case the matrix $C^F$ is no longer block diagonal, but diagonal:
\begin{equation}
 C^F_{ij} = \xi^F_{ij} + 
 \frac{\left< N_i\right>^2}{\bar f_i^2} \, \delta_{ij}^K 
         \sim \sigma_{Ti}^2 \, \delta_{ij}^K ~,
\end{equation}
where $\sigma_{Ti}^2 = \sigma_{Fi}^2 + \left< N_i \right>^2/{\bar f_i}^2$ 
is the total variance in the pixel, including the \lya intrinsic variance.

The expressions for $E_\alpha$ and $A_{\alpha \beta}$ are now equal to
\begin{equation} 
 E_\alpha 
    = 2 \sum\limits_{k=0}^{N_H} 
     \sum\limits_{i=0}^{N_F} \sum\limits_{j=0}^{N_F} 
     M^\alpha_{jk} (C^F)^{-1}_{ij} \, d^F_i 
    = 2 \sum\limits_{k=0}^{N_H} 
     \sum\limits_{i=0}^{N_F} M^\alpha_{ik} \, w_i \, d^F_i ~, 
\end{equation}
and
\begin{equation} 
A_{\alpha \beta} 
  = 2 \sum\limits_{k=0}^{N_H} 
     \sum\limits_{i=0}^{N_F} \sum\limits_{j=0}^{N_F} 
     M^\alpha_{jk} (C^F)^{-1}_{ij} \, M^\beta_{ik} 
  = 2 \sum\limits_{k=0}^{N_H} 
     \sum\limits_{i=0}^{N_F} M^\alpha_{ik} \, w_i \, M^\beta_{ik} ~,
\end{equation}
where we have defined the pixel weight $w_i = \sigma_{Ti}^{-2}$. 

We now see that the estimator in equation \ref{eq:xiA} corresponds
to this case if we use only the diagonal part of $A_{\alpha \beta}$:
\begin{equation}
 \hat \xi_\alpha 
     = A^{-1}_{\alpha \alpha} E_\alpha 
     = \frac{\sum\limits_{k=0}^{N_H} \sum\limits_{i=0}^{N_F} 
        M^\alpha_{ik} w_i \, d^F_i }
        {\sum\limits_{k=0}^{N_H} \sum\limits_{i=0}^{N_F} 
	M^\alpha_{ik} \, w_i \, M^\alpha_{ik} } 
     = \frac{\sum\limits_{k=0}^{N_H} \sum\limits_{i \in \alpha} w_i \, d^F_i }
        {\sum\limits_{k=0}^{N_H} \sum\limits_{i \in \alpha} w_i } ~. 
\end{equation}

If wide redshift bins are used with a substantial evolution of 
the signal being present within the bin, the estimator needs to be
modified. In particular, equation \ref{eq:X} should now be
\begin{equation}
 X_{ij} = M^\alpha_{ij} \, \xi_\alpha 
	\, \left(\frac{1+z_{ij}}{1+z_\alpha} \right)^{\gamma} ~,
\end{equation}
where we have assumed that the evolution of the cross-correlation is described
by an amplitude varying as a power-law with the scale factor with an
index $\gamma$.
The simple estimator used in Section \ref{sec:method} needs to be modified
as well, 
by multiplying the weights in equation \ref{eq:w_delta} by the same factor.

  We have not applied this correction for the weights, because the
cross-correlation signal does not evolve strongly with redshift.
In an Einstein-de Sitter universe, or at sufficiently high redshift to
make the dark energy component negligible, the growth 
factor is proportional to the scale factor, and the value of $\gamma$ is 
simply related to the evolution of the bias parameter:
\begin{equation}
 \left( \frac{1+z_{ij}}{1+z_\alpha} \right)^\gamma = 
	\left( \frac{b(z_{ij})}{b(z_\alpha)} \right)^2 \, 
 \left( \frac{1+z_{ij}}{1+z_\alpha} \right)^{-2} ~.
\end{equation}
From the evolution of the line-of-sight \lya forest power spectrum measured
in \cite{2006ApJS..163...80M}, we can assume that the amplitude of the 
\lya clustering evolves with $\gamma_F=3.8$, implying that $b_F$ evolves
as $\left(1+z\right)^{2.9}$. Assuming that the DLA bias is constant over the
redshift range, we obtain a value of $\gamma_X=0.9$ for the cross-correlation.
This leads to a very small change over the narrow redshift range of our
DLA sample.

\end{document}